\documentclass[12pt,preprint]{aastex}

\slugcomment{Accepted for publication in ApJ}

\shorttitle{Mid--Infrared Star Formation Rate Indicators}
\shortauthors{Calzetti et al.}

\begin{document}

\title{The Calibration of Mid--Infrared Star Formation Rate
Indicators.\altaffilmark{1}}

\author{D. Calzetti\altaffilmark{2,3}, R. C. Kennicutt\altaffilmark{4},
C. W. Engelbracht\altaffilmark{5}, C. Leitherer\altaffilmark{3}, B. T. Draine\altaffilmark{6}, L. Kewley\altaffilmark{7},  J. Moustakas\altaffilmark{8},  M. Sosey\altaffilmark{3},  D.A. Dale\altaffilmark{9}, 
K. D. Gordon\altaffilmark{5},  G.X. Helou\altaffilmark{10}, D.J. Hollenbach\altaffilmark{11},  
L. Armus\altaffilmark{10},  G. Bendo\altaffilmark{12}, C. Bot\altaffilmark{10}, 
B. Buckalew\altaffilmark{10},  T. Jarrett\altaffilmark{10},  A. Li\altaffilmark{13},
M. Meyer\altaffilmark{3},  E.J. Murphy\altaffilmark{14}, M. Prescott\altaffilmark{5}, 
M. W. Regan\altaffilmark{3}, G. H. Rieke\altaffilmark{5}, H. Roussel\altaffilmark{15} , 
K. Sheth\altaffilmark{10},  J. D. T. Smith\altaffilmark{5}, M. D. Thornley\altaffilmark{16} ,  
F. Walter\altaffilmark{15}  }

\altaffiltext{1}{Based on observations obtained with the Spitzer Space
Telescope, which is operated by JPL, CalTech, under NASA Contract
1407, and with the NASA/ESA Hubble Space Telescope at the Space
Telescope Science Institute, which is operated by the Association of
Universities for Research in Astronomy, Inc., under NASA contract
NAS5-26555. } 
\altaffiltext{2}{Dept. of Astronomy, University of Massachusetts, Amherst, MA 01003; calzetti@astro.umass.edu}
\altaffiltext{3}{Space Telescope Science Institute, Baltimore, Maryland}
\altaffiltext{4}{Institute of Astronomy, Cambridge University, Cambridge, U.K.}
\altaffiltext{5}{Steward Observatory, University of Arizona, Arizona}
\altaffiltext{6}{Princeton University Observatory, Peyton Hall, Princeton, New Jersey}
\altaffiltext{7}{Institute for Astronomy, University of Hawaii, Hawaii}
\altaffiltext{8}{Department of Physics, New York University, New York}
\altaffiltext{9}{Dept. of Physics and Astronomy, University of Wyoming, Wyoming}
\altaffiltext{10}{Spitzer Science Center, Caltech, California}
\altaffiltext{11}{NASA/Ames Research Center, California}
\altaffiltext{12}{Astrophysics Group, Imperial College, London, U.K.}
\altaffiltext{13}{Dept. of Physics and Astronomy, University of Missouri, Missouri}
\altaffiltext{14}{Dept. of Astronomy, Yale University, Connecticutt}
\altaffiltext{15}{Max Planck Institut f\"ur Astronomie, Heidelberg, Germany}
\altaffiltext{16}{Dept. of Physics and Astronomy, Bucknell University, Pensylvania}

\begin{abstract}
With the goal of investigating the degree to which the mid--infrared emission traces the star 
formation rate (SFR), we analyze Spitzer 8~$\mu$m and 24~$\mu$m data of star--forming regions 
in a sample of 33 nearby galaxies with available HST/NICMOS images in the  
Pa$\alpha$ ($\lambda$1.8756~$\mu$m) emission line. The galaxies are drawn from the SINGS sample, 
and cover a range of morphologies and a factor $\sim$10 in oxygen abundance. Published data on 
local low--metallicity starburst galaxies and Luminous Infrared Galaxies are also included in 
the analysis.  Both the stellar--continuum--subtracted 8~$\mu$m emission and the 24~$\mu$m
emission correlate with the extinction--corrected Pa$\alpha$ line emission, although neither 
relationship is linear.  Simple models of stellar populations and dust extinction and 
emission are able to reproduce the observed non--linear trend of the 24~$\mu$m emission 
versus number of ionizing photons, including the modest deficiency of 24~$\mu$m emission in the 
low metallicity regions, which results from a combination of decreasing dust opacity and dust 
temperature at low luminosities.  Conversely, the trend of the 8~$\mu$m emission as a function of 
the number of ionizing photons is not well reproduced by the same models. The 8~$\mu$m 
emission is contributed, in larger measure than the 24~$\mu$m emission,  by dust heated by 
non--ionizing stellar populations, in addition to the ionizing ones, in agreement with previous 
findings. Two SFR calibrations, one using the 24~$\mu$m emission and  the other using a combination 
of the  24~$\mu$m and H$\alpha$ luminosities \citep{kenn06}, are presented. No calibration is 
presented for the 8~$\mu$m emission, because of its significant dependence on both metallicity and
 environment. The calibrations presented here should be directly applicable to systems dominated 
by on--going star formation.
\end{abstract}


\section{Introduction}

The multi--wavelength galaxy surveys of  unprecedented 
angular resolution recently made available by combined 
space (HST, Spitzer) and ground--based observations  
are providing for the first time the tools to cross--calibrate 
star formation rate (SFR) indicators at different wavelengths, and to test 
the physical assumptions underlying each indicator. 

Easy accessibility has traditionally favored the use of the ultraviolet (UV) stellar continuum 
and of the optical nebular recombination lines as SFR indicators, the former 
mainly in the intermediate--high redshift regime (as it gets redshifted into the 
optical observer frame) and the latter mostly in low--redshift surveys. Both 
indicators only probe the stellar light that {\em emerges from a galaxy unabsorbed 
by dust}. The UV is heavily affected by dust attenuation, and numerous efforts 
have attempted to find general tools to mitigate the effects of dust on this 
important SFR indicator \citep[e.g.,][]{calz94, kenn98, meur99, hopk01, sulli01, buat02, buat05,
bell03, hopk04, sali07}. Cross--calibrations with optical recombination lines and other indicators have 
also attempted to account for the $\sim$10 times or more longer stellar timescales probed by 
 the UV relative to tracers of ionizing photons  
 \citep[e.g.,][]{sulli01, kong04, calz05}. Among the recombination lines, 
H$\alpha$ is the most widely used, due to a combination of its intensity and a 
lower sensitivity to dust attenuation than bluer nebular lines. Although to a much 
lesser degree than the UV, the H$\alpha$ line is still affected by dust attenuation, 
plus is impacted by assumptions on the underlying stellar absorption and 
on the form of the high end of the stellar initial mass function 
\citep[e.g.,][]{calz94, kenn98, hopk01, sulli01, kewl02, rosgon02}.

Infrared SFR indicators are complementary to UV--optical indicators, because they measure 
star formation via the dust--absorbed stellar light that emerges beyond a few $\mu$m. 
Although SFR indicators using the infrared emission had been calibrated during  
the IRAS times \citep[e.g.,][]{pers87, rowa89, sauv92}, interest in the this wavelength 
range  had been rekindled  in more recent times 
by the discovery of submillimeter--emitting galaxy populations at cosmological 
distances \citep[e.g.,][]{smai97,hugh98,barg98,eale99,chap05}. 
In dusty starburst galaxies, the bolometric infrared luminosity L$_{IR}$ 
in the $\sim$3--1100~$\mu$m window is directly proportional to the
SFR \citep[ e.g.,][]{kenn98}. However, even assuming that
most of the luminous energy produced by recently formed stars is
re-processed by dust in the infrared, at least two issues make the use
of this SFR indicator problematic. (1) Evolved, i.e., non-star-forming,
stellar populations also heat the dust that emits in the IR
wavelength region, thus affecting the calibration of SFR(IR) in a
stellar-population-dependent manner \citep[e.g.,][]{pers87,helo86,kenn98}. (2) 
In  intermediate/high redshift studies, the bolometric infrared luminosity is often 
extrapolated from measurements at 
sparsely sampled wavelengths, most often in the sub--mm and radio observer's 
frame \citep[e.g., ][]{smai97,chap05}, and such extrapolations 
are subject to many uncertainties.  

The interest in calibrating monochromatic mid-infrared SFR diagnostics stems from 
their potential application to both the 
local Universe  and intermediate and high redshift galaxies observed with Spitzer and
future infrared/submillimeter missions \citep{daddi05,wu05}. One such application 
is  the investigation of the scaling laws of star formation in the 
dusty environments of galaxy centers \citep{kenn98b,kenn06}. 
The use of monochromatic  (i.e., one band or wavelength) infrared  emission 
for measuring SFRs  offers one definite  advantage over the bolometric infrared luminosity:  it 
removes the need for  highly uncertain extrapolations of the dust spectral energy distribution 
across the full wavelength range.  Over the last few years, a number of efforts have gone 
into investigating the potential use of monochromatic infrared emission for measuring SFRs. 

Early studies employing ISO data have not resolved whether the warm dust and aromatic
bands emission around 8~$\mu$m can be effectively used as a SFR
indicator, since different conclusions have been reached by different
authors.  \citet{rous01} and \citet{forst04} have shown that the emission 
in the 6.75~$\mu$m ISO band correlates with the number of ionizing photons 
(SFR) in galaxy disks and in the nuclear regions of galaxies. Conversely, 
\citet{bose04} have found that the mid--IR emission  
in a more diverse sample of galaxies (types Sa through Im--BCDs) correlates more closely 
with tracers of evolved stellar populations not linked to the current star formation.  Additionally,
\citet{haas02} find that the ISO 7.7~$\mu$m emission is correlated with the 
850~$\mu$m emission from galaxies, suggesting a close relation between the ISO band 
emission and the cold dust heated by the general (non--star--forming) stellar population. 
This divergence of results highlights the multiplicity of sources for the emission at 
8~$\mu$m \citep[e.g.,][]{peete04,tacco05}, as well as the limits in the ISO angular resolution 
and sensitivity for probing a sufficiently wide range of galactic conditions. 

The emission in the
8~$\mu$m and other MIR bands is generally attributed to Polycyclic Aromatic
Hydrocarbons \citep[PAH, ][]{lege84,sell84,alla85,sell90}, large molecules transiently heated
by single UV and optical photons in the general radiation field of galaxies or near B stars
 \citep{li02,haas02,bose04,peete04,wu05,matt05}, and which can be
destroyed, fragmented, or ionized by harsh UV photon fields
\citep{boul88,boul90,helo91,houc04,pety05}.  Spitzer data of the nearby galaxies 
NGC300 and NGC4631 show that 
8~$\mu$m emission highlights the rims of
HII regions and is depressed inside the regions, indicating that the PAH dust is heated in the PDRs surrounding HII regions and is destroyed within the regions \citep{helo04,bend06}. 
Analysis of the mid--IR emission from the First Look 
Survey \citep{fang04} galaxies shows that the correlation between the Spitzer 8~$\mu$m band emission 
and tracers of the ionizing photons is shallower than unity \citep{wu05}, in agreement 
with the correlations observed for HII regions in the nearby, metal--rich, star--forming galaxy NGC5194 
 \citep[M51a][]{calz05}. 
 
The 24~$\mu$m emission is a close tracer of SFR in the dusty 
center of NGC5194 \citep{calz05} and in NGC3031 \citep{gonz06}. The general 
applicability of this monochromatic indicator has so far  been  
explored only for a small number of cases, mostly bright galaxies 
\citep[e.g.,][]{wu05,alon06}. A potential complication is that most of the energy 
from dust  emerges at wavelengths longer than $\sim$40--50~$\mu$m 
\citep[see ][ and references therein]{dale06}. Thus the mid--IR does not trace the 
bulk of the dust emission, and, because it lies on the Wien side of the blackbody spectrum,
could be sensitive to the dust temperature rather than linearly correlating
with source luminosity.

This study investigates the use of the Spitzer IRAC 8~$\mu$m and MIPS
24~$\mu$m monochromatic luminosities as SFR indicators for star
forming regions in a subsample of the SINGS galaxies \citep[SINGS, or the
Spitzer Infrared Nearby Galaxies Survey, is one of the Spitzer Legacy
Programs, ][]{kenn03}.  Star--forming regions  in 
galaxies represent a first stepping--stone for characterizing SFR indicators, as they
can be considered simpler entities than entire galaxies. 

We also extend our analysis to include both new and published 
integrated (galaxy--wide) data on local low--metallicity 
starburst galaxies \citep{enge05} and Luminous Infrared Galaxies \citep[LIRGs,][]{alon06}. 
These  data are used to explore whether the relationships 
derived for the star--forming regions that constitute our main sample are applicable to 
starburst--dominated  galaxies as a whole.  A future paper will investigate the viability of the 
mid--infrared luminosities as SFR tracers for more general classes of galaxies \citep{kenn06b}. 

The Spitzer observations are coupled with near--infrared 
HST/NICMOS observations centered on the Paschen--$\alpha$ hydrogen
emission line (Pa$\alpha$, at 1.8756~$\mu$m), and with ground--based
H$\alpha$ observations obtained by the SINGS project.  The hydrogen
emission lines trace the number of ionizing photons, and the Pa$\alpha$ line 
is only modestly impacted by dust extinction. Furthermore, the Pa$\alpha$ and 
H$\alpha$ lines are sufficiently separated in wavelength that reliable extinction
corrections can be measured \citep{quil01}. Because of its relative insensitivity to 
dust extinction (less than a factor of 2 correction for the typical extinction in our galaxies,  
A$_V\lesssim$5~mag), Pa$\alpha$ represents a nearly unbiased tracer of the current
SFR over a timescale of about 10~Myr  \citep{kenn98}. 
The access to  Pa$\alpha$  images  to use as a yardstick for calibrating 
the mid--infrared emission is  the basic motivation for the present work. 

The present paper is organized as follows: Section~2 introduces the
sample of local star--forming galaxies from SINGS; Section~3 presents the data, 
while the measurements used in the analysis are presented in Section~4. Section~5 
briefly introduces the low metallicity starburst galaxies from \citet{enge05} and the 
LIRGs from \citet{alon06}. The 
main findings are reported in Section~6, and the comparison with models is made in 
Section~7. Discussion and a summary are given in Sections~8 and 9, respectively. Details on 
the models of dust absorption and emission are in the Appendix.

\section{Main Sample Description}

\begin{figure}
\figurenum{1}
\plotone{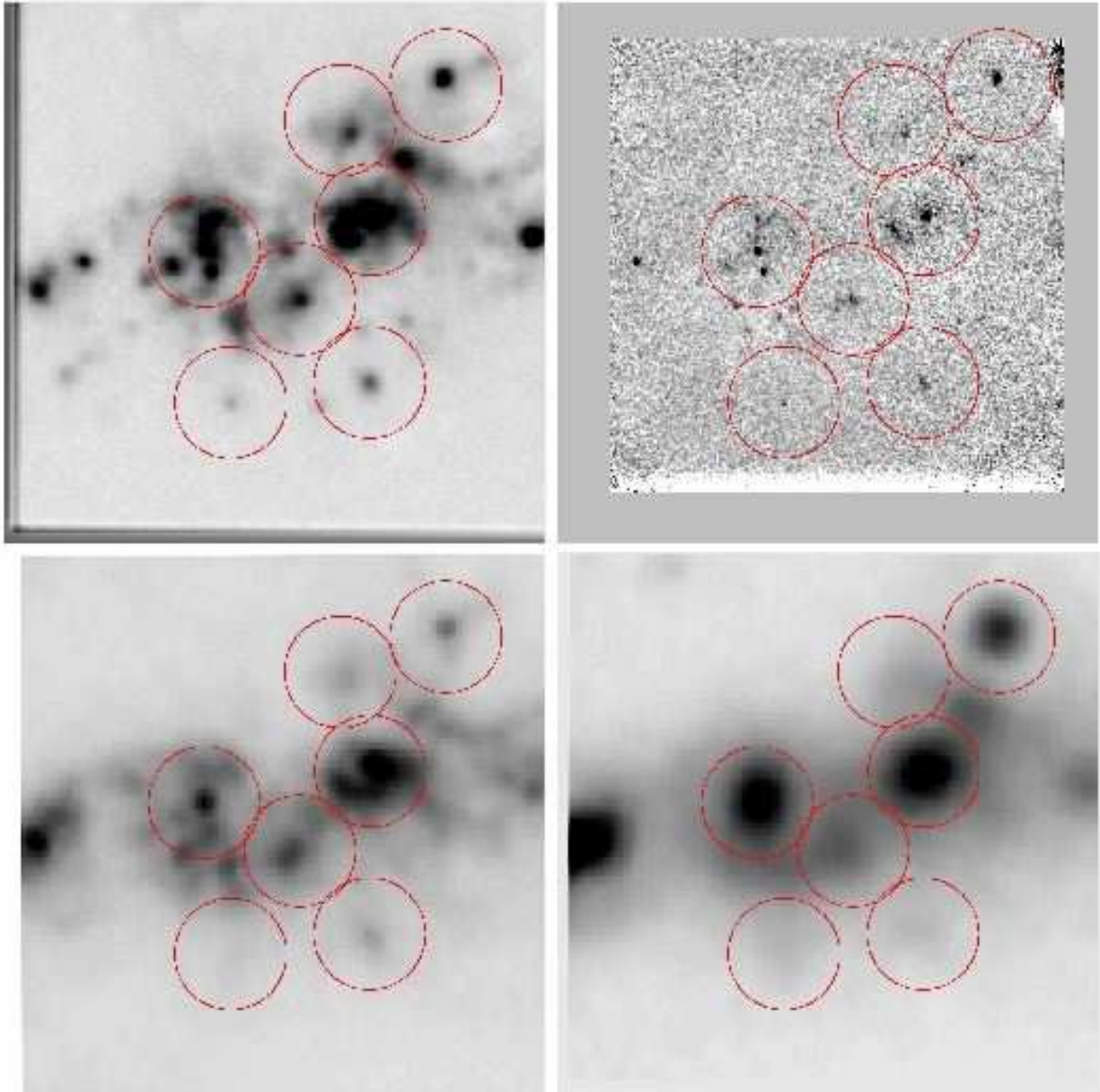}
\caption{Example of aperture selection in one of our fields. The central $\sim$1$^{\prime}$ of the 
galaxy NGC0925 is  shown at H$\alpha$ (top--left), Pa$\alpha$ (top--right), 
stellar--continuum--subtracted 8~$\mu$m (bottom--left), and 24~$\mu$m (bottom--right).  The 
13$^{\prime\prime}$ apertures used for photometric measurements are shown as red circles, 
and correspond to physical sizes of $\sim$580~pc. The field 
shown is one of the two obtained in the central region of this galaxy.  North is up; East is left. 
\label{fig1}}
\end{figure}

The SINGS sample of 75 galaxies \citep{kenn03} was used as our
baseline sample for which HST observations in the infrared were either
obtained as part of our project or retrieved from the HST archive (see details in 
section~3.2). The only criterion required for a SINGS galaxy to be observed with the HST 
was to have a redshifted Pa$\alpha$ emission within the transmission 
curve of one of the NICMOS narrowband filters. A total of 39 galaxies, or 52\% of the SINGS 
sample, were observed in the Pa$\alpha$ line (example in Figure~\ref{fig1}). 
The HST/NICMOS--observed galaxies are fully representative of the SINGS sample as a whole, 
in terms of morphological types, range of metallicity, and SFRs. 

The infrared data of 4 of the 39 galaxies show non--recoverable
problems (see section~3.2 for additional explanation); two more galaxies, M81DwA 
and DDO154 do not show either optical line emission or mid--IR dust emission 
in the region imaged in the near--infrared with HST.  All six galaxies were
discarded from the current analysis, thus leaving a net sample of 33 galaxies. Table~\ref{tab1}
lists the main characteristics of the 39 galaxies, separating the
discarded ones from the remainder of the sample.

The 33 galaxies are divided in three groups according to their oxygen
 abundance: high metallicity galaxies (12$+$log(O/H)$>$8.35), medium
metallicity galaxies (8.00$<$12$+$log(O/H)$\lesssim$8.35), and low
metallicity galaxies (12$+$log(O/H)$\lesssim$8.00). The two sets of disk-averaged oxygen 
abundance values listed in Table~\ref{tab1}  differ systematically by about 
0.6~dex \citep{mous07}. As described by \citet{mous07}, the set of lower numbers for the 
oxygen abundance is roughly tied to the electron temperature abundance scale \citep{pily05}, 
while the higher abundance set is based on stellar populations plus 
photoionization modelling \citep{kobu04, kewl02a}.  The difference between the two 
scales is due to a as--yet unidentified systematic zeropoint offset, and the `true' 
oxygen abundance should lie somewhere between the two listed values; however, 
the relative ranking of abundances on either of the scales should be fairly accurate.  
On this basis, we  assign a galaxy into a metallicity bin based on the average of the two values. 
Metallicity  gradients across galaxies are likely of little impact in our analysis. The observations  
probe the inner $\approx$0.8--5.1~kpc, depending on the distance; 
typical metallicity variations over these region sizes are less than $\sim$0.3~dex  
for our spiral galaxies \citep{mous07}, and therefore are not expected to play a 
significant role in our results.

Within the area imaged by the HST/NICMOS 
for each galaxy in the main sample (Table~\ref{tab1}), regions of  
star formation are identified and their fluxes measured over typical sizes 
of $\sim$200--600~pc  (section~4 and Figure~\ref{fig1}). These regions 
are termed here `HII~knots', and they are far simpler units, in terms of stellar 
population and star formation history, than whole galaxies.  
The HII knots in this study cannot be considered individual
HII regions in the strict meaning of the term. Limitations in angular
resolution, as discussed in section~4, force us to consider areas within
galaxies which may be populated by multiple HII regions. The main
requirement is for such areas  to be local peaks of {\it current}
star formation, as determined from hydrogen line or infrared emission. The
ionizing populations in these regions can be approximated as having comparable ages,
and more evolved stellar populations do not tend to dominate the
radiation output. Although caution should be used when deriving a star formation 
rate for quasi--single--age populations, the investigation of simpler, star--formation--dominated
structures should offer better insights than whole galaxies on the
strengths and weaknesses of the mid--infrared SFR indicators of
interest here.  

Details on the low--metallicity starburst galaxies from \citet{enge05} and on the 
LIRGs from \citet{alon06} are given in section~5.

\section{Observations and Data Reduction}

\subsection{Spitzer IRAC and MIPS Imaging Data}

Spitzer images for the  galaxies in Table~\ref{tab1} were obtained with both IRAC
(3.6, 4.5, 5.8, and 8.0~$\mu$m) and MIPS (24, 70, and 160~$\mu$m), as
part of the SINGS Legacy project, between $\sim$March 2004 and
$\sim$August 2005. A description of this project and the observing
strategy can be found in \citet{kenn03}.

Each galaxy was observed twice in each of the four IRAC bands,  with 
a grid covering the entire galaxy and the surrounding sky. The observing strategy
allowed a separation of a few days between the two observations to enable
recognition and exclusion of asteroids and detector artifacts. Total
exposure times in each filter are 240~s in the center of the field,
and 120~s along the grids' edges. The SINGS IRAC pipeline was used
to create the final mosaics, which exploits the sub-pixel dithering to
better sample the emission, and resamples each mosaic into
0.75$^{\prime\prime}$ pixels \citep{rega04}. The measured 8~$\mu$m PSF
FWHM is, on average, 1.9$^{\prime\prime}$, and the 1~$\sigma$
sensitivity limit in the central portion of the 8~$\mu$m mosaic is
1.2$\times$10$^{-6}$~Jy~arcsec$^{-2}$.

As the interest in this paper is in using the dust emission at mid--infrared
wavelengths (8~$\mu$m and 24~$\mu$m) as SFR tracers, we 
need to remove the stellar continuum
contribution from the 8~$\mu$m images. This contribution is, in
general, small in high metallicity, dusty galaxies
\citep[e.g.,][]{calz05}, but can become significant in lower
metallicity, and more dust--poor galaxies. `Dust--emission' images at
8~$\mu$m are obtained by subtracting the stellar contribution using
the recipe of \citet{helo04}:
\begin{equation}
f_{8 \mu m,\ dust}(\nu) = f_{8 \mu m}(\nu) - \beta f_{3.6 \mu m} (\nu),
\end{equation}
where the coefficient $\beta$ is in the range 0.22--0.29, as determined from
isolated stars in the galaxies' fields. Visual inspection of the
stellar--continuum subtracted images suggests that this approach is
fairly accurate in removing stellar emission; occasional foreground
stars located along the galaxies' lines of sight are in general 
removed by this technique. Although the 3.6~$\mu$m images
can include, in addition to photospheric emission from stars, a
component of hot dust emission, this component is unlikely to
have an impact beyond a few percent on the photometry of the
dust--only 8~$\mu$m images \citep{calz05}. 

MIPS observations of the galaxies were obtained as scan maps, 
with enough coverage to include surrounding background in
addition to the galaxy.  The reduction steps for MIPS mosaics are
described in \citet{gord05} and \citet{bend06}. At 24~$\mu$m, the PSF FWHM is
$\sim$5.7$^{\prime\prime}$, and the 1~$\sigma$ detection limit is
1.1$\times$10$^{-6}$~Jy~arcsec$^{-2}$.  The MIPS images
are considered `dust' images for all purposes, as contributions from
the photospheric emission of stars and from nebular emission 
are negligible (a few percent) at these wavelengths.

\subsection{HST Imaging Data}

The main advantage of using near--infrared narrowband imaging, rather
 than spectroscopy, is the potential of capturing, in principle, all of the 
 light in the Pa$\alpha$ line, thus enabling a more secure measurement of 
 the total line emission from the targets. The HST/NICMOS narrowband 
 filters of interest here have $\sim$1\% band--passes, that can easily 
 accommodate gas line emission with a few hundred km/s shift relative to 
 the galaxy's systemic velocity. 

Most of the HST/NICMOS observations for the galaxies in our sample
come from the HST SNAP program 9360 (P.I.: Kennicutt). For 9 of the
galaxies, archival HST data were used, from programs GO-7237 and
SNAP-7919.

Observations for SNAP-9360 were obtained with the NIC3 camera, in the
narrowband filters F187N, F190N (Pa$\alpha$ emission line at restframe
wavelength $\lambda$=1.8756~$\mu$m and adjacent stellar continuum), 
and the broadband filter
F160W. The NIC3 camera has a field of view of 51$^{\prime\prime}$, and
observations were obtained with 4 dithered pointings along a square pattern 
with 0.9$^{\prime\prime}$ sides, to better remove cosmic
rays and bad pixels. Thus, NICMOS observations imaged the central
$\lesssim$1~arcmin of each galaxy.  The NIC3 0$^{\prime\prime}$.2 pixels
undersample the NICMOS PSF, although this is not a concern for the
diffuse ionized gas emission. On--target total exposure times were
640~s, 768~s, and 96~s, for F187N, F190N, and F160W, respectively.

The data were reduced with the STScI IRAF/STSDAS pipeline {\it calnica}, which
removes instrumental effects, bad pixels, and cosmic rays, and
produces images in count--rate units. The removal of the
quadrant-dependent `pedestal' was done with the IRAF/STSDAS routine
{\it pedsub}. The four dithered exposures were combined with the IRAF/STSDAS 
mosaicing pipeline {\it calnicb}. 

For our analysis, only the two narrowband images are used, and the
emission~line--only images are obtained by subtracting the
continuum--only images, rescaled by the ratio of the filters'
efficiencies, from the line$+$continuum image. Program 9360 was
executed after the NICMOS Cryocooler System (NCS) had been installed on the
HST, 
providing a detector quantum efficiency about 30\% higher in the H-band than during 
pre--NCS (i.e., pre--2002) operations\footnote{The Near Infrared 
Camera and Multi-Object Spectrometer Instrument Handbook, version 9.0, E. Barker et al. 
eds., 2006, STScI}. This is an important difference when comparing depths of SNAP--9360 with 
those of  the archival NICMOS images, which  were obtained pre--NCS. 
The average 1~$\sigma$
sensitivity limit of the continuum--subtracted image is
6.4$\times$10$^{-17}$~erg~s$^{-1}$~cm$^{-2}$~arcsec$^{-2}$. In units
that will be easier to relate to the analysis performed in this paper,
our 1~$\sigma$ limit for a specific Pa$\alpha$ luminosity measured in a
13$^{\prime\prime}$--diameter aperture is
2.83$\times$10$^{37}$~erg~s$^{-1}$~kpc$^{-2}$; in a 
50$^{\prime\prime}$--diameter aperture, the 1~$\sigma$ limit is 
1.04$\times$10$^{38}$~erg~s$^{-1}$~kpc$^{-2}$ .

The archival NICMOS data from HST snapshot  program 7919 are described in
\citet{boek99}.  Here we summarize the main differences with SNAP--9360. 
Data for the SNAP--7919 were obtained with a single
pointing (and a single integration) of the galaxy's center with the
NIC3 camera. One narrowband filter (F187N or F190N depending on
redshift) and the broadband F160W filter were used, for $\sim$768~s
and 192~s, respectively. We re-processed the archival images through
{\it calnica}, to improve the removal of instrumental effects and of
cosmic rays by using a more recent version of the calibration pipeline
than the one used in \citet{boek99}; the quadrant--dependent pedestal
was removed with {\it pedsub}. As in \citet{boek99}, the rescaled
broadband filter is used for removal of the underlying stellar
continuum from the image containing the Pa$\alpha$ emission line. The images from
SNAP--7919 are deeper than in SNAP--9360, with an average 1~$\sigma$ sensitivity
limit of the continuum--subtracted image of
3.5$\times$10$^{-17}$~erg~s$^{-1}$~cm$^{-2}$~arcsec$^{-2}$.

Broadband filters may not provide the optimal underlying stellar
continuum signature, especially if uneven dust extinction in the
galaxy produces color variations within the filter's bandpass. To
check the impact of this potential effect, we have compared
observations of galaxies in common between the SNAP--9360 and
SNAP--7919 programs: NGC3184, NGC4826, NGC5055, and NGC6946 (images of
NGC0925 are also present in both programs, but the pointings are only partially
overlapping, and are sufficiently different that both images are used in
our analysis, see Table~\ref{tab1}). For SNAP--9360, two
narrowband images are available, thus yielding a `cleaner' line
image. Comparison of continuum--subtracted images in both programs for
regions in the common galaxies yields differences in the Pa$\alpha$
photometry in the range 10\%--30\%, which is in general well within
our random uncertainty for the Pa$\alpha$ measurements (section~4.2). 

The NICMOS archival data for NGC5194 (HST program 7237) are described
in \citet{scov01} and \citet{calz05}. The main difference with the data in 9360 is
that the NGC5194's image is a 3$\times$3 NIC3 mosaic that spans the
central 144$^{\prime\prime}$ arcsec$^2$. Each pointing was observed in both
F187N and F190N, with 128~s exposure times. The sensitivity is
variable, being lower at the seams of the 9 images that form the
mosaic. The average 1~$\sigma$ sensitivity limit of the
continuum--subtracted image for this galaxy is
1.8$\times$10$^{-16}$~erg~s$^{-1}$~cm$^{-2}$~arcsec$^{-2}$.

\begin{figure}
\figurenum{2}
\plotone{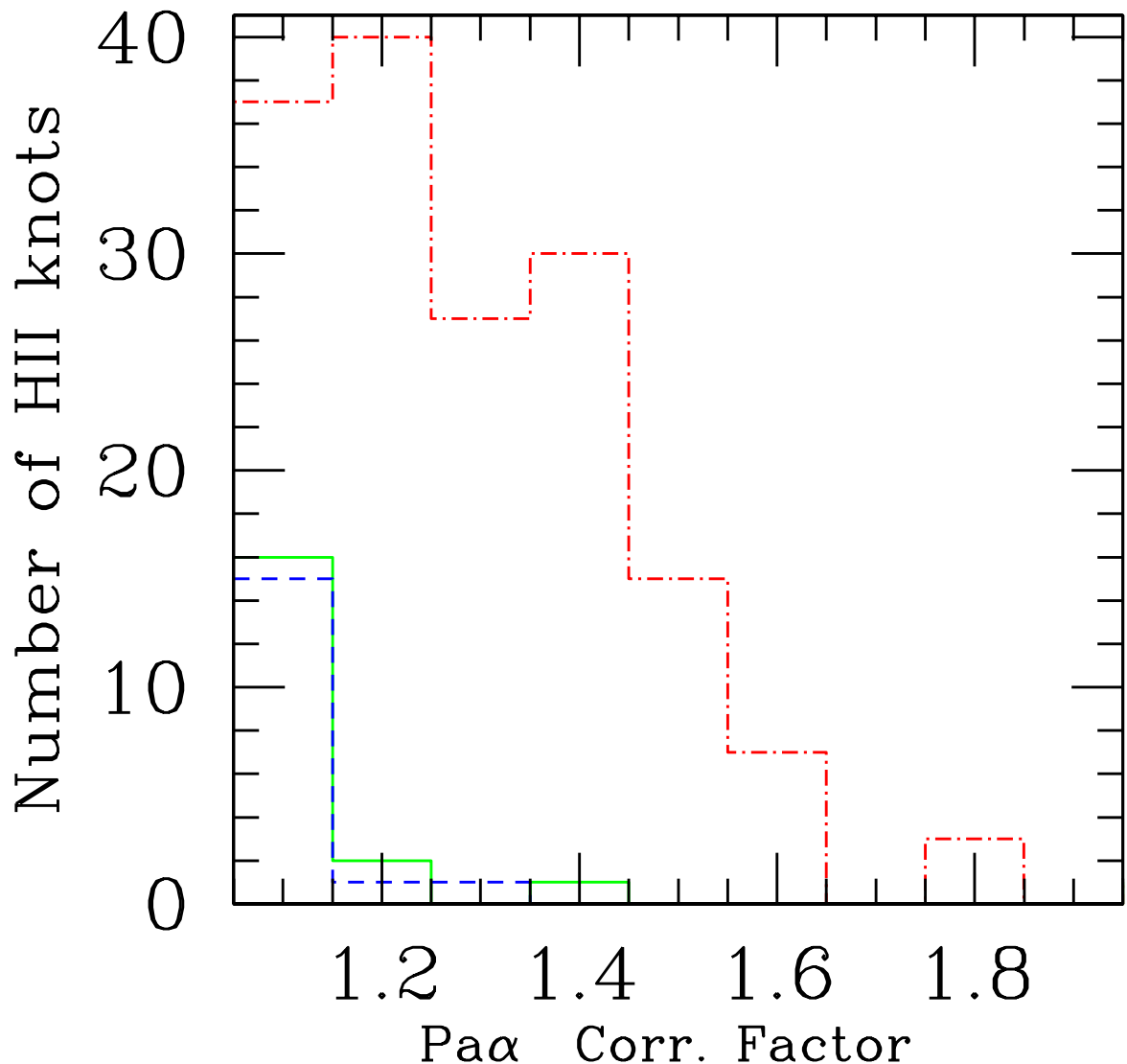}
\caption{Histogram of the multiplicative factors applied to the observed Pa$\alpha$ line emission 
to correct for the effects of dust extinction, as determined using the H$\alpha$/Pa$\alpha$ ratio (section~3.2). The vertical axis shows the number of HII~knots to which each correction factor 
is applied.  The high (12$+$log(O/H)$>$8.35), medium
 (8.00$<$12$+$log(O/H)$\lesssim$8.35), and low (12$+$log(O/H)$\lesssim$8.00)  
metallicity regions  (section~2) are shown as three separate 
histograms, coded as red dot--dash line, green continuous line, and blue dash line, 
respectively. Most corrections are less than 50\%.
\label{fig2}}
\end{figure}

The HST/NICMOS observations are the shallowest in our sample when 
compared to the other images, and
represent the true limitation to our analysis. On the other hand,
Pa$\alpha$ measurements offer an opportunity to obtain a nearly unbiased measure of
the number of ionizing photons produced in a region, as it is only
weakly affected by dust extinction. An extinction as large as 5~mag at
V produces an extinction of 0.73 magnitudes at Pa$\alpha$, i.e.,
roughly a change of a factor of 2 in the line intensity (Figure~\ref{fig2}), for 
foreground screen dust geometry. Still, we
combine the Pa$\alpha$ measurements with complementary measurements at
H$\alpha$ to correct the line emission for the effects of dust. We
adopt a metallicity--dependent  intrinsic ratio 
H$\alpha$/Pa$\alpha$=7.82, 8.45, and 8.73 for the high, 
medium, and low metallicity 
subsamples, respectively, which correspond to electron temperatures  
T$_e$=7,000~K, 10,000~K, and 12,500~K for the HII~knots 
\citep[for n$_e$=100~cm$^{-3}$][]{oste06,garn04}. We also adopt an extinction 
curve\footnote{The extinction curve k($\lambda$) is defined through the following equation:
 F$_{obs}(\lambda)$=F$_{int}(\lambda)$\ 10$^{-0.4 k(\lambda) E(B-V)}$, where 
 F$_{obs}$ and F$_{int}$ are the observed and intrinsic fluxes, and E(B$-$V) 
is the color excess.} with differential value k(H$\alpha$)$-$k(Pa$\alpha$)=2.08 \citep{fitz86,land84}. 

Four of the  galaxies  discarded from our sample
(Table~\ref{tab1}) present an array of problems mainly in their NICMOS
observations. The F187N image of NGC0024 is heavily affected by cosmic
ray persistence, which has caused the effective noise level of the
frame to be about 7 times higher than nominal; the net result is that
the faint emission from the galaxy is undetectable. The NICMOS frames
of NGC1291 missed the galaxy because of guide star problems. 
The F187N images of
NGC4631 show a faint flat--field imprint (generally a sign of residual
pedestal) that, coupled with the large dynamical range of the emission
from this edge--on galaxy, produces a very uneven background. For NGC3034 (M82)§, 
problems related  to non-linearity 
corrections and saturation for this bright target exist for the NICMOS, IRAC, 
and MIPS images, making photometry in the center of this object highly 
unreliable at the present time.

The HST archive was also mined for H$\alpha$ images for those cases
where (a) coverage was similar between NIC3 and optical images, and (b) the
narrowband filter  provides a better rejection of the [NII] emission line than 
the ground--based images. WFPC2 images that met these criteria were available for
NGC1512, NGC4736, NGC4826, and NGC5055.  The line emission was observed through the 
narrowband filters F656N or F658N, and the underlying
continuum through F547M, F555W, and/or F814W  (equivalent to medium--V, V,  
and I, respectively). For NGC4736, NGC4826, and NGC5055, the [NII]/H$\alpha$ values 
listed in Table~\ref{tab2} come from the comparison of the fluxes in the HST and ground--based 
(see below) narrowband filters; the [NII] contamination in the HST filters is minimal, and has been 
used to guide our extrapolation of the best nitrogen--to--H$\alpha$ ratio to attribute to each 
galaxy. This value has been used for those areas in the ground--based images not covered 
by the HST. 

\subsection{Ground--based Optical Imaging Data}

R--band and H$\alpha$--centered narrowband images were obtained for
most of the galaxies as part of the SINGS ancillary data program,
either at the 2.1--m KPNO telescope or at the 1.5--m CTIO telescope
\citep{kenn03}. Exposure times were typically around 1800~s for the
narrowband filters, and a few hundred seconds for R. Standard
reduction procedures were applied to all the images.  Standard stars
observations were obtained during each observing run to derive
photometric calibrations.

The rescaled broadband images were subtracted from the narrowband
images to obtain emission--line--only images. The [NII] contamination
within the filter bandpass is removed using [NII]/H$\alpha$ values
measured either from large--aperture ($\approx$50$^{\prime\prime}$)
SINGS optical spectroscopy \citep{mous07} or retrieved from the literature 
(Table~\ref{tab2}), and accounting 
for changes in the filter transmission between the wavelengths of 
H$\alpha$ and the two [NII] emission lines.  High
metallicity galaxies for which [NII]/H$\alpha$ ratios are not
available from either source, or cases which have optical spectra
dominated by a central non--thermal source \citep[Seyfert~2 or LINER,][]{mous07} 
are assumed to have
[NII]/H$\alpha\sim$0.5. Within each galaxy, a constant [NII]/H$\alpha$ is
adopted, although the ratio can change significantly from individual
HII regions to the more diffuse component \citep{hoope03}. Radial
variations of [NII]/H$\alpha$ within a galaxy are less of a concern
here, as only the central region of each galaxy is imaged.

Typical 1~$\sigma$ sensitivity limits of the final H$\alpha$ images
are 1--2$\times$10$^{-17}$~erg~s$^{-1}$~cm$^{-2}$~arcsec$^{-2}$, i.e.,
they are a factor 3--10 deeper than the Pa$\alpha$ images. This, coupled with 
the fact that the H$\alpha$ is, intrinsically, about 8 times brighter than Pa$\alpha$, 
implies that our H$\alpha$ measurements will have higher signal--to--noise ratio 
than the Pa$\alpha$ ones for A$_V\lesssim$4~mag.

Narrowband and R--band images of DDO053, M81DwB, Holmberg9, and NGC4625 
were obtained using a CCD imager on the
Steward Observatory Bok 2.3~m telescope, as part of the 11HUGS project
\citep{kenn06c}.  Narrowband and R--band images of NGC5408 were obtained at the 
CTIO 0.9--m telescope, also as part of the 11HUGS project. 
Images were taken using a 70 \AA\ narrowband
filter centered at 6580 \AA\ and an R-band filter and a Loral 2kx2k 
 CCD detector.  Exposure times were 1000~s in H$\alpha$ and 200~s
in R, and reach comparable depth to the KPNO images because of the
high throughputs of the filter and the CCD detector.
Data reduction followed similar procedures as described above. 

Ground--based H$\alpha$ images for NGC3627, NGC4736, NGC4826, and 
NGC5055 were provided by the SONG collaboration \citep{sheth02,helf03}, as SINGS did not 
repeat these observations.  The data were obtained at the KPNO
0.9--m telescope, with an observing strategy and filter selection
similar to those of SINGS. The main difference between the SINGS and
SONG H$\alpha$ images is the total exposure time (and the depth of the images), 
being in the latter case 3--5 times shorter than in the former. For this reason, the 
ground--based SONG images were used in conjunction with the  
HST H$\alpha$ images for photometric measurements in NGC4736, NGC4826, and NGC5055.

\section{Photometric Measurements}

\subsection{Aperture Photometry}

For each galaxy, the H$\alpha$, stellar--continuum--subtracted 8~$\mu$m, and 24~$\mu$m
images were registered to the same coordinate system of the Pa$\alpha$ image, 
before performing measurements. Photometric measurements at all four wavelengths of
local 24~$\mu$m and  H$\alpha$ peaks were performed on the common field
of view of the four images. Emission peaks at 24~$\mu$m
(and 8~$\mu$m)  have generally corresponding H$\alpha$ peaks;  the opposite, however, 
is not always true, and there are some cases of H$\alpha$ emission peaks without 
corresponding mid--IR emission. Thus, both 24~$\mu$m and H$\alpha$ images were
used independently to locate local peaks of star formation.

The size of the aperture used for photometric measurements is dictated
by the lowest angular resolution image, the MIPS 24~$\mu$m image, with
a PSF FWHM$\sim$6$^{\prime\prime}$. We chose apertures with
13$^{\prime\prime}$ diameter as a compromise between the desire to
sample the smallest possible scale compatible with HII regions and the
necessity to have reasonable aperture corrections on the
photometry (Figure~\ref{fig1}). For the chosen aperture size, corrections to infinite 
aperture are 1.045, 1.05, and 1.67 at 3.6~$\mu$m, 8~$\mu$m, and 24~$\mu$m, 
respectively, for point sources
 \citep[SSC IRAC Handbook and MIPS Handbook, respectively; ][]{reac05,enge07,jarr06}, 
and are assumed to be small or negligible in the Pa$\alpha$ and H$\alpha$ images  \citep{calz05}. 

In the case of the IRAC 3.6~$\mu$m and 8~$\mu$m emission,  extended emission has a 
different  aperture correction than point sources. Best current estimates  \citep{jarr06} indicate that 
our  aperture choice requires an additional correction factor 1.02 at 3.6~$\mu$m and 
0.90 at 8~$\mu$m, for extended sources. As our sources are neither totally extended nor 
point--like, actual aperture corrections are likely to be closer to a value of unity than those 
reported here. 

The fixed aperture corresponds to different spatial scales in
different galaxies, as distances between $\sim$0.5~Mpc (spatial scale $\sim$30~pc) and
$\sim$20~Mpc ($\sim$1.26~kpc) are covered.  In order to allow comparison among
luminosities measured over areas that differ by a factor as much as
$\sim$40 (for the typical distance range 3--20~Mpc), we report all 
measurements as luminosities per unit of physical area
 (luminosity surface density, LSD)  S$_{Pa\alpha}$, S$_{H\alpha}$, S$_{8 \mu m,\ dust}$, 
and S$_{24 \mu m}$, in units of erg~s$^{-1}$~kpc$^{-2}$. 
Luminosities at mid--infrared wavelengths are expressed  as $\nu$L($\nu$).

The use of luminosity surface densities  removes most dependence of our measurements with 
distance, as  the  LSDs are, for our purposes, equivalent to fluxes. 
Notable exceptions are the cases where the area covered by our aperture contains only 
one HII region, with intrinsic size  smaller than our adopted fixed aperture's size; in these cases 
the LSDs will be artificially decreased by the larger area of the aperture relative to the values they 
would have if we selected apertures matched to the intrinsic size of each HII~region/complex. 
The latter choice is not easily applicable to our sample due to the angular resolution 
limitations of some of the data. Furthermore,  we will see in section~6 that  this effect 
does not appear to have an important impact on our results. 

Photometry for a total of 220 separate HII~knots 
is obtained in the 33 galaxies. Of these, 179 are in the 23 high
metallicity galaxies, including 11 non--thermal nuclei (Seyfert~2 or LINERs as 
retrieved from NED\footnote{The exact classification of galactic nuclei is beyond the scope 
of the present work; we restrict ourself to well--known non--thermal sources as described in
 the literature, as these are the sources that most deviate from the general trends described in 
 the following sections.}; no aperture was laid on top of the active nucleus of the edge--on 
galaxy NGC5866). In the five medium metallicity and five low metallicity
galaxies, 22 and 19 regions are measured, respectively, including 4
regions (one each in IC2574, Holmberg~IX, M81DwB, and NGC6822) that are
strongly emitting in the mid-infrared, but are undetected
in both our Pa$\alpha$ and H$\alpha$ data. These line--undetected objects are
detected in the optical continuum bands and are extended; thus they are likely  
background sources. Heavily obscured sources, like  
those discussed in \citet{presc07}, should represent about 3\% of the 24~$\mu$m sources, but 
we find none; we attribute this lack of heavily obscured sources in our sample to the small 
spatial region subtended by the NICMOS FOV within each galaxy. The 11 non--thermal sources and 
the 4 background sources (Figures~\ref{fig3}--\ref{fig4}) will be excluded from all subsequent 
statistical analysis. 

Crowding of emission peaks within each frame prevents the use of  
`annuli' around individual apertures to perform background subtraction 
from the photometric measurements. Background removal is thus achieved 
by subtracting a mode from each frame, as described in \citet{calz05}. 

`Integrated' values of H$\alpha$, Pa$\alpha$, 8~$\mu$m, and 24~$\mu$m
luminosity surface density are also derived for each galaxy within the area
imaged by the NICMOS/NIC3 camera. These integrated values are therefore
the LSD of each galaxy within the central
$\sim$50$^{\prime\prime}$, except for NGC5194, where 
the central $\sim$144$^{\prime\prime}$ are measured (Table~\ref{tab2}). The integrated 
values mix the emission from the star forming regions (measured
with the smaller apertures) with areas of little or no star
formation, thus providing some insights into the impact of the complex
galactic environment on SFR calibrations.

\subsection{Uncertainties in the Photometric Measurements}

The uncertainties assigned to the photometric values at each
wavelength and for each galaxy are the quadrature combination of four
contributions: Poisson noise, variance of the background, photometric calibration
uncertainties, and variations from potential mis-registration of the
multiwavelength images. The variance on the image background is
derived in each case from the original--pixel--size images. The impact
of potential background under-- or over--subtractions varies from
galaxy to galaxy, and also depends on the relative brightness of  
the background and the sources. The effect of potential misregistrations 
have been evaluated
for the case of NGC5194 by \citet{calz05}. Because of the large
apertures employed for our photometry, this contribution is either
small (a few \%~ of the total uncertainty) or negligible.

For the Spitzer 8~$\mu$m and 24~$\mu$m images, calibration
uncertainties are around 3\% and 4\%, respectively \citep{reac05,enge07}. 
This, added in quadrature to the other uncertainties, produces overall uncertainties
 in the measurements  that range between 15\% and a factor of two, with the
median value being around 22\%. The superposition of the PSF wings in
adjacent apertures produces an additional effect in the 24~$\mu$m
measurements, that is evaluated and removed on a
case--by--case basis \citep[see example in][]{calz05}.

For the HST images, photometric calibrations are generally accurate to
within $\sim$5\%, for narrowband filters. The faintness of the
Pa$\alpha$ emission, and therefore the impact of the background
variance and stellar continuum subtraction is what mostly dominates
the photometric uncertainty on the Pa$\alpha$ emission line measurements, with
values between 15\% and a factor of roughly 2, with a median value of
60\%. For the extinction--corrected Pa$\alpha$ luminosities, the
uncertainty on the attenuation A$_V$ increases the Pa$\alpha$
uncertainty by a factor of 1.22.

For the ground--based H$\alpha$ images, which are the deepest images
in our set, the main sources of uncertainty are: photometric
calibrations, stellar continuum subtraction, and the
correction for the [NII] contribution to the flux in the narrow--band
filter. These translate into uncertainties in the final photometric
values between 10\% and 50\% (with occasional factor--of--2
uncertainty). The median uncertainty for the H$\alpha$ luminosities is
20\%. Although less deep, the HST H$\alpha$ images are characterized
by more stable photometry, better continuum subtraction, and smaller
[NII] contamination; uncertainties on the final luminosities are in
the range 5\%--10\%.

For a few of the galaxies of Table~2, some special circumstances are present or special 
treatment was required. For NGC2841, the very faint line emission produces large, and highly 
uncertain, A$_V$ values. For NGC5033, no H$\alpha$ image is available;  the uncorrected 
Pa$\alpha$ can be up to 70\% underestimated for the largest A$_V$ measured in our sample
(A$_V\sim$4~mag), and, therefore, this galaxy is excluded from all fits reported below. 

In Holmberg~IX, H$\alpha$ emission is detected in two of the three selected regions;  for one of these 
two regions, 24~$\mu$m emission is also detected, at the $\sim$2.5~$\sigma$ level.  
A strong 24~$\mu$m detection is present in the third region, together with the only 8~$\mu$m 
detection in the field; because of the absence of hydrogen line emission and of the extended nature 
of the broad band emission, this source is identified with one of the background 
sources discussed in section~4.1. For the two regions  with H$\alpha$ emission, only upper 
limits can be derived for the Pa$\alpha$ and  8~$\mu$m emission. The presence of H$\alpha$ emission 
provides a lower limit to the Pa$\alpha$ line intensity  for the zero extinction case (after including
the uncertainty on the H$\alpha$ measurement itself). We have taken
the range between this lower limit and the upper limit measured from the HST/NICMOS images to be
our fiducial range of values for Pa$\alpha$, and therefore we report
the middle values (in logarithmic scale)  as measurements, rather than use the 
actual upper limits. 

In NGC5408, the brightest, and most extended, line--emitting region 
is only partially imaged by  NICMOS. The Pa$\alpha$ image is
therefore used only to derive a typical A$_V$ value for the region,
using small-aperture photometry and the matching H$\alpha$
measurements. The A$_V$ value derived in this way is then applied to the H$\alpha$
emission of the entire, extended, region, for which a
larger--than--nominal, 17.1$^{\prime\prime}$ diameter, aperture is
used, not only for H$\alpha$, but also for the 8~$\mu$m and 24~$\mu$m
emission. The other two regions in this galaxy are treated with the nominal
procedure described in section~4.1.

\section{Starburst Galaxies}

Our baseline sample of 220 HII~knots is augmented with 10 local low--metallicity starburst galaxies 
and 24 LIRGs from \citet{enge05} and \citet{alon06}, respectively,  in order to verify that trends and 
correlations 
observed for star--forming regions within galaxies can also be applied to galactic--scale ($\sim$kpc) 
star formation.  In this context, starbursts are defined as galaxies with a central, connected star forming 
region whose energy dominates the light output in the wavebands of interest.  

The low--metallicity starbursts and the LIRGs also expand the mid--IR and line emission LSD 
parameter space of the low--  and high--metallicity HII~knots,  respectively, by more than an order of magnitude at the high end.

\subsection{Low--Metallicity Starburst Galaxies}

As part of the HST/NICMOS SNAP--9360, about 40 nearby 
starburst galaxies were observed. Of these, 13 also have Spitzer imaging 
 as part of the MIPS and IRS GTO observations \citep{enge05}. The main characteristics 
 and measurements for 10 of these galaxies are listed in Tables~\ref{tab3} 
 and \ref{tab4}. The three remaining galaxies, NGC3079, NGC3628, and NGC4861, are omitted 
 from the present analysis for the following reasons. For NGC4861, the 
 HST/NICMOS pointing targeted the relatively quiescent center of this galaxy, rather than the 
 peripheral giant HII region. The other two galaxies, NGC3079 and NGC3628, have 
 extended optical line and mid--IR emission: about 40\% and 60\% of the emission is outside  of the 
 field--of--view imaged by HST/NICMOS; corrections for the fraction of light in the Pa$\alpha$ line 
 outside of the observed frame would be thus substantially larger than the typical uncertainties in the 
 measurements. 
 
 The data for the galaxies in Table~\ref{tab3} were reduced in the same fashion as the SINGS galaxies 
 discussed in sections~2--3. In particular, the HST/NICMOS images, which are presented here 
 for the first time, were treated following the same procedure as section~3.2. The main difference 
 between the HII~knots in the SINGS galaxies and the local starbursts is in the photometry: 
 integrated flux values  encompassing the {\em entire central starburst} (the dominant source of emission 
 at the wavelengths of interest) are derived for the latter sample. 
 The integrated measurements at 8~$\mu$m and 24~$\mu$m are from  \citet{enge05}, and are 
 reported in Table~\ref{tab4}. 
 
 The Pa$\alpha$ measurements (Table~\ref{tab4}) are performed using the aperture sizes 
 listed in  Table~\ref{tab3}, and  are corrected for the Galactic foreground 
 extinction (fourth column of Table~\ref{tab3}),  but not for internal extinction. 
 We expect the internal extinction to represent a small 
 effect  on the Pa$\alpha$ flux in these mostly low metallicity galaxies (compare with Figure~\ref{fig2}). 
 An exception may be represented by SBS0335-052, for which \citet{houc04} measure 
 A$_{9.7~\mu m}\sim$0.5~mag. 
 If the region of silicate absorption is coincident with the region of line emission, this would correspond 
 to A$_{Pa\alpha}\sim$2~mag. Given the uncertainty in the spatial co-location of the dust-hidden source detected by \citet{houc04} and the main source(s) of the Pa$\alpha$ emission and the fact that the introduction of an extinction correction for one of the galaxies does not impact our conclusions, we 
 do not perform the correction. 
 
 \subsection{Luminous Infrared Galaxies}
 
HST/NICMOS 
Pa$\alpha$ data and extinction corrections, as well as information on the physical extent of the star forming area for each of the 24 LIRGs used in this analysis,  are 
presented in \citet{alon06}; the reader is referred to that work for details.  
Infrared measurements 
at 25~$\mu$m from IRAS and distances for each galaxy are from \citet{sand03} and \citet{sura04}.  
At the time of this writing, no 8~$\mu$m emission measurements are available for these 
galaxies. 
The LIRGs' metallicities are characteristic of our high--metallicity HII~knots sample \citep{alon06}. 
Photometry for these galaxies, as in the case of the local starbursts (section~5.1), includes 
the entire line--emitting and IR--emitting galactic region, thus the measurements are 
integrated galaxy values. 

\section{Analysis and Results}

\begin{figure}
\figurenum{3}
\plottwo{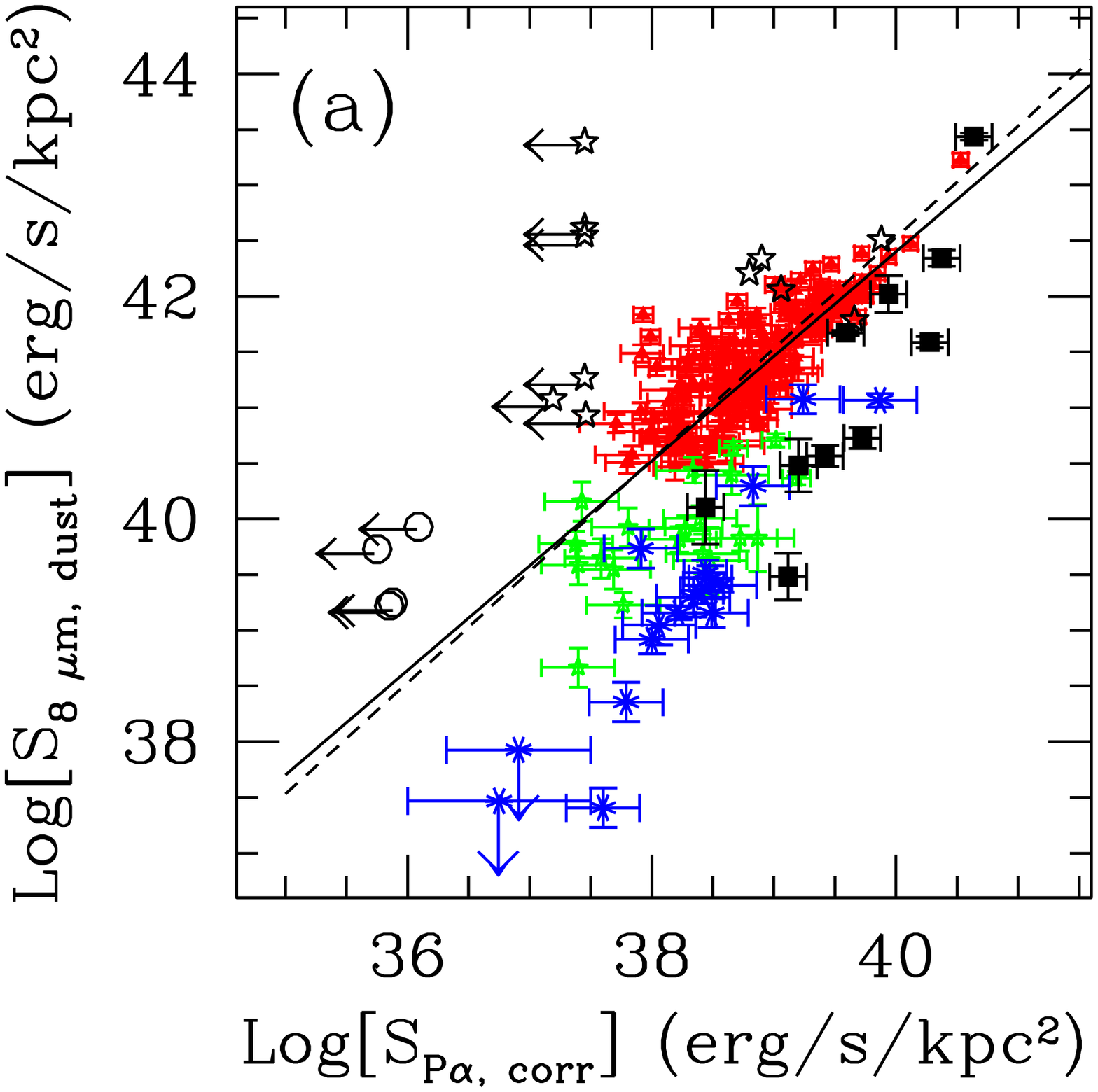}{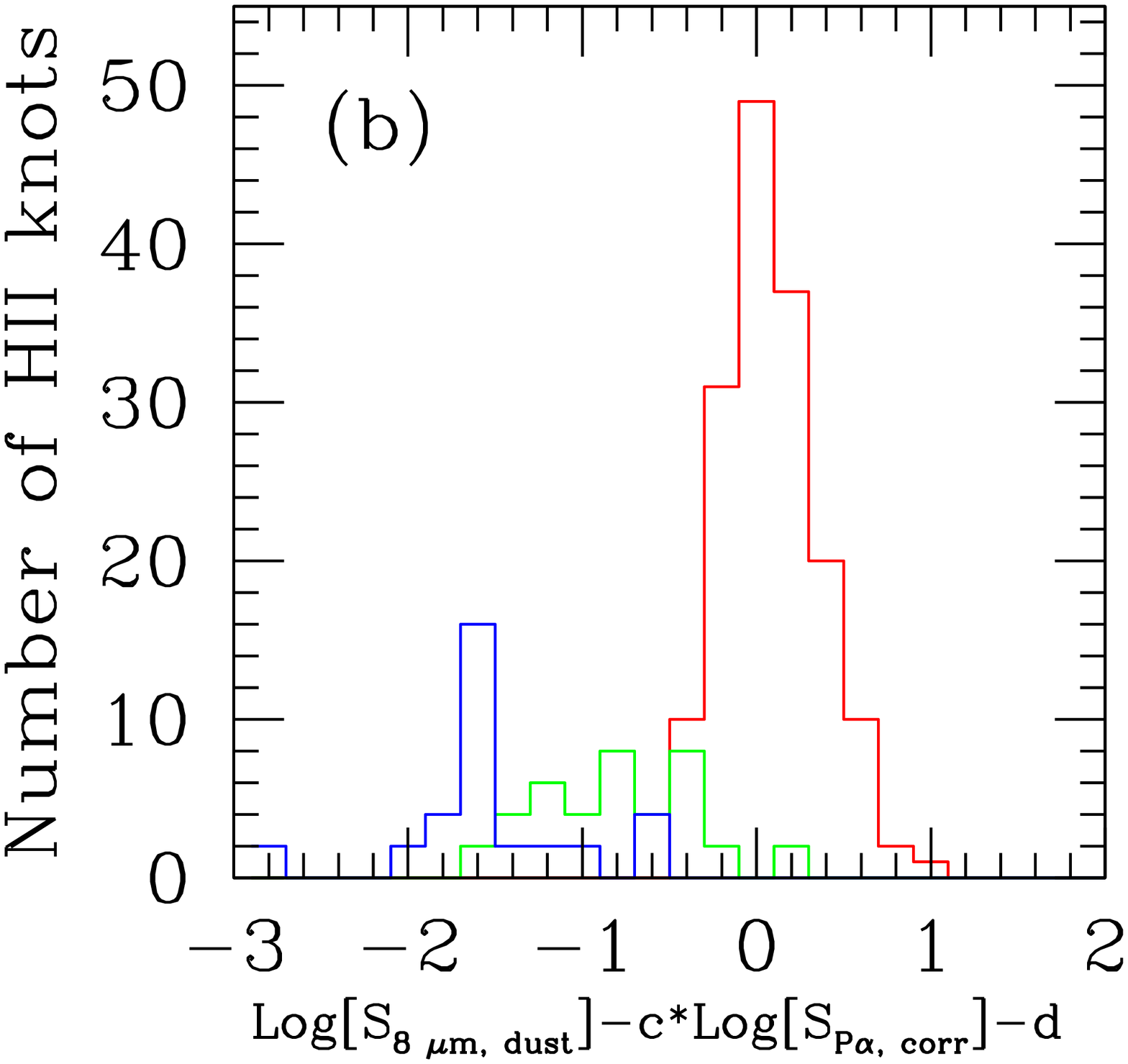}
\caption{(a) Luminosity surface density at 8~$\mu$m, S$_{8~\mu m,\ dust}$, as a
function of the extinction--corrected LSD at
Pa$\alpha$, S$_{Pa\alpha, corr}$, for the 220 HII~knots  in the 33
galaxies for which photometric measurements have been obtained.  LSDs are averaged over  
13$^{\prime\prime}$ photometric apertures.  The 8~$\mu$m 
emission is stellar~continuum--subtracted (section~3.1). Data
points are divided into three metallicity bins: high (red filled triangles), medium
(green stars), and low (blue asterisks) oxygen abundance (section~2). Filled black squares mark 
the local starbursts from the sample of \citet{enge05} (section~5.1).  3~$\sigma$ error bars are 
indicated for each data point. Open black star symbols indicate the location 
of the non-thermal sources (Sy2s or LINERs, section~4.1) 
and open black circles indicate extended background  sources. 
The best fit line through the high metallicity (red) datapoints is shown as a continuous line, while 
the dashed  line is  the linear fit through the same datapoints with fixed slope of 1.  (b)
Histogram of the deviation of the HII~knot data in panel~(a) from the best fit
line through the high metallicity data (the continuous line in
panel~(a)). 
The values of the best--fit coefficients are c=(0.94$\pm$0.02) and d=(4.80$\pm$0.85) 
(equation~2). Three separate histograms are 
shown, for high (red), medium (green), and low (blue) metallicity data.  The medium and low
metallicity histograms have been multiplied by a factor 2 to make them visible.
\label{fig3}}
\end{figure}

\begin{figure}
\figurenum{4}
\plottwo{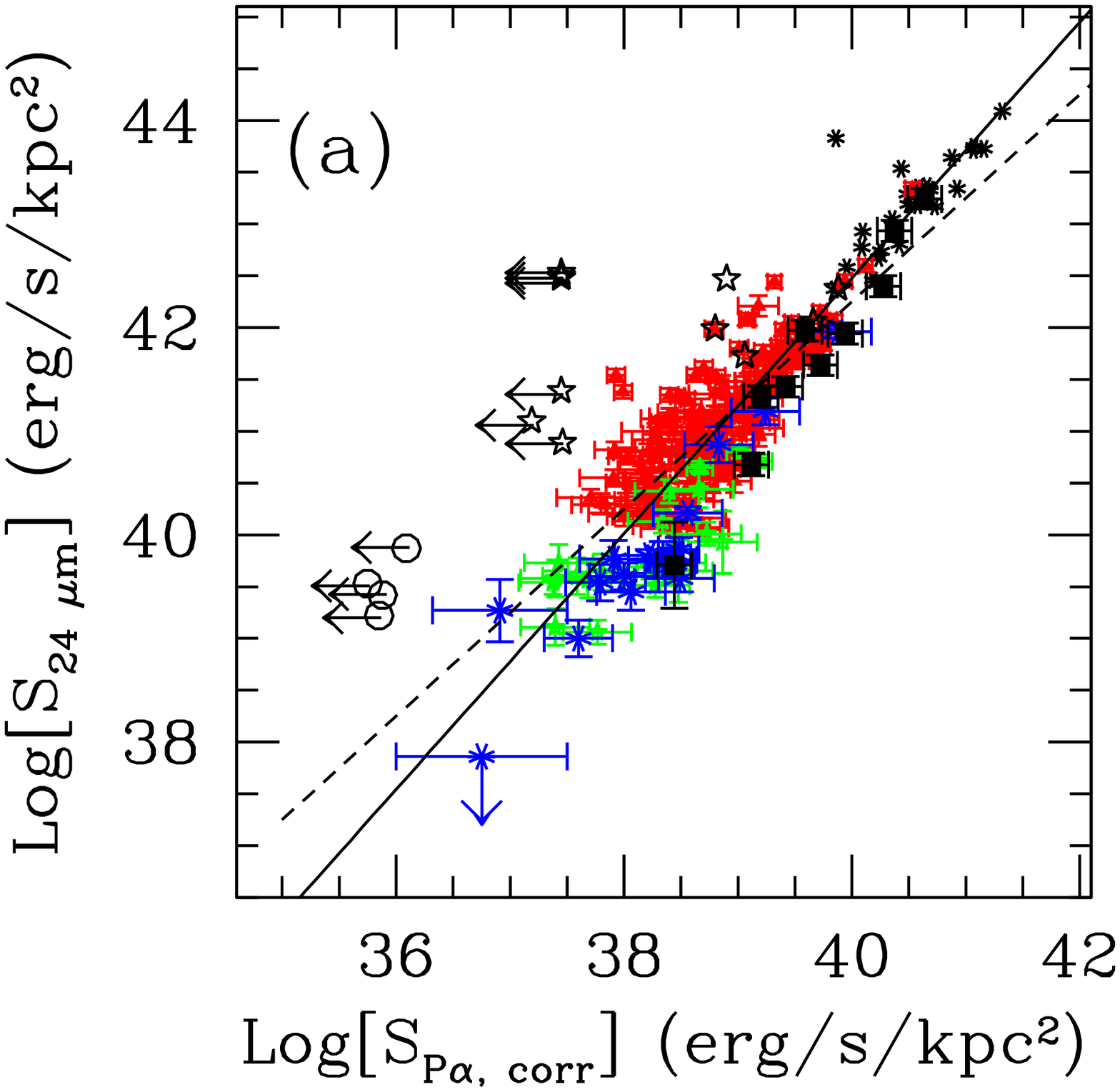}{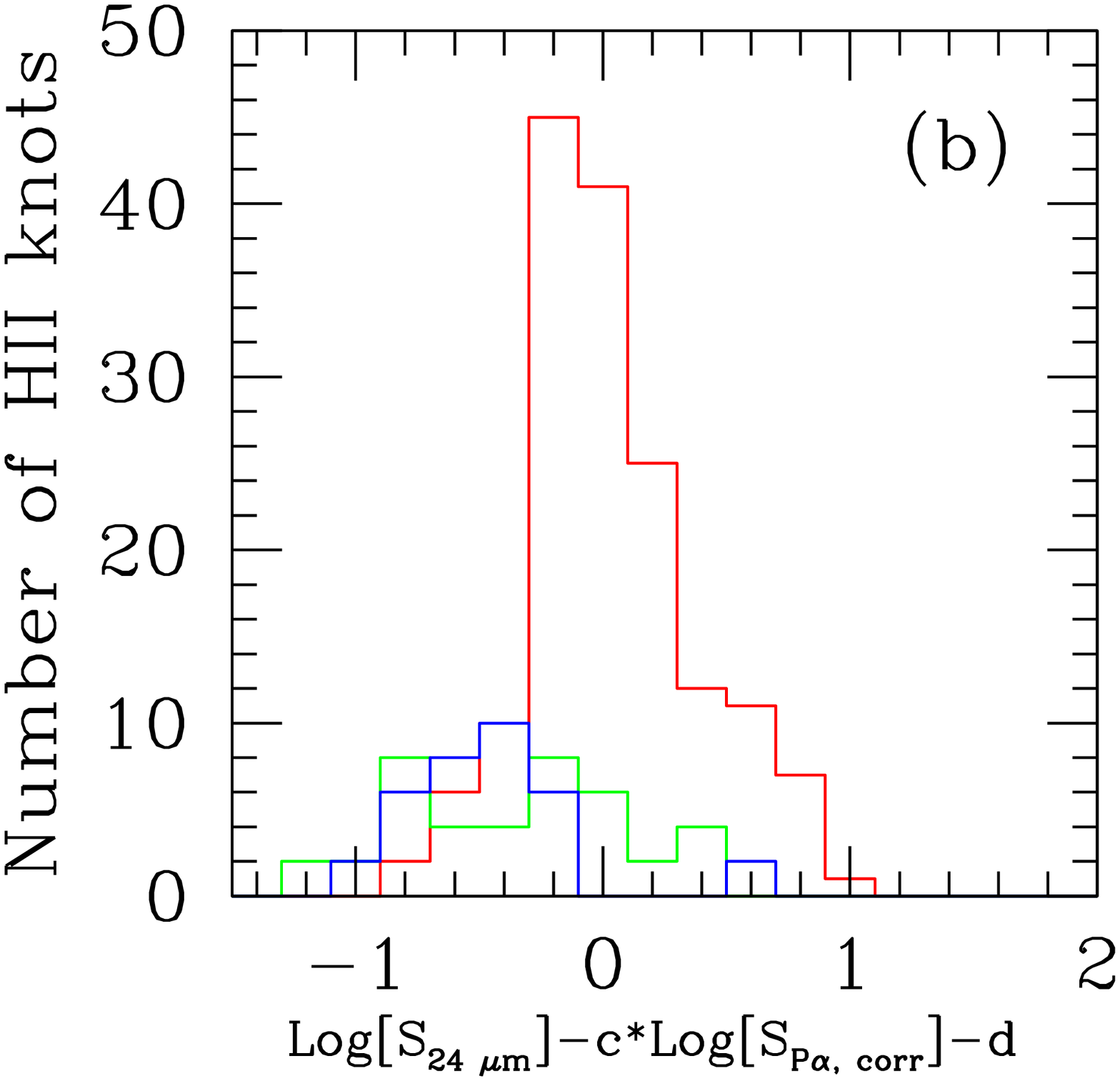}
\caption{(a) and (b). The same as Figure~\ref{fig3}, for the luminosity surface density
at 24~$\mu$m, S$_{24~\mu m}$.  In addition to the same datapoints as Figure~\ref{fig3}, 
panel~(a) also reports the Luminous InfraRed Galaxies (LIRGs) from the sample of \citet{alon06} (black asterisks; section~5.2).  The values of the parameters ({\em c, d}) in the horizontal axis of of 
panel~(b), are given in equation~3, and are c=(1.23$\pm$0.03) and d=($-$6.88$\pm$0.97).
\label{fig4}}
\end{figure}

Photometric measurements for the 220 HII~knots, the local low--metallicity starbursts, and 
the LIRGs are shown in 
Figures~\ref{fig3}--\ref{fig4}, where the infrared LSD in the two mid--IR wavebands 
 is shown as a function of the extinction--corrected Pa$\alpha$ LSD. 

One characteristic immediately apparent in
Figures~\ref{fig3}--\ref{fig4} is the overall correlation between the  
infrared LSDs and the Pa$\alpha$ LSD (panels [a]), although 
the scatter is non negligible in both cases (panels [b]). The
correlations appear especially significant for the high metallicity
HII~knots (the most numerous subsample among those under analysis here), and 
span a little over two orders of magnitude in Pa$\alpha$ LSD.  Bi-linear least--square fits 
through the high--metallicity data points yield:
\begin{equation}
Log\   S_{8~\mu m,\ dust} = (0.94\pm 0.02)\ Log\ S_{Pa\alpha, corr}  + (4.80\pm 0.85),
\end{equation}
\begin{equation}
Log\  S_{24~\mu m} =  (1.23\pm 0.03)\ Log\ S_{Pa\alpha, corr} - (6.88 \pm 0.97), 
\end{equation}
where S$_{Pa\alpha, corr}$ is the extinction--corrected Pa$\alpha$ LSD. 
Equation~3 accounts effectively for the trend of the LIRGs, although these data were 
not used in the fitting procedure.

The scatter of the datapoints about the best fit lines of
equations~2--3 are approximately the same, with $\sigma$=0.3~dex
(panels (b) of Figure~\ref{fig3}--\ref{fig4}). Thus the 1~$\sigma$
scatter is about  a factor of 2 for the high metallicity regions.

\begin{figure}
\figurenum{5}
\plottwo{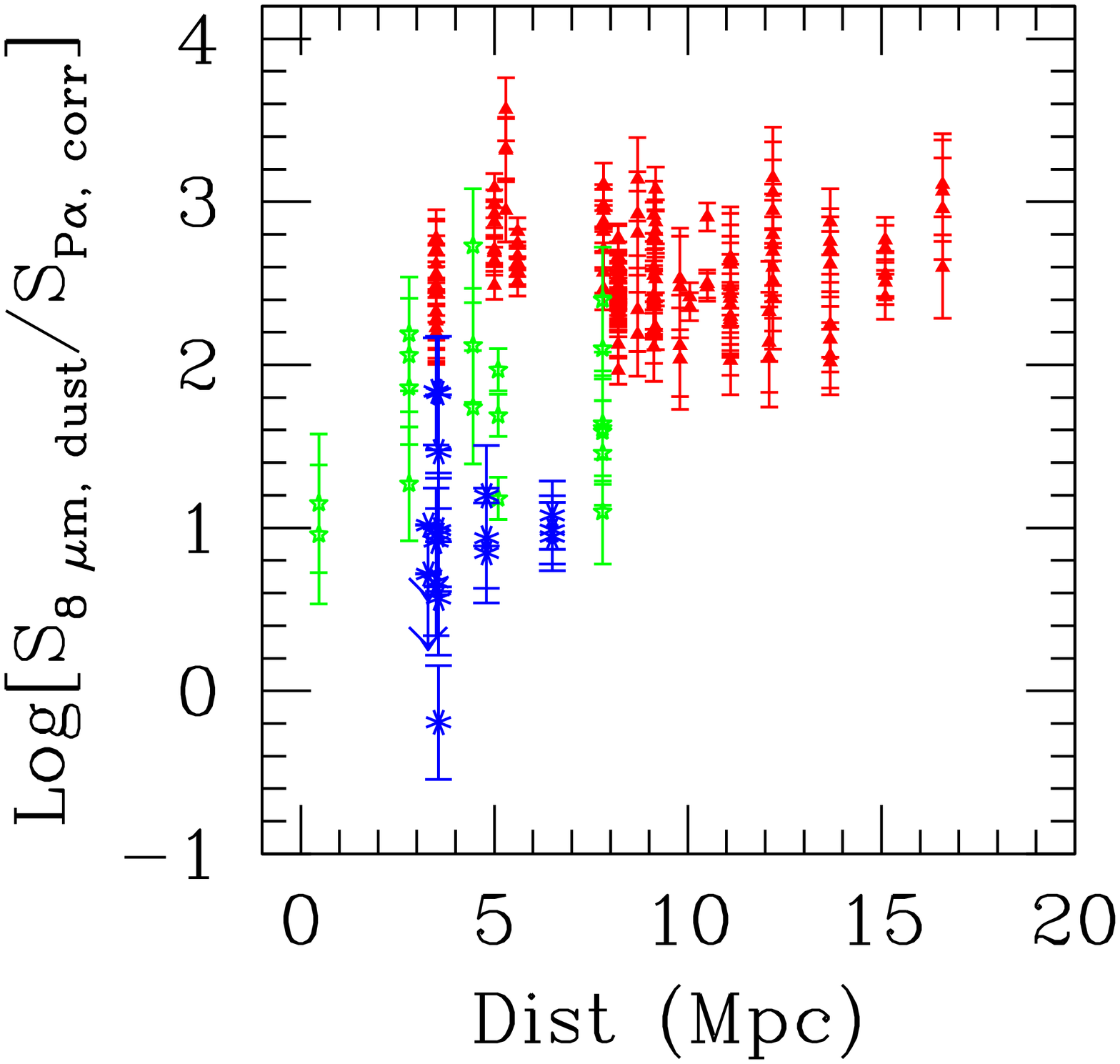}{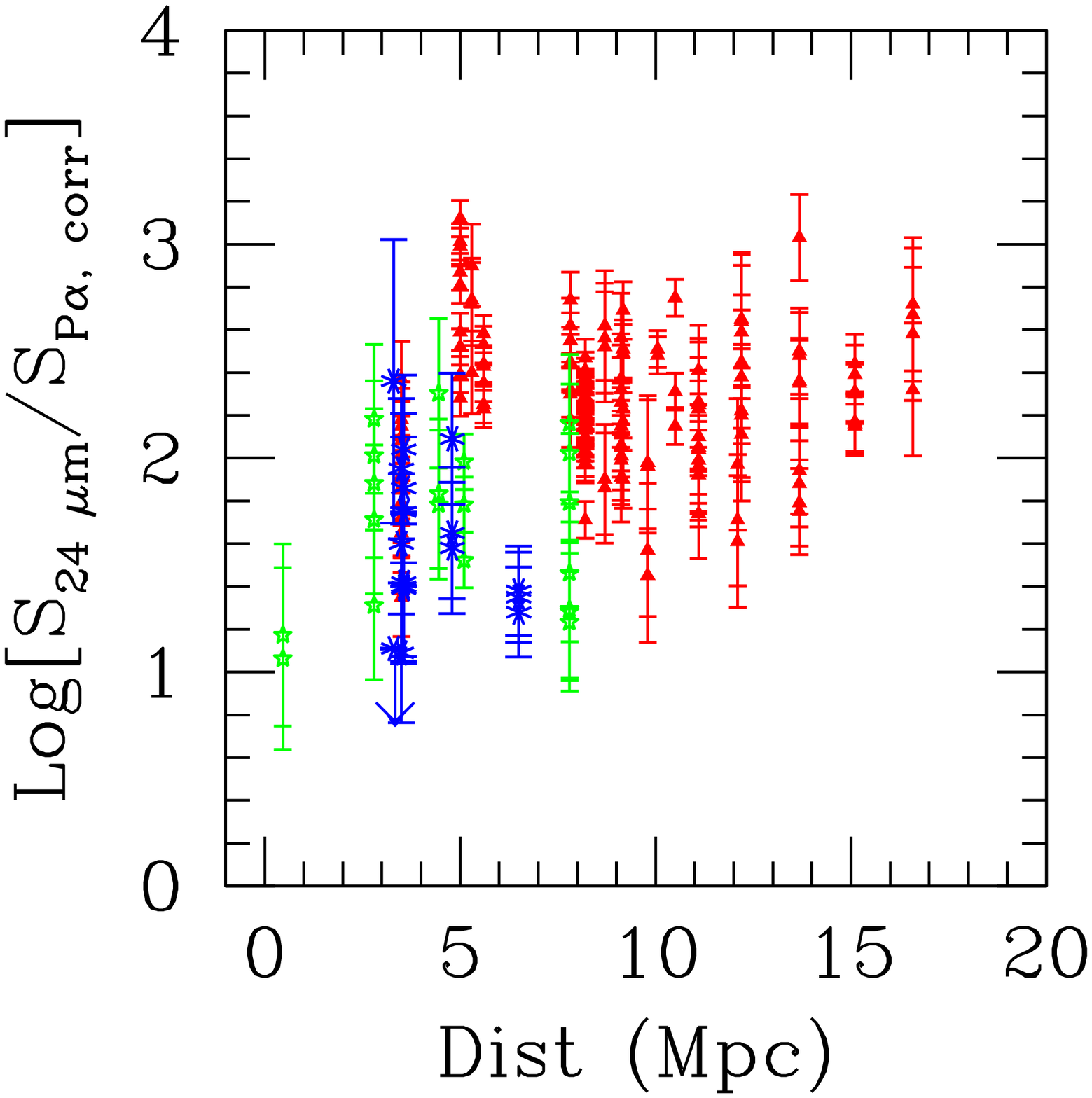}
\caption{The ratio of the mid--IR to the extinction--corrected Pa$\alpha$ LSDs, as a 
function of the galaxies' distances, for the 8~$\mu$m (left) and the 24~$\mu$m (right)  HII~knot 
measurements, respectively. Background sources and non--thermal sources are not included. 
In both panels, the high--metallicity datapoints are consistent 
with no correlation of the ratio as a function of distance, implying that the correlations between 
the mid--IR emission and the Pa$\alpha$ emission are not driven by distance effects. 
The 8~$\mu$m emission of the medium and low--metallicity HII~knots remains on average deficient relative to that of the high--metallicity data also when only galaxies at similar distances are 
considered (thus flux measurements are performed in similar--size regions). With the possible 
exception of NGC~6822, which is at a distance of only 0.47~Mpc and shows lower--than--average 
values for its metallicity bin, the observed 8~$\mu$m emission deficiency in metal--poor regions is 
not an effect of a bias in the size of the regions that are being measured. 
\label{fig5}}
\end{figure}

A potential source of bias in equations~2 and 3 
is the large range of distances that our sample covers, about a
factor of 6 for the high metallicity galaxies. Our fixed photometric
aperture of 13$^{\prime\prime}$ diameter thus probes regions that are
about 30 times different in area between the nearest and the farthest
targets in the high~metallicity subsample, i.e., from 0.04~kpc$^2$ at 3.5~Mpc 
to 1.12~kpc$^2$ at 17~Mpc (for the most distant galaxy in our sample, NGC4125, located at 21~Mpc,  
only the central Sy2 nucleus is detected and is excluded from the  
analysis). Although we remove the background from 
each photometric measurement, uncertainties in this subtraction will affect the
farthest targets more strongly than the closest ones, if HII
regions/complexes have constant sizes of $\approx$100--200~pc. Furthermore, 
we may expect that our fixed aperture photometry may dilute the LSDs of the 
more distant regions, for the extreme hypothesis that only one HII~region is contained in 
each aperture.  We have tested the impact of these effects by looking at the distribution of the
ratios S$_{8~\mu m,\ dust}$/S$_{ Pa\alpha, corr}$ and S$_{24~\mu m}$/S$_{ Pa\alpha, corr}$ 
as a function of galaxy distance (Figure~\ref{fig5}). For the high metallicity subsample, 
non--parametric (both Spearman and Kendall) tests show that the data are uncorrelated  with 
the galaxy's distance, suggesting that there is no obvious bias in our analysis. 

Both the 8~$\mu$m and 24~$\mu$m LSDs of medium and
 low metallicity regions are deficient relative to the extrapolation of the best
fit lines for the high metallicity regions (Figure~\ref{fig3} and \ref{fig4}). The deficiency is far more
pronounced in the case of S$_{8 \mu m,\ dust}$, a fact already noted in a number of previous 
investigations \citep[e.g., ][]{enge05,gall05,hogg05,rose06,drai07}. 
A potential source of concern in this case is that the high metallicity subsample has a higher 
mean distance than the medium and low metallicity ones (Figure~\ref{fig5}). 
\citet{helo04} have shown that the 8~$\mu$m emission is brighter at the edges of an HII~region 
(i.e., in the PDR) 
than at its center. Our fixed aperture photometry could therefore underestimate the 8~$\mu$m flux from 
the low~metallicity regions, if the apertures are not large enough to sample the entire 
area surrounding the HII~knot. However, Figure~\ref{fig5}  shows that the 8~$\mu$m emission 
is deficient in the medium and low metallicity subsamples relative to the high metallicity one 
even when galaxies at comparable distances are considered.
The only potential exception is NGC~6822, 
the closest galaxy to the Milky Way in our sample, which, at a distance of 0.47~Mpc, could suffer 
from the effect of having too a small aperture applied to the 8~$\mu$m emission measurements;  
indeed its mean value is lower (although not statistically significantly) than the average of the 
other data in the same metallicity bin. 

\begin{figure}
\figurenum{6}
\plotone{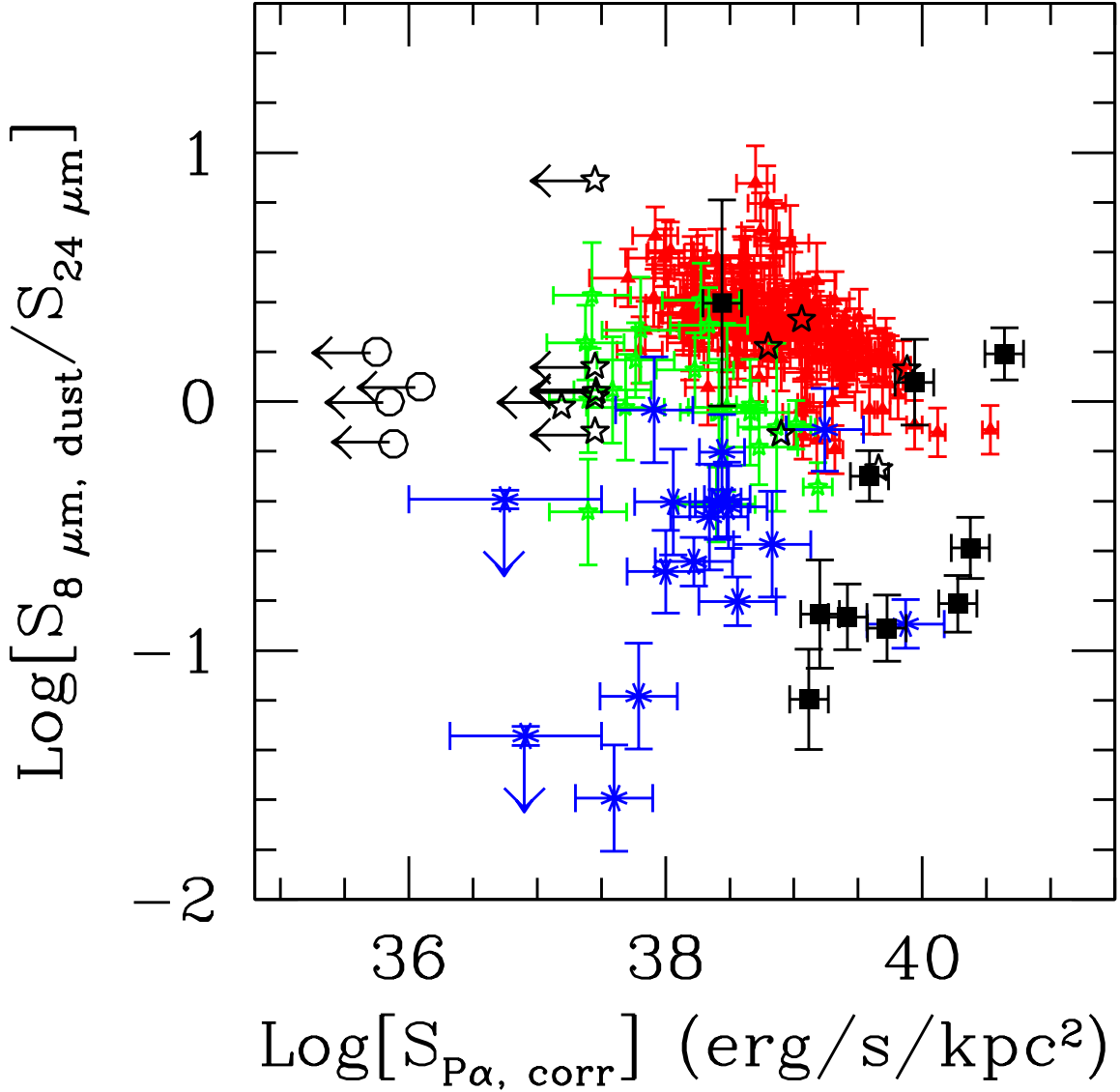}
\caption{Ratio of the 8~$\mu$m to 24~$\mu$m LSD as a
function of the extinction--corrected Pa$\alpha$ LSD for the 220 
HII~knots and for the local starbursts. Symbols and colors are as in Figure~\ref{fig3}.  
The deficiency of the 8~$\mu$m emission, relative to the 24~$\mu$m emission for the 
low--metallicity datapoints is independent of the average  ionizing photon rate in the region. 
The decrease of the 8~$\mu$m to 24~$\mu$m LSD ratio as a function of increasing Pa$\alpha$ LSD 
for the high metallicity points indicates 
 that the component of thermal equilibrium dust contributing to the 24~$\mu$m emission 
is increasing in strength \citep[the dust is in thermal equilibrium and `warmer' at higher ionizing 
photon densities, see, ][]{helo86,drai06}.  A contribution to the decrease of the 8~$\mu$m 
emission due to increased destruction rate of the carriers for increasing starlight intensity 
(Pa$\alpha$ LSD) may also be present \citep{boul88}. 
\label{fig6}}
\end{figure}

The trend of the S$_{8 \mu m,\ dust}$/S$_{24 \mu m}$ ratio as a function of S$_{Pa\alpha, corr}$ 
(Figure~\ref{fig6}) highlights the decrease of the 8~$\mu$m LSD for 
decreasing metallicity, and also shows that the effect is independent of the 
number of ionizing photons in the region.  The latter suggests that: (1) our aperture 
sizes are large enough to encompass both the HII regions and the surrounding PDRs, as noted 
above; and (2) in these large regions the dependence of the 8~$\mu$m--to--24~$\mu$m ratio 
 on the luminosity surface density of the  HII~region/complex that heats the dust is a small effect 
 relative to the effect of metallicity.  
The decrease of the 8~$\mu$m to 24~$\mu$m LSD ratio as a function of increasing Pa$\alpha$ LSD 
in the high metallicity points (i.e., at roughly constant metallicity) indicates 
 that the component of thermal equilibrium dust contributing to the 24~$\mu$m emission 
is increasing in strength \citep[the dust is in thermal equilibrium and `warmer' at higher ionizing 
photon densities, see ][]{helo86,drai06}. An additional contribution may also come from 
an increased  destruction rate of the 8~$\mu$m dust emission for increasing starlight intensity \citep{boul88}. 

\begin{figure}
\figurenum{7}
\plottwo{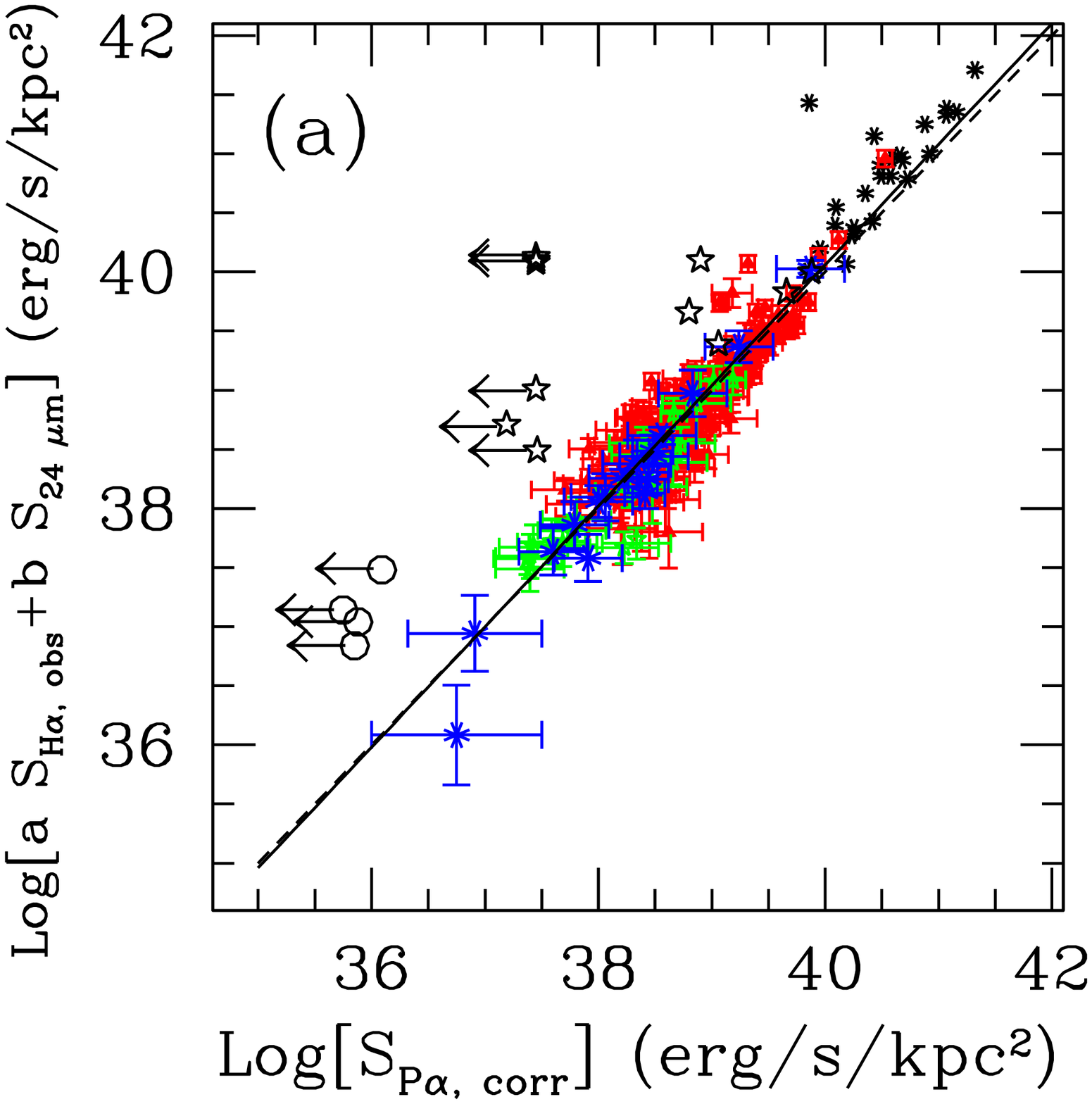}{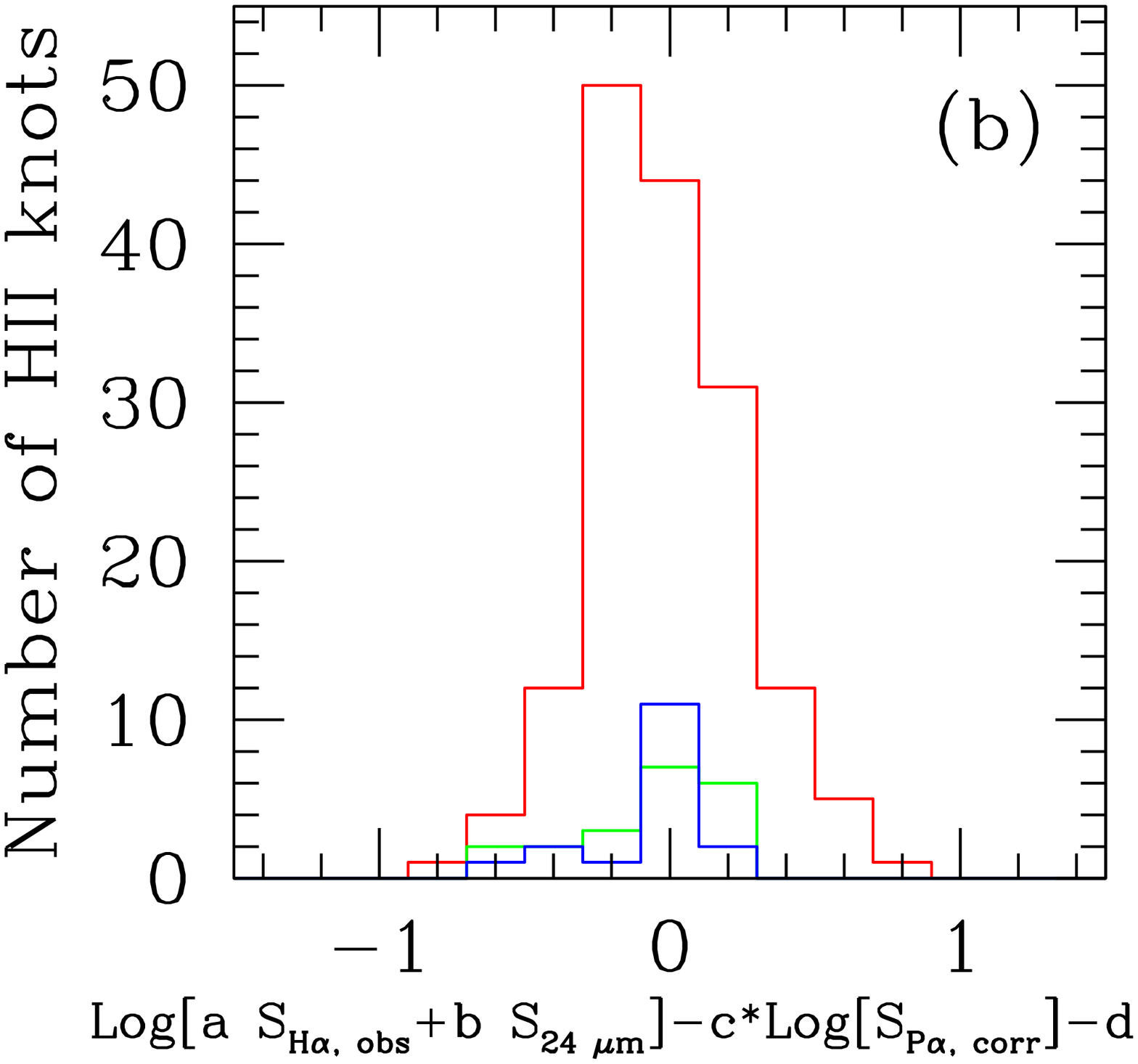}
\caption{(a) and (b). As Figure~\ref{fig4}, for the linear combination
of H$\alpha_{obs}$ and 24~$\mu$m LSD (equation~4). 
Symbols are as in Figures~\ref{fig3} and \ref{fig4}. 
(Panel~a): Data for the HII~knots and the LIRGs. 
The best fit line through the high metallicity HII~knots (continuous line) is not significantly 
different from a linear relation with slope of unity (dashed line). (Panel~b): Histogram of the deviation of the HII~knot data in panel~(a) from the best fit
line through the high metallicity data (the continuous line in
panel~(a)). Unlike Figures~\ref{fig3} and \ref{fig4}, the histograms of the medium and low 
metallicity datapoints have not been 
multiplied by a factor 2. The values of the x--label parameters ({\em c, d)} are derived from 
equation~4 and are c=(1.02$\pm$0.02) and d=($-$0.74$\pm$0.97). 
\label{fig7}}
\end{figure}

As suggested by \citet{kenn06}, the combination of measurements at 
H$\alpha$ and 24~$\mu$m can provide insights into both the unobscured and 
obscured regions of star formation. We have combined linearly the observed H$\alpha$ 
and 24~$\mu$m  LSDs and scaled them to the  Pa$\alpha$ LSD.  
The best fit line through the data is:
\begin{equation}
Log\ S_{Pa\alpha, corr} = (0.98\pm 0.02)\  Log\ (a S_{H\alpha, obs} + b S_{24~\mu m}) + (0.73\pm 0.93)
\end{equation}
where $a$ is the intrinsic Pa$\alpha$/H$\alpha$ ratio, thus is dictated by atomic physics and 
is only moderately dependent on metallicity ($a$=0.128, 0.118, and 0.114 for the high, medium, and low 
metallicity data, respectively; see section~3.2). The coefficient $b$ for the 
24~$\mu$m LSD has been empirically rescaled to bring the sum of the optical and IR LSDs in 
agreement with the Pa$\alpha$ one ($b$=0.0040, 0.0037, and 0.0036 for the high, medium, and 
low metallicity datapoints, respectively; Figure~\ref{fig7}). The best fit from equation~4 gives 
$b/a$=0.031$\pm$0.006, and this ratio is independent of metallicity. 
Equation~4 is, within the uncertainties, 
consistent with a linear relation with null intercept  between the two quantities, as expected if 
the right--hand--side expression is a measure of the ionizing photon rate,  like 
S$_{Pa\alpha, corr}$. The linearity of the relation is by construction, as the requirement is to approach 
unity as much as possible for all the combined data, but the null intercept has not 
been fixed a priori; furthermore, the ratio b/a was left as a free parameter in the analysis, and its 
constant value is a result (not an input). 

Interestingly, the high metallicity datapoints show approximately the same dispersion
around the mean trend of equation~4 as they do for equations~2 and 3, with a 
1~$\sigma\sim$0.3~dex.  In the case of the combined optical/mid--IR, the dispersion 
is the same whether the high metallicity datapoints alone or all datapoints are included in 
the statistical analysis (panel~(b) of Figure~\ref{fig7}). Conversely,  for the two mid--IR LSDs 
the dispersion  
is measured for the high metallicity datapoints only, and increases substantially (on one side) 
when the medium and low metallicity datapoints are included in the statistics 
(panels~(b) of Figures~\ref{fig3} and \ref{fig4}). These considerations do 
not include the LIRGs, that in Figure~\ref{fig7} show evidence of having  
higher combined optical/mid--IR LSDs than inferred from the extrapolation of equation~4.  A possible 
explanation for this effect will be discussed in Section~7.

As already discussed in \citet{kenn06}, the sum on the right--hand--side 
of equation~4 can be interpreted as a representation of the dust extinction corrected
H$\alpha$ luminosity or LSD. as: 
\begin{equation} 
S_{H\alpha, corr} = S_{H\alpha, obs} + (0.031\pm 0.006) S_{24~\mu m}.  
\end{equation} 
The proportionality coefficient for the 24~$\mu$m luminosity is $\sim$20\%
smaller than that derived for NGC5194 alone \citep{kenn06}, which is within the 
1~$\sigma$ uncertainty. This small difference is likely due to the larger variety of galaxies 
used in the present work which provides a dynamical range in  luminosity surface density about 
an order of magnitude larger than in the NGC5194 case.  

The proportionality coefficient for the 24~$\mu$m emission in equations~4 and 5, b/a=0.031, is independent of metallicity. This 
suggests  that in the S$_{24 \mu m}$ versus S$_{Pa\alpha, corr}$ plane the observed deviations
of the medium and  low metallicity data from the best fit for the high--metallicity datapoints 
are simply due to the progressively lower dust content of the ISM for 
decreasing metallicity  (section~7). No other effect  
beyond the simple increase in the medium's transparency is required. Indeed, most of the 
contribution to S$_{H\alpha, corr}$ comes from the observed H$\alpha$ emission at low 
S$_{Pa\alpha, corr}$ LSDs  (low dust systems) 
and, vice-versa, it is mainly contributed by the 24~$\mu$m emission at the high LSD end of our sample 
(dusty systems). 

\begin{figure}
\figurenum{8}
\plotone{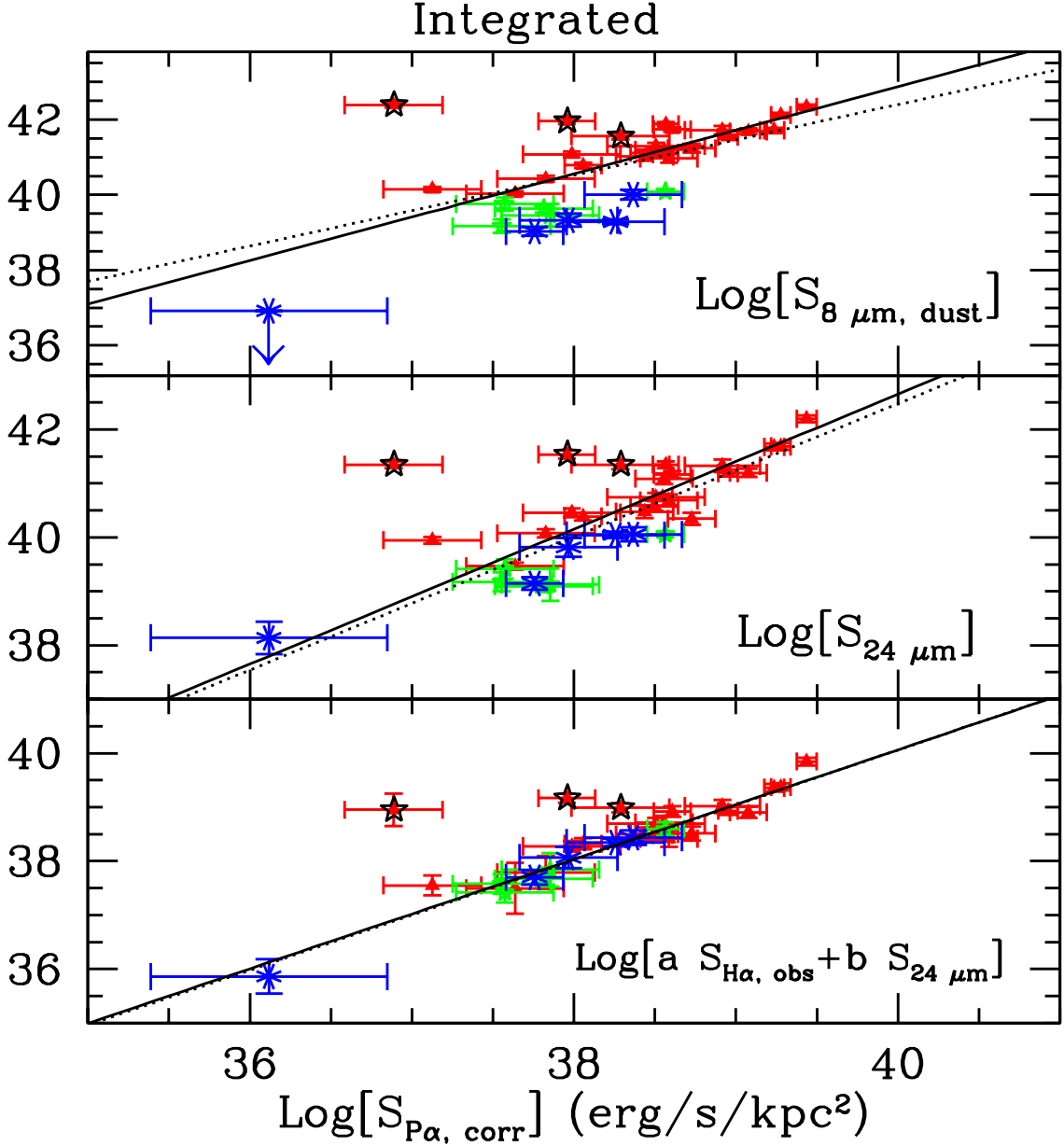}
\caption{ Mid--IR or combined--optical/mid--IR 
LSD as a function of the extinction--corrected Pa$\alpha$ LSD, averaged 
over the entire central region covered by the HST/NICMOS observations, for each of the 
star--forming galaxies from Table~1 (section~4.1). 
The three panels show on the vertical 
axis, from top to bottom, the 8~$\mu$m, 24~$\mu$m, and the linear combination of H$\alpha$ and 
24~$\mu$m LSDs  in logarithmic scale, as indicated by the label at the bottom--right corner of each 
plot; the vertical axis has the same units as the horizontal axis, erg~s$^{-1}$~kpc$^{-2}$. 
Each HST/NICMOS image is  about 50$^{\prime\prime}$ in size, except for NGC5194,  where a 
region of 144$^{\prime\prime}$ in size has been observed. 
 Color coding of each galaxy is the same as the HII~knots in Figure~\ref{fig3}. The continuous lines are
  the best linear fit through the high metallicity (red triangles) data, after excluding the Sy2--dominated fluxes of NGC4569, NGC4736, and NGC5195 (marked as black stars), thus leaving 19 independent datapoints.   The dotted lines are the best fits through the high 
  metallicity HII~knots  from Figures~\ref{fig3}, \ref{fig4}, and \ref{fig7} for S$_{8 \mu m,\ dust}$, 
  S$_{24 \mu m}$, and a~S$_{H\alpha, obs}+$b\ S$_{24 \mu m}$, respectively.
\label{fig8}}
\end{figure}

Similar correlations as those seen for the HII~knots  within galaxies 
exist between the integrated LSDs of the galaxies' centers (section~4.1). 
Figure~\ref{fig8} shows the trends for the 33 star--forming galaxies in our main sample. For
the combined optical/mid--IR LSDs, a linear fit 
through the integrated datapoints of the high metallicity galaxies are
consistent, within 1~$\sigma$, with the best fit lines through the individual HII~knots, both in slope and intercept (third panel of Figure~\ref{fig8}). For the 24~$\mu$m LSD, the 
slope of the linear fit is consistent (again within 1~$\sigma$) with that of the individual 
HII~knots, and the intercept is consistent 
(within 0.1~$\sigma$) with the value expected 
by simply rescaling the HII~knots' mean LSD for the larger area used in the 
integrated measures. The results for both the 24~$\mu$m and the combined 
optical/mid--IR integrated measures suggest that within the central areas covered by the 
NICMOS observations any diffuse 24~$\mu$m emission contributing to the 
measured LSD is matched by diffuse Pa$\alpha$ LSD with comparable intensity.  This, of course, 
does not 
mean that diffuse 24~$\mu$m emission is not present; indeed, such diffuse emission has been 
observed in the SINGS galaxies \citep{dale06}. Our result simply implies that such diffuse 24~$\mu$m emission traces the diffuse ionized emission, at least within the central galaxy regions sampled 
by our data.

A more complicated scenario appears for the 8~$\mu$m LSD: a best fit line through the 
high~metallicity integrated regions  produces a higher slope (1.16$\pm$0.09) than 
derived for the individual HII~knots. The difference is marginally significant (2.2~$\sigma$), but  
implies that the 8~$\mu$m LSD is higher by about a factor of 2 over what is expected from a simple 
rescaling of areas at the high luminosity end\footnote{For the `integrated'  diffuse emission, the  
extended source aperture correction provided by \citet{jarr06} has been used.}. 
A visual inspection of the images shows that the 
galaxies with low 8~$\mu$m LSDs generally have line and mid--IR emission which is centrally 
concentrated or coming from thin, almost edge--on, disks or annuli located in the central 
50$^{\prime\prime}$, while at the high 8~$\mu$m LSD
end galaxies tend to have a more homogeneous distribution of HII~knots.

\section{Comparison with Models}

To help clarify the nature of some of the characteristics of  the observed correlations, this 
section is devoted to the  comparison of our data with  simple models that exploit the energy 
balance between the stellar light absorbed by dust at 
UV, optical and near--infrared wavelengths, and the light emitted by the dust in the mid-- and 
far--infrared. The details of the 
models are presented in the Appendix, in addition to a discussion on limitations to their use and 
applicability. Here we provide a brief summary of those models. 

The basic approach adopts a range of plausible stellar populations for our HII~knots 
(and starburst galaxies), in terms of star formation histories, stellar population ages, and 
metallicities 
\citep[2005 update of Starburst99\footnote{http://www.stsci.edu/science/starburst99/}, ][]{leit99}. Simple assumptions are also made for the ISM structure and metal content. 
The intrinsic stellar populations are then dust--attenuated according to empirical recipes 
 \citep{calz94,meur99,calz00,calz01} to provide a `predicted'  
 infrared emission, S$_{IR}$. As the stellar populations probed in our analysis range from 
 groupings of a few to several HII~regions for the HII~knots to populations with extended star 
 formation histories in the case of  starbursts and LIRGs, both  instantaneous bursts and constant 
 star formation populations are included. The 
 total infrared emission will, in general, depend not only on the adopted stellar population, 
 but also on the  extinction curve and the dust geometry. Since for the last two parameters, we 
 make a simplifying assumption and use the prescription of 
 \citet{calz01}; the impact of varying the dust geometry is discussed in section~A.4. 
 For the spectral energy distribution (SED) of the infrared emission, S$_{IR}$, we adopt the  model 
 of \citet{drai06}, according to which the fraction of IR 
 power emerging in the IRAC 8~$\mu$m and MIPS~24~$\mu$m bands is a function of the starlight 
 intensity. We determine (section~A.2) the range of starlight intensities corresponding to the model 
 stellar populations we are considering, so to obtain a direct  correlation between 
 the Pa$\alpha$ LSD and the fraction of IR light emerging in the two mid--IR bands.  
 Since our HII~knots follow the well known correlation between SFR and extinction 
\citep[Section~A.1 and][]{wang96,heck98,hopk01,calz01,mous06}, which we parametrize as a 
relation between color excess E(B$-$V) (section~3.2) and the ionizing photon rate per unit area 
$\Sigma_{ion}$, we use this relation to link the stellar population models to the dust attenuation model,
 and eliminate one degree of freedom in our models. 
 Model parameters that we allow to vary are  the star formation history of the 
 stellar populations (bursts or constant star formation), their age (0--10~Myr for instantanous 
 bursts, the range chosen to ensure presence of significant ionizing photon rate, \citet{leit99}; 
 6--100~Myr for constant star formation), the mass  
 (10$^3$--10$^8$~M$_{\odot}$) or  
 SFR (4$\times$10$^{-5}$--4~M$_{\odot}$~yr$^{-1}$) 
 of the stellar cluster(s) associated with the HII~knot or starburst galaxy, and the metallicity of both the 
 population and  the interstellar  medium (0.1--1~Z$_{\odot}$\footnote{We adopt
  the oxygen abundance 12+log(O/H)=8.7 as solar metallicity value \citep{alle01}, which we take here 
 as representative of our high--metallicity HII~knots.}).   Figures~\ref{fig9}--\ref{fig11} show the basic results from the 
 comparison between the  models described so far and our data for the 8~$\mu$m, 24~$\mu$m, 
 and combined optical/mid--IR emission from HII~knots and star--forming galaxies.
 
The larger--than--unity  slope of the 24~$\mu$m versus Pa$\alpha$ LSD (in log--log 
scale, Figure~\ref{fig9}) is a natural outcome of the models in the high luminosity surface density 
regime, Log(S$_{Pa\alpha, obs}$)$>$39, and is an effect of the `hotter' IR SEDs  
for increasing starlight intensity. In other words, regions with higher 
Pa$\alpha$ LSD emit proportionally more of their infrared energy into the 
24~$\mu$m band, because the peak of the IR SED moves towards shorter wavelengths (higher 
`effective' dust temperatures, see Appendix and \citet{drai06}). 
The models also predict a slightly larger than unity value for the slope of the 
8~$\mu$m LSD correlation, which is steeper than that  of the HII~knot data 
(Figure~\ref{fig10}), but is roughly consistent with the slope of the integrated measures. 

The models account well for the linear relation of the combined optical/mid--IR LSD  
with the Pa$\alpha$ LSD (Figure~\ref{fig11}), for luminosity surface densities 
S$_{Pa\alpha, corr}<$10$^{40}$~erg~s$^{-1}$~kpc$^{-2}$. At high luminosity surface density, the   
models  for the combined LSDs depart from a linear relationship, as increased starlight 
intensities are expected to raise the temperatures of the larger grains so that the fraction 
of the absorbed energy re--radiated at 24~$\mu$m (which is, at these high LSDs, the dominant 
contribution to equation~5) increases. 
The LIRGs data, that populate the high LSD regime in our plot, do indeed confirm 
observationally the deviation from the extrapolation of the best fit line; they show a 
steeper--than--one slope, in qualitative agreement with the models'  expectations (Figure~\ref{fig11}). 

At the high luminosity end (LIRGs and brighter), an additional effect that can contribute to the deviation 
from the slope of unity observed in Figure~\ref{fig11} and the steeper--than---unity slope of 
Figure~\ref{fig9} is the competition between the dust and the gas for the absorption of some of the 
ionizing photons. In the high luminosity regime, star formation occurs in environments of increasing 
density, e.g., ultracompact HII regions \citep{rigb04}, and the dust absorbs the ionizing photons before 
they can excite the gas. In this regime, standard extinction--correction methods become progressively less effective at recovering the intrinsic Pa$\alpha$ emission, and will produce an underestimate of the 
hydrogen emission line LSD at constant 24~$\mu$m LSD (section~A4). The impact of this effect 
on our data is unclear (and currently not included in our models), although it may be relatively small 
as the bulk of the observed trends is fully accounted for by our baseline model. 

Instantaneous burst populations and constant star formation populations produce mostly degenerate 
models for all three mid--IR quantities (Figures~\ref{fig9}--\ref{fig11}). A young, 4~Myr~old,  instantaneous burst population in the 
mass range $\sim$10$^3$--10$^8$~M$_{\odot}$ provides similar model lines as a constant 
star formation model forming stars since 100~Myr and with SFR in the range 
4$\times$10$^{-5}$--4~M$_{\odot}$~yr$^{-1}$.  

However, 
even the high--metallicity HII~knots in Figures~\ref{fig9}--\ref{fig10} show a fairly large dispersion 
around the mean trends described above, with a clear increase of the dispersion around 
the mean S$_{8~\mu m,\ dust}$ and S$_{24~\mu m}$ values for 
S$_{Pa\alpha, corr}\le$10$^{39}$~erg~s$^{-1}$~kpc$^{-2}$. Furthermore, in this Pa$\alpha$ 
LSD regime, most 
of the 8~$\mu$m and 24~$\mu$m emission from the high--metallicity HII~knots is located 
above the baseline model lines, i.e., the models underpredict the mean values of the mid--IR emission 
(Figures~\ref{fig9}--\ref{fig10}). The `downward' curvature 
of the models is a direct product of the 
increasing transparency of the interstellar medium for decreasing ionizing photon rate density and, 
from equation~A2, decreasing dust amount. With a more transparent medium, 
proportionally less IR radiation is produced. The medium is still thick to Lyman continuum 
photons, and the ionized hydrogen emission lines are still a measure of the total number of 
ionizing photons in the region. An additional parameter is thus required to account 
for both the large scatter of the datapoints around the mean trends and the large number of 
high--metallicity datapoints above the model lines for the 
S$_{8 \mu m,\ dust}$ and S$_{24 \mu m}$ LSD plots.  {\em This second parameter appears to be 
 the age of the stellar population}. Ageing bursts  between  $\sim$0.01~Myr and $\sim$8~Myr  
produce a decreasing number of ionizing photons, while at the same time  
remaining luminous at UV--optical wavelengths (the major contributors to the IR emission). 
Figures~\ref{fig9}--\ref{fig11} show that the `flaring' of the high--metallicity HII~knots datapoints 
around the mean value for decreasing Pa$\alpha$ LSD is compatible with the `flaring' of the 
ageing burst models.  Such ageing populations can also account for the data points above the 
mean trends in Figures~\ref{fig9} and \ref{fig10}.

The presence of ageing bursts is a sufficient (and physically expected), but 
not a necessary, condition to account for the dispersion in the data. 
As briefly discussed in the Appendix (section~A.4), different assumptions from our 
default one about the average dust geometry can also produce a higher mid--IR emission 
than our fiducial model lines. For instance, presence of ultracompact HII regions within our HII~knots 
will produce higher IR emission at fixed S$_{Pa\alpha, corr}$ than expected from the models. 
This is a consequence of the higher 
opacity of such regions, for which the use of the H$\alpha$/Pa$\alpha$ ratio to recover the intrinsic 
line fluxes will lead to an underestimate of the intrinsic Pa$\alpha$ luminosity in the region.  Recently, 
\citet{dale06} have shown that for local star--forming galaxies the UV/IR ratio is heavily determined 
by the morphology of the 24~$\mu$m dust emission, in particular by the `clumpiness' of such 
emission, which therefore determines the escape fraction of UV photons from star--forming regions. 
A clumpy configuration of dust is, however, well described by the empirical recipes of dust extinction 
and attenuation used in the present work \citep{calz94,meur99,calz01}.

For the combined optical/mid--IR LSD, the models are degenerate as a function of metallicity 
(Figure~\ref{fig11}). 
This is not surprising if the main driver of the discrepancy  between the high and low 
metallicity S$_{24 \mu m}$ at 
fixed Pa$\alpha$ LSD  is the larger medium transparency, i.e., lower dust column density, 
 in the lower metallicity data (equations~A2 and A4). This is indeed the case (Figure~\ref{fig9}): the 
 separation at low Pa$\alpha$ LSD between the solar metallicity  and the 1/10th solar metallicity 
model lines is mostly due to the metallicity scaling factor in equations~A2 and A4, and, to a much 
smaller extent, to  the difference in metallicity of the two stellar populations. The 1/10th metallicity 
model line in Figure~\ref{fig9} provides the lower envelope to the data; most of the 
galaxies in our sample are above 1/10th solar in metallicity, and thus are expected to lie  
above this model line. 

This result lends credence to the use of a combination of S$_{24 \mu m}$ 
and S$_{H\alpha, obs}$ \citep[equation~5 and][]{kenn06} as an effective tool for 
measuring the ionizing photon rates, and, ultimately, SFRs, at least up to Pa$\alpha$ 
LSDs $\approx$10$^{40}$--10$^{41}$~erg~s$^{-1}$~kpc$^{-2}$. In this framework, S$_{24 \mu m}$  probes the obscured star formation, and the only metallicity effects are those induced by reduced opacity; conversely, S$_{H\alpha, obs}$ probes that part of the star formation unabsorbed by the 
dust, independent of the gas metallicity.  The behavior of the models in Figure~\ref{fig11} shows   
little difference between different parameters choices, at least within our data uncertainties, and they reproduce the main trend of the data reasonably well.

The discrepancy observed between the high metallicity and low metallicity S$_{8 \mu m,\ dust}$ data 
at fixed Pa$\alpha$ LSD requires one additional ingredient, together with the increased 
transparency of the medium. \citet{drai06} have suggested that  
the fraction of low--mass PAH molecules present in the dust mixture decreases for 
decreasing metallicity. In the Appendix, we show that the two ingredients (increased medium transparency and decrease of low--mass PAH molecule fraction) provide comparable 
contributions to the depression of the 8~$\mu$m emission, and the two together produce the expected 
lower envelope to the datapoints in Figure~\ref{fig10}. 

\section{Discussion}

The scope of this study has been to investigate the extent of the regime of 
applicability of mid--IR emission as a SFR tracer, to use models to reproduce the main 
characteristics of the data, and to investigate reasons for any limitation we have encountered.
The general trend of mid--infrared  luminosity surface densities to correlate
with the ionizing photon rates or with SFR tracers 
had already been found by a number of authors 
\citep[for some of the most recent results, see][]{rous01,forst04,bose04,calz05,wu05,alon06}. 

\subsection{The Combined Optical/Mid--IR SFR Indicator}

Of the three indicators investigated here, the linear combination of
the observed H$\alpha$ and the 24~$\mu$m emission is the one
most tightly correlated with the extinction--corrected Pa$\alpha$
emission. The linear relation between the combined optical/mid--IR emission 
and the SFR as traced by S$_{Pa\alpha, corr}$ is common to all galaxies investigated, 
independent of their metallicity. The most
straightforward interpretation \citep{kenn06} is that the 24~$\mu$m
emission traces the dust--obscured star--formation, while the observed
H$\alpha$ emission traces the unobscured one. The combination of the
two, thus, recovers all the star formation in a region. This interpretation is 
confirmed by the models investigated in the previous section, which 
also suggest the trend to be relatively independent of the characteristics of 
the underlying star--forming population. The deviations 
from the linear relation (i.e., from a slope of 1 in log--log scale, Figure~\ref{fig11})
 observed at  luminosities larger than 
 S$_{Pa\alpha, corr}>$10$^{40}$--10$^{41}$~erg~s$^{-1}$~kpc$^{-2}$ are also 
 consistent with the models' expectations:   as the 24~$\mu$m emission starts dominating 
the luminosity budget, the same physical mechanism  producing the 
S$_{24~\mu m}$--versus--S$_{Pa\alpha, corr}$ trend also produces  that of the  
combined optical/mid--IR indicator. We speculate that this mechanism (see below) is  
the emission from grains with approximately steady temperatures, rather than 
transiently--heated grains, that come into play at high dust temperatures, and 
which causes the ratio of 24~$\mu$m/IR to deviate from a constant value at low starlight intensities 
to one that increases with the starlight intensity. 

Using our baseline best--fitting model of 100~Myr constant SFR, for solar 
metallicity and the stellar initial mass function (IMF)  described in section~A.2, 
the conversion between SFR and H$\alpha$ luminosity is: 
\begin{equation}
SFR(M_{\odot}~yr^{-1}) = 5.3 \times 10^{-42} L(H\alpha)_{corr} (erg~s^{-1}).
\end{equation}
Variations of $\pm$20\% over the constant in this relation are present for younger 
ages and metallicities down to $\sim$1/5th solar. The $\sim$50\% difference between 
the calibration in Equation~6 and that of  \citet{kenn98} is mainly due to differences in 
the stellar IMF assumptions (59\%), with a small contribution in the opposite direction 
coming from different assumptions on the stellar populations (100~Myr in our case versus 
infinite age in \citet{kenn98}, which gives a 6\% decrease to the discrepancy given by the 
different IMFs).  
Using equations~5 and 6:
\begin{equation}
SFR(M_{\odot}~yr^{-1}) = 5.3\times 10^{-42} [L(H\alpha)_{obs} + (0.031\pm0.006) L(24~\mu m)],
\end{equation}
where the luminosities are in erg~s$^{-1}$, and L(24~$\mu$m) is
expressed as $\nu$L($\nu$).  This calibration does not change if the
luminosities are measured over a substantial area of the galaxy (in
our case the inner $\sim$0.8 to $\sim$5.1 kpc), rather than in smaller
regions hugging the HII complexes that produce the ionizing radiation
(Figures~\ref{fig7} and \ref{fig8}). However, the potential non--linearity at large 
LSDs is an important caveat. 

\subsection{The 24~$\mu$m SFR Indicator}

Conversely, neither the 8~$\mu$m emission nor the 24~$\mu$m emission
alone are linearly correlated with the number of ionizing photons that
are measured in a region. The 
non--linearity at high 24~$\mu$m luminosity is  a direct consequence of the 
increasing dust temperature for more actively star forming 
objects \citep{li01,drai06,dale01}; higher dust temperatures correspond to higher fractions 
of the infrared emission emerging at mid--IR wavelengths. 
Following \citet{drai06}, in the regime of low 
stellar intensities (low SFRs in our actively star--forming regions, or roughly 
S$_{Pa\alpha, corr}<10^{39}$~erg~s$^{-1}$~kpc$^{-2}$), most of the 
24~$\mu$m emission comes from single photon transient heating of small grains.
In this case, the 24~$\mu$m photon flux is directly proportional to the
stellar UV photon flux (or any other photon capable of single photon heating).  Thus, the
24~$\mu$m emission counts stellar UV/optical photons, while the Pa$\alpha$ 
counts the Lyman continuum photons;  since there is  proportionality
between the two types of photons, the expectation is for a linear scaling between 24~$\mu$m 
emission and Pa$\alpha$ emission. However, this regime corresponds to the Pa$\alpha$ LSD range  
where the decrease of dust opacity also decreases non--linearly with the amount of stellar energy 
re-processed by dust in the infrared. At high Pa$\alpha$ LSDs,  hence high stellar intensities, 
the dust absorbing most of the stellar photons is warm. There is thus an increasing contribution to 
the 24~$\mu$m emission from larger, warm grains (the Wien side of the 
emission from grains), which leads to a non--linear  dependence of the 24~$\mu$m flux on the 
stellar flux. This is in agreement with the conclusions of \citet{smith07}, which observe a 
decrease of the PAH/24~$\mu$m luminosity ratio for increasing 24~$\mu$m/70~$\mu$m 
luminosity ratio;  this dependence is highly suggestive of an increasing contribution of warm 
dust to the 24~$\mu$m emission. 

The observed non--linearity  in the 24~$\mu$m versus Pa$\alpha$ relation, 
S$_{24~\mu m}\propto$S$_{Pa\alpha, corr}^{1.23}$,  also argues against the case that the high 
LSDs values measured of our apertures may be due to the cumulative contribution of many faint HII~regions, rather than a few, increasingly bright HII~regions. In the case of many faint HII~regions 
(low stellar 
intensities, and, therefore, single photon heating) we should expect the 24~$\mu$m LSD to 
scale linearly with the Pa$\alpha$ LSD at the high end. The observed non--linear behavior argues 
in favor of the high luminosity end to be contributed mainly by intrinsically bright regions, although 
presence of apertures with many faint HII~regions cumulatively giving a high LSD may still be 
present and  contribute to the scatter of the datapoints around the mean trend.  

At low metallicities, the deviation from a linear correlation is due 
to lower opacities for decreasing metal content, and thus column densities \citep{walt07}. The effect 
has been well known since the early IRAS observations \citep{helo88}: as the metallicity decreases, regions become proportionally more transparent and emit less in the infrared as a
larger fraction of the radiation escapes the area unabsorbed by
dust. This accounts for the underluminosity of the medium in low metallicity regions  
in correspondingly intense ionizing fields. Using the 24~$\mu$m luminosity as a SFR tracer is thus
subject to many caveats, including that lower metallicity sources 
will generally be more transparent than their metal-rich counterparts,
and the infrared emission will typically underestimate 
their SFR by a factor $\sim$2--4. 

The non--linear correlation between S$_{24~\mu m}$ and S$_{Pa\alpha, corr}$  
requires some care for deriving SFR calibrations. From equation~3, and using equation~6, 
we derive a SFR density (SFR per unit area) calibration:
\begin{equation}
\Sigma_{SFR} (M_{\odot}~yr^{-1}~kpc^{-2}) = 1.56\times 10^{-35} [S_{24~\mu m} \ (erg~s^{-1}~kpc^{-2})]^{0.8104} .
\end{equation}
In order to derive a calibration for SFRs, we convert our LSDs into luminosities, and the resulting 
best fit through the high--metallicity datapoints produces:
\begin{equation}
SFR(M_{\odot}~yr^{-1}) = 1.27\times 10^{-38} [L_{24~\mu m} \ (erg~s^{-1})]^{0.8850} .
\end{equation}
The exponents of equations~8 and 9 are the same within the 3~$\sigma$ error (the combined 
1~$\sigma$ uncertainty is 0.03). Equation~9 is closer to a linear relation than equation~8, 
 because  we add a distance effect when using luminosities (which depend on the distance squared). 
 A large  sample of regions with comparable distances may be needed to fully sort out 
 intrinsic  effects from  distance--related effects. 
 Both relations are derived from best bi-linear fitting of data on HII~knots, but their extrapolations 
 account for the observed properties of LIRGs as well. Additionally, when considering more extended galactic regions or starburst galaxies, equation~8 does not change significantly (Figures~\ref{fig4} 
 and \ref{fig8}). Equations~8 and 9 may thus be applicable to galaxies in general whose energy 
 output is  dominated by recent star formation.  

Equation~9 is very similar to that of \citet{alon06}, who have 
derived a SFR calibration for the 24~$\mu$m luminosity using a sample of Ultraluminous Infrared 
Galaxies, LIRGs, and NGC5194. The difference in the calibration constant between our equation~9 
and the calibration of \citet{alon06} is entirely due to the slight difference in exponent between the 
two relations, and the different SFR--L(H$\alpha$) calibrations used here and in that work. 
\citet{gonz06} find a lower exponent, $\sim$0.77, than the one in equation~9, about a 
4~$\sigma$ difference; however, their result is based on line emitting regions in just two 
galaxies, NGC5194 and NGC3031.

\subsection{The 8~$\mu$m Emission}

The analysis of the HII~knots in M51 has shown a general, non--linear correlation between the 
8~$\mu$m and the Pa$\alpha$ emission \citep[with exponent 0.79,][]{calz05}. The present study similarly 
recovers a non--linear behavior for the high--metallicity data, albeit less extreme than in the M51 case:  
S$_{8 \mu m,\ dust}\propto$ S$_{Pa\alpha, corr}^{0.94}$. Our simple Z=Z$_{\odot}$ models also 
predict a non--linear correlation between the 8~$\mu$m and Pa$\alpha$ emission, but with an 
exponent slightly {\em above} unity. Therefore, the gap between 
observations and expectations is even wider than a simple deviation from a linear correlation; 
with our uncertainties, the discrepancy is at the 10~$\sigma$ level.  This level of discrepancy 
remains unchanged when other uncertainties, e.g., on the dust modelling and on the correlation 
between dust attenuation and number of ionizing photons (see Appendix), are included. In contrast, 
the same simple models are quite successful at explaining the observed trend of S$_{24~\mu m}$. We conclude that the 8~$\mu$m emission as measured within our apertures must include 
additional contributions that  are not included in our simplified models. 

Mechanisms that can produce a lower--than--expected slope in a correlation between 
S$_{8 \mu m,\ dust}$ and S$_{Pa\alpha, corr}$ include the potential contamination of our 
measurements by the diffuse emission from the general galactic radiation field \citep{li02,haas02,bose04,peete04,wu05,matt05} and/or 
destruction/fragmentation of the  8~$\mu$m emission carriers \citep{boul88,boul90,helo91,houc04,pety05}. In the case of destruction or fragmentation of 
the PAH emitters at 8~$\mu$m, the brightest HII regions will show a deficiency in the 8~$\mu$m 
luminosity relative to the fainter regions. In  the case where non--ionizing populations, as well 
as ionizing ones, heat the 8~$\mu$m dust carriers, 
the contribution of the former to the S$_{8 \mu m,\ dust}$ measurements within our apertures will 
become proportionally larger as the HII regions become fainter (decreasing S$_{Pa\alpha, corr}$), 
again flattening the observed trend.  Finally, if the volume filling factor of the 8~$\mu$m 
luminosity originating in the PDRs evolves differently from that of the HII regions 
 (or other inhomogeneities in the 8~$\mu$m emission distribution are present) as the 
HII region's luminosity increases \citep{forst04,helo04}, the net result will also be a 
lower--than--expected exponent between S$_{8 \mu m,\ dust}$ and S$_{Pa\alpha, corr}$. 

The correlation exponent predicted by models of the 8~$\mu$m LSD versus Pa$\alpha$ LSD 
is better matched by the data of large--scale (multiple--kpc) measurements of the centers of the 
high--metallicity galaxies (Figure~\ref{fig8}).
There is, however, an offset between models and observations, in the sense that the data imply 
about 65\%--100\% more 8~$\mu$m emission than expectations from models. This result  argues in 
favor of  one of the mechanisms described above: the presence within our apertures 
of diffuse 8~$\mu$m emission unrelated to the current star formation. Indeed the result can be 
explained if  stellar populations other than those related to the current star formation can 
heat the carriers of the 8~$\mu$m emission, and their contribution becomes proportionally 
larger than that of the ionizing populations as the size of the sampled region within each galaxy 
increases.

The underluminosity of the 8~$\mu$m emission at low metallicities 
\citep{bose04,enge05,hogg05,gall05,rose06,madd06,wu06,drai07} 
is not a function of the intensity of the radiation field (Figure~\ref{fig6}). Metal--poor 
regions as bright (in ionizing photon density) as metal--rich region have, 
neverthless, 8~$\mu$m LSDs that are almost an order of 
magnitude lower.  The most metal--poor regions can be as much as a 
factor of 30 fainter at 8~$\mu$m than their metal--rich counterparts at fixed Pa$\alpha$ LSD. 
In addition to the decrease of the dust opacity with metallicity (the same effect present for the 
24~$\mu$m), the data require a second ingredient to account for the 
underluminosity at 8~$\mu$m. The second ingredient,  from the models of \citet{drai06}, 
is the decrease of the mass fraction of the low--mass polycyclic aromatic hydrocarbon  
molecules in the dust mixture for decreasing metallicity. This decrease affects selectively more 
the 8~$\mu$m emission than the 24~$\mu$m emission, as the former has a larger fraction 
of its flux contributed by single--photon--excitation of PAHs. The models, indeed, account reasonably 
well for the observed deficiency of the 8~$\mu$m flux in metal--poor objects \citep{drai07}. 
A possible mechanism to reduce the mass fraction of PAH molecules in the
low--metallicity galaxies is destruction by the hard radiation field in those
galaxies \citep[e.g.,][]{madd06,wu06}, although these molecules have proven
surprisingly robust \citep{enge06}.  An alternative possibility to destruction
is that the lowest metallicity galaxies may not have formed the carriers of
the aromatic features in the first place \citep{enge05,dwek04}.

The correlation that exists between Log[S$_{8~\mu m,\ dust}$] and
Log[S$_{Pa\alpha, corr}$] for the metal--rich regions (equation~3) is statistically as 
significant as those existing for the 24~$\mu$m and for the combined optical/mid--IR 
luminosity. However, there are many caveats in  using such correlation to trace SFRs 
with the 8~$\mu$m emission. First and foremost, stellar populations other than those 
which are currently forming stars contribute to the mid--infrared emission: thus, 
a calibration of the 8~$\mu$m emission as a SFR tracer will depend on the galactic 
area probed, and, possibly, on the ratio between current and past (or recent--past) 
star formation.  A second limitation is the extreme sensitivity of the 
 8~$\mu$m emission to metallicity, which is about an order of magnitude larger than 
what  observed for the 24~$\mu$m emission. 

In all cases, the presence of Active Galactic Nuclei in unresolved galaxies will complicate the use 
of the mid--IR band emission for SFR determinations. Galaxies in the SINGS sample that 
contain central non--thermal sources (Sy2s or LINERs)  show prominent nuclear emission 
in the mid--IR bands; in our case, the 8~$\mu$m and 24~$\mu$m emission from the non--thermal 
nuclei tends to be as bright as some of the brightest HII~knots in their host galaxies (Figures~\ref{fig3} and 
\ref{fig4}).

\section{Summary and Conclusions}

The analysis of the mid--infrared emission, at the Spitzer's IRAC 8~$\mu$m and MIPS 
24~$\mu$m wavelengths, from a set of local galaxies drawn from the SINGS sample has 
shown that their  viability as SFR indicators is subject to a number of caveats.  The calibrations 
given in this work, provided the caveats are taken into account, should however be useful for 
measurements of SFRs in actively star--forming galaxies using their 
observed mid--infrared emission, in the absence of prominent central AGNs. 

The most robust of the indicators analyzed in this study is the one proposed by \citet{kenn06}, 
which combines the observed H$\alpha$ and 24~$\mu$m luminosities as probes of the total 
number of ionizing photons present in a region. Here we present a calibration (equation~7) for that 
relation that is based on measurements of 220 HII~regions/complexes in 33 nearby galaxies. 
Comparisons with models suggest that the calibration should be applicable 
to large systems or galaxies whose energy output is dominated by young stellar populations, 
up to at least luminosity surface densities of S$_{Pa\alpha, corr}>$10$^{40}$--10$^{41}$~erg~s$^{-1}$~kpc$^{-2}$. 

Second--best is the 24~$\mu$m emission alone, which, although robust (within the limits of our 
analysis) for metal--rich objects, shows substantial deviations, at the level of factors 2--4,  from 
the mean trend for 
decreasing metallicity. Models can account for the observed deviations as an effect of the 
increased transparency of the medium for lower metal abundances (less dust means less 
infrared emission). The calibration presented here (equations~8 and 9) is non--linear and reflects 
both observations and expectations from models. The calibration is appropriate for metal--rich 
HII~regions or starbursts, and caution should be exercised when applying it to extended 
star--forming galaxies \citep{kenn06b} or to metal--poor targets.  In particular, galaxies with a 
prominent contribution from non--star--forming populations to the integrated light may receive a 
yet unquantified contribution to the 24~$\mu$m emission from a diffuse component. This diffuse 
component would not be directly related to star formation and may be in the single--photon--heating 
regime; its presence would add to the integrated 24~$\mu$m emission and would alter the calibration of 
this emission for SFR measurements.

We do not present a calibration of the SFR based on the 8~$\mu$m emission, as emission
at this wavelength shows strong dependence not only on metallicity (more than an order 
of magnitude), but also on the size (about a factor of 2) and, possibly, star formation history of the 
region being measured.  It should be remarked, however, that when measurements are limited 
to regions of star formation with a narrow spread in metallicity around the solar value, the 8~$\mu$m 
emission shows a correlation that is almost linear with the ionizing photon rate. 

The mean trends between the mid--IR emission (at 8~$\mu$m, 24~$\mu$m, 
and combined optical/mid--IR) and the ionizing photon tracer Pa$\alpha$ are  well accounted for 
by either   a 4~Myr old burst of star formation or  a $>$100~Myr old constant star formation 
 model; this provides a measure of the degeneracies in the population models, but also a measure 
 of the general applicability of the calibrations discussed here to both star--forming regions within 
 galaxies and whole star--formation--dominated galaxies.
  
In all cases, the spread around the mean trends has a r.m.s. of $\sim$0.3~dex (each side).  
This spread is  well accounted for by allowing  stellar populations with a range of ages, 
between 0~Myr and 8~Myr, to produce the observed emission. The 
older stellar populations can account for  HII~knots with high mid--IR emission relative to the ionizing 
photons density, although more complex dust geometries than those analyzed here can also contribute
to the spread. 
 
Our analysis has concentrated on HII~knots and starburst galaxies, where star--forming stellar 
populations dominate over more evolved populations in the bolometric output of the galaxy.  
However, the contribution of non--ionizing stellar populations to the heating of the dust emitting in the 
mid--IR region needs to be fully quantified, in order to test the applicability of (or derive 
modifications for) the SFR calibrations presented in this work to more general environments, 
such as quiescently star forming galaxies. This is the subject of a future investigation which employs
 the SINGS  and other star--forming galaxies as testbeds \citep{kenn06b}.

\acknowledgments

The authors would like to acknowledge the anonymous referee for the 
speedy report and for the many constructive comments that have helped 
improve the manuscript. 

This work has been partially supported by the NASA HST grant GO--9360
and by the JPL, Caltech, Contract Number 1224667. It is part of SINGS,
The Spitzer Infrared Nearby Galaxies Survey, one of the Spitzer Space
Telescope Legacy Science Programs.

This work has made use of the NASA/IPAC Extragalactic Database (NED),
which is operated by the Jet Propulsion Laboratory, California
Institute of Technology, under contract with the National Aeronautics
and Space Administration.

\appendix

\section{Models of Dust Absorption and Emission}

\subsection{The Energy Balance}

To aid the interpretation of the observed correlations between the mid-infrared emission 
and the Pa$\alpha$ emission (a tracer of the number of ionizing photons), we build a simple model 
of dust attenuation of the spectral energy distribution of a young stellar populations at UV--to--nearIR wavelengths, with the prescription that  the dust absorbed energy is re--emitted in the mid/farIR 
with a spectral energy distribution that follows the \citet{drai06} models. Ideally, the `perfect'  
estimator of SFR should correlate linearly with the SFR itself. We have seen in  section~6  
that this is not the case for S$_{24~\mu m}$ and  S$_{8~\mu m,\ dust}$; a legitimate question is whether 
the nature of the observed non--linearity is such to jeopardize a reliable use of the mid--IR emission 
as a SFR indicator.  Thus, our main interest in this section is to 
understand whether a heuristic model can account for the observed trends as a function of 
both luminosity surface density and metallicity, and is able to:
\begin{itemize}
\item reproduce the observed mid--IR emission level and its trend as a function of the Pa$\alpha$ 
LSD with simple prescriptions for the stellar emission spectrum, dust geometry, and ISM 
metallicity in the regions; 
\item provide a physical explanation for the non--linear correlations between the mid--IR 
LSDs and the line emission LSD; 
\item  match overall expectations on the age of the stellar populations in the `HII knots'.  
\end{itemize}

For the simplest assumption that the dust extinction is described as an effective foreground 
attenuation A($\lambda$) \citep{calz94,meur99}, the bolometric  (3--1000~$\mu$m) 
infrared luminosity L$_{IR}$ is given by:  
\begin{equation}
L_{IR} = \int_0^{\infty} F_S(\lambda) [1-10^{-0.4 A(\lambda)}] d\lambda,
\end{equation}
where  
F$_S$($\lambda$) is  the stellar light SED. We use 
the `foreground attenuation' as a working hypothesis, since it enables us to account for both 
absorption and scattering (in and out of the line of sight) of the stellar light by dust with a simple expression \citep{calz94,calz00}. 
Potential limitations to this assumption will be discussed in section~A.4. 

Our general results indicate a correlation between the mid--infrared emission and the number of ionizing photons in HII knots. Similar correlations have been observed for whole galaxies 
\citep[e.g., ][]{rous01,forst04,wu05,alon06}, and 
are qualitatively not dissimilar from the correlation between the bolometric infrared emission and the 
SFR  derived for galaxies dominated by young stellar populations 
using IRAS data \citep{pers87,rowa89,deve90,sauv92}. Equation~A1 shows that such correlations 
are mediated by the dust attenuation A($\lambda$); indeed, starburst  and star--forming galaxies have 
been shown to be redder and to suffer generally higher dust attenuation as their SFRs increase 
\citep{wang96,heck98,hopk01,calz01,mous06}. 
This general trend is also shown by the HII~knots in our sample. A plot 
of the color excess 
E(B$-$V)\footnote{A($\lambda$)=k($\lambda$)\ E(B$-$V), where k($\lambda$) is the extinction  
curve, section~3.2} , derived from the H$\alpha$/Pa$\alpha$ ratios of individual HII knots,  as a function of the Pa$\alpha$ LSD is shown in Figure~\ref{fig12}, together with the best  fit through the datapoints. The data in Figure~\ref{fig12} are for the high metallicity subsample, the group with the largest number 
of datapoints and the widest range in color excess in our sample. If we introduce the  ionizing 
photon rate per unit area, $\Sigma_{ion}$, in units of s$^{-1}$~kpc$^{-2}$ as   
derived directly from S$_{Pa\alpha, corr}$ for case B recombination,  then  in solar 
metallicity regions E(B$-$V) is related to the ionizing  photon rate via 
the best  fit (bi--linear on the log of the quantities):
\begin{equation}
E(B-V) = 0.21  \  \biggr({\Sigma_{ion}\over 10^{51} s^{-1} kpc^{-2}}\biggl)^{(0.61\pm0.03)}, 
\end{equation}
where E(B$-$V) is in magnitudes. The correlation has  5.7~$\sigma$ significance, and is 
independent of the specifics of the stellar population (age, burst or constant star formation, etc.). 

Some physical insights on the meaning of equation~A2 can be gained by looking at the 
implications of the scaling laws of star formation. The Schmidt Law \citep{kenn89,kenn98} provides 
a way to relate the SFR density to the gas density in 
galaxies. Furthermore, in our own Galaxy, there is a tight correlation between the color excess E(B$-$V) 
and the gas (HI$+$H$_2$) column density \citep{bohl78}. By combining the Schmidt Law, as derived locally for regions in NGC5194 \citep{kenn06}, and the extinction--gas~column~density correlation, we get a relation between the color excess and the SFR density: 
\begin{equation}
E(B-V) = 10.30\  \gamma\   \Sigma_{SFR}^{0.64},
\end{equation}
where  $\Sigma_{SFR}$ is in units of M$_{\odot}$~yr$^{-1}$~kpc$^{-2}$ and the factor $\gamma$ 
accounts for the fact that not 
all the gas (and the dust) is in front of the stars. Equation~A3 can be readily related to equation~A2, since the two exponents, 0.61 in the first and 0.64 in the second equation are formally the same 
number within the 1~$\sigma$ uncertainty of our fitting procedure  (see, also, 
Figure~\ref{fig12}). By fitting the data of Figure~\ref{fig12} with a straight 
line of fixed slope 0.64 (equation~A3), we get an estimate of $\gamma=$0.4--0.45 for the constant star 
formation models, depending on the time elapsed since the onset of star formation (6--100~Myr). 
This value of $\gamma$ is very close to the mean 
value of 0.5 expected in the case the star formation is located on average at the mid--point of the 
gas column density.  

Metallicity variations in the stellar populations produce small variations in equation~A2. However, 
observations of the extinction to gas column density correlation for the Large and Small Magellanic Clouds suggest roughly linear scaling  with metallicity \citep{koor81,bouc85}. This is likely to 
be the dominant dependence of equation~A2 on metallicity, and we model it as a linear 
dependence on $Z$, the interstellar medium's metallicity in solar units: 
\begin{equation}
E(B-V) = 0.21  \  Z\  \biggr({\Sigma_{ion}\over 10^{51} s^{-1} kpc^{-2}}\biggl)^{0.61}, 
\end{equation}

Although non--parametric (Spearman and Kendall) rank tests indicate that the data in 
Figure~\ref{fig12} deviate from the null 
(uncorrelated) hypothesis by 5.7~$\sigma$, the spread around the mean trend is still 
significant. The 90\% boundary is located at (-0.4~dex, $+$0.5~dex) along the vertical axis. 
We will see in section~A.4 that the large spread in Figure~\ref{fig12} has a small impact on the main 
conclusions of this Appendix.  That said, it should be important to keep in mind that although the trends 
have a physical base in the combination of the Schmidt Law with the extinction--gas~column~density correlation (Heckman, T.M. 2004, private communication; Tremonti, C.A., 2006, private communication), 
individual variations remain important. 

Equations~A2 and A4 will be used in combination with equation~A1 in the following sections to 
provide a heuristic description of the relation between ionizing photon rates (star formation rates) and monochromatic dust emission in our regions, in order to understand the broad trends 
observed in the data.  Thus, equations~A2 and A4 are derived from data and will be used to 
explain data, and the potential of circular argument should be addressed. The data in Figure~\ref{fig12} are completely independent of the data presented in Figures~\ref{fig3}, \ref{fig4}, and \ref{fig7}. The color 
excess E(B$-$V) is derived from ionized gas emission, while the monochromatic dust emission is 
mainly due to stellar continuum emission. Equation~A1 also requires independent 
assumptions on the underlying stellar population, and the dust geometry relative to 
both the stellar population and the ionized gas \citep[see next section,][]{calz94}. Therefore,  the 
use of the  extinction--ionizing~photons correlation of equations~A2 and A4 is unlikely to 
automatically enforce a fit to the data in  Figures~\ref{fig3}, \ref{fig4}, and \ref{fig7}. Finally, we 
will see in section~A.4 that equations~A2 and A4 are not necessary to account for 
the observed data for Pa$\alpha$ LSDs above 
$\sim$3$\times$10$^{38}$~erg~s$^{-1}$~kpc$^{-2}$.

\subsection{Model Parameters}

For F$_S$($\lambda$) in equation~A1, we adopt the stellar population models   
of Starburst99  \citep[2005 update; ][]{leit99}. Since we are exploring the infrared emission of 
both HII knots and star--formation--dominated galaxies, we consider both bursts of star formation and 
constant star formation. 
Burst models are explored in the age range 0--10~Myr, as older ages do not produce enough 
ionizing photons to provide Pa$\alpha$--detectable HII knots in our data. As bursts age, the number of ionizing photons produced by the massive, short--lived  stars, decreases, while the UV--optical stellar luminosity (the main contributor to the infrared emission) remains comparatively high, since it is contributed by lower--mass, longer--living stars. Thus, aging stellar populations produce high 
luminosity ratios L(IR)/L(Pa$\alpha$). For constant star formation,  
the representative case of a 100~Myr--duration model is considered.  The difference 
between this model and a much younger case is the increase of optical and infrared light in the older model, due to the 
accumulation of previous--generations, low--mass, long--lived stars. The light from these older stars 
contributes to L(IR), but not to the ionizing photons budget. The net effect is that 
the 100~Myr constant star formation model produces about twice 
as much infrared emission as a, e.g.,  6~Myr old case, for the same dust opacity. The variation 
between the 100~Myr and 1~Gyr models is about 6\%.  For the stellar IMF, we adopt 
the Starburst99 default, which consists of two power laws, with slope $-$1.3 in the range 
0.1--0.5~M$_{\odot}$ and slope $-$2.3 in the range 0.5--120~M$_{\odot}$.  The SFRs derived 
from this IMF are a factor 1.59 smaller than those derived from a Salpeter IMF in the range 
0.1--100~M$_{\odot}$, for the same number of ionizing photons.

For the wavelength dependence of the dust attenuation in equation~A1, 
A($\lambda$)=k($\lambda$)\ E(B$-$V), we adopt the starburst obscuration curve of \citet{calz01} 
\citep[see, also, ][]{calz94,meur99,calz00}, which prescribes an attenuation of the stellar 
continuum a factor 0.44 that of the emission lines.  

The fraction of the infrared luminosity that emerges in the IRAC 8~$\mu$m and MIPS 24~$\mu$m bands, 
L(8)/L(IR) and L(24)/L(IR),  is  from the model of \citet{drai06}, which updates  the model of 
\citet{li01}. At solar metallicity, the two papers provide consistent prescriptions for our case. 
The fraction of infrared light emitted in either 8~$\mu$m or 24~$\mu$m band
 increases non--linearly as a function of the starlight intensity (Figure~15 of \citet{drai06}); the ratios are 
flat at the low intensity end (single photon heating of dust)  and rise sharply at higher intensities, 
 producing  
`hotter' infrared SEDs, i.e., peaked at shorter wavelengths. At solar metallicity, the dependence of 
L(8)/L(IR) as a function of the starlight intensity is much shallower than that of L(24)/L(IR); the 
former varies by 50\% over the full range of intensities analyzed here, while 
the latter changes by more than a factor of 10. The results of \citet{drai06} are expressed 
as a  function of the parameter U, the starlight intensity in units of the solar neighborhood value. 

We  relate the typical stellar radiation field in our HII~knots to U using the individual HII regions detected 
in the HST Pa$\alpha$ image of the nearby (and, thus, better resolved), 
actively star--forming galaxy NGC6946. From the image we measure the sizes, defined as 
the radius of the circular area containing 80\% of the Pa$\alpha$ emission in each region,  and 
line emission values of the HII regions. Total luminous energies are derived by associating 
the range of stellar models described above to the measured line intensities, and energy densities  
are derived by dividing for the measured volumes. By extrapolating the observed quantities, we find 
that burst  models with mass 10$^6$~M$_{\odot}$ and ages $<$6~Myr   or  
constant star formation models with SFR=1~M$_{\odot}$~yr$^{-1}$ 
(L$_{Pa\alpha, corr}\lesssim$10$^{40}$~erg~s$^{-1}$), with  radii R$\approx$100~pc,  
 produce average starlight intensities U$\sim$400--1200. The factor $\sim$3 variation in U accounts 
 for two uncertainties in our derivation: (1)  star forming regions have hotter SEDs than 
 the local interstellar medium \citep{math83},  and (2) our stellar models and regions' sizes cover 
 a range of values/characteristics. The presence of dust absorption within the HII region and  the 
conversion of some of the energy to free--free emission also contribute to the uncertainty (and some 
decrease) in U.  However, the results described in the next section are fairly insensitive 
 to the actual value of U, within the range of 3--4 uncertainty detailed above.  The relation  
 between number of ionizing photons and starlight intensity derived above suggests that, for 
 constant radii of the HII regions, the 
 transition from single--photon heating to thermal equilibrium heating for the 24~$\mu$m emission 
 begins to occur at emission line intensities S$_{Pa\alpha, corr}\approx$10$^{39}$~erg~s$^{-1}$~kpc$^{-2}$. 

The impact of metallicity variations  is explored by considering how the infrared 
emission changes under the two conditions that the regions/galaxies have solar metallicity 
(Z=Z$_{\odot}$) and one-tenth solar metallicity (Z=1/10~Z$_{\odot}$). The latter value is at the 
low end of the metallicity range  
in our sample; our goal is, indeed, to model the lower envelope to the data. The metallicity enters in 
two components of equation~A1.  The first component is the stellar population model, for which we use 
two metallicity values: solar and 1/5 solar (Starburst99 does not provide models for our default 
1/10~Z$_{\odot}$, and 1/5~Z$_{\odot}$ is the closest metallicity value for the stellar populations). 
The second component is the dependence of the color excess on 
metallicity, which we describe in a linear fashion via equation~A4.  Additional metallicity--dependent 
ingredients are the conversion from ionizing photon rate to hydrogen nebular line luminosity (at the $\sim$15\% level; see section~3.2), and the fraction of infrared light emerging in the 
8~$\mu$m and 24~$\mu$m bands. We use, again, the model of \citet{drai06} for the latter, with the 
assumption that  variable metallicities have a major impact on the mass fraction of the polycyclic 
aromatic hydrocarbons in the dust mixture\footnote{Based on \citet{drai07}, a factor 9--10 
difference in the mass fraction of small PAH is a reasonable assumption for metallicities 
between Z$\sim$0.1 and Z$>$0.5.}; this, in turn, affects the dust emission in the mid--infrared 
bands, mainly  the IRAC 8~$\mu$m band, which results predominantly from single--photon--heating 
at low values of the starlight intensity. For reference, in the low--U regime, L(24)/L(IR) changes 
by about 60\% and L(8)/L(IR) changes by an order of magnitude for a factor 10 variation in the 
abundance of small PAH; the differences become negligible at the high--U end \citep{drai06}.  

\subsection{General Model Trends}

In the extreme case that we have no dust in our region, equation~A1 will be null, and there will 
be no relation between SFR (or ionizing photon rate) and dust emission. In the presence of 
even small amounts of dust, however, the relation established by equation~A1 will produce a general 
trend, with details that depend on a host 
of parameter assumptions. At the lowest levels of dust extinction, the relation between L(IR) and 
the number of ionizing photons is non-linear, with an asymptotic exponent of 1.64 for our 
equations~A1--A4;  a linear relation  is established only 
once the integral on the right--hand--side of  equation~A1 corresponds to most of the stellar energy. 
In other words, once the region contains enough dust that most of the stellar energy is absorbed and 
re--emitted in the IR, larger numbers of ionizing photons will linearly correspond to larger IR 
luminosities. 

The impact of the stellar population parameters on the characteristics of the  
 8~$\mu$m and 24~$\mu$m emission  has already been shown in Figures~\ref{fig9}--\ref{fig11}.  Stellar 
populations undergoing constant star formation since 100~Myr  in the range SFR=4$\times$10$^{-5}
$--4~M$_{\odot}$~yr$^{-1}$ have equivalent characteristics to bursts of star formation of constant 
age 4~Myr and mass in the range M=10$^3$--10$^8$~M$_{\odot}$. 
For the burst models, the impact of age variation is investigated 
at constant cluster mass (10$^6$~M$_{\odot}$) and in the two cases of constant or age-variable 
(equation~A2)
color excess. The general effect of increasing the age is to move the model curves towards smaller 
values of the ionizing photon rates for roughly constant IR emission, almost independently of 
assumptions on the variation (or constancy) of the color excess.

The introduction of the age--variable extinction (via equation~A2), however, attempts to mimic
 the observation that aging HII regions tend to be  less dust extincted than the 
younger HII regions, whether because the more evolved populations have shed the native 
cocoon or have migrated away from it through secular motions \citep{mayy96}. This also introduces 
a modest dependence of the IR SED and the fraction of infrared light emerging in the 
8~$\mu$m and 24~$\mu$m bands on age, via changes in the starlight intensity that aging 
HII regions produce.  However, the model does not include additional effects, such as the 
expansion of HII regions as they evolve, that will affect the IR SED via the decrease of the dust 
temperature.

Not surprisingly, metallicity variations have a strong effect on the observed mid--IR emission as 
a function of the number of ionizing photons (Figure~\ref{fig13}).  For the 24~$\mu$m emission, most 
of the effect comes from the fact that the ISM is more transparent at lower metallicities. For the 
8~$\mu$m emission, roughly equal contributions are given by the more transparent medium and by 
the decreased mass fraction of low--mass PAH molecules in the dust mixture (Figure~\ref{fig13}, 
left). 

The fiducial model for the IR SED as a function of starlight intensity \citep[from][]{drai06} is 
compared with the more extreme assumption that the IR SED is constant for all analyzed 
regions/galaxies. For the assumption L(24)/L(IR)=const=0.3 (Figure~\ref{fig13}, right), 
the model line has a different curvature from our fiducial 
one, starting with higher Log(S$_{24 \mu m}$) values  at low Pa$\alpha$ LSD, 
overshooting the datapoints in the LSD range Log(S$_{Pa\alpha, corr}$)=39--40, 
and converging to a slope of unity at higher LSD values.  Our fiducial model 
appears to better reproduce the slope of $\sim$1.23 of the best fit to the datapoints (in the log-log 
diagram), and this slope is entirely due to the relation between L(24)/L(IR) and the starlight 
intensity. For the L(8)/L(IR) ratio, the change between a constant IR SED and a starlight--intensity--dependent SED is small, as the ratio changes by about 50\% in the full range of LSD  
under consideration in this work. 

\subsection{Model Uncertainties and Limitations}

The simple assumptions on stellar populations and dust geometry made in the 
previous section lead to a number of limitations. Furthermore, the data show in some cases 
(e.g., Figure~\ref{fig12}) significant dispersion around the mean trend. The impact of 
 considering different assumptions or including the full range (90th percentile) of data dispersion  is 
 briefly discussed here.
 
 The 90th percentile region for the E(B$-$V) versus Pa$\alpha$ LSD correlation is shown in 
 Figure~\ref{fig12} by dotted lines that enclose 90\% of the data around the mean correlation. The same 
 region is reproduced on the 24~$\mu$m versus Pa$\alpha$ LSDs in Figure~\ref{fig14} (left). We do not 
 show the case for the 8~$\mu$m LSD because the results are similar for both mid--IR 
 bands. The dispersion 
  in the color excess for fixed Pa$\alpha$ LSD has minor impact on our baseline conclusions for 
  the mid--IR emission, with significant impact (factor $\sim$6 peak--to--peak variation) concentrated 
  towards the low--intensity, and low 
  extinction, regions. This effect is readily understood by recalling that once the color excess  
  E(B$-$V)$>$0.5~mag,  
  over 90\% of the UV emission is  converted by dust into IR emission. Thus, the dependence of 
  E(B$-$V) on the 
  Pa$\alpha$ LSD for S$_{Pa\alpha, corr}\gtrsim$10$^{39}$~erg~s$^{-1}$~kpc$^{-2}$ has little 
  impact on the amount of IR emission produced for stellar SEDs that emit most 
  of their power in the UV. 
  
  The independence of the IR luminosity on the color excess for high
   Pa$\alpha$ LSDs is also shown for the case where  E(B$-$V) is independent of  
  S$_{Pa\alpha, corr}$ 
  (i.e., equations~A2--A4 are not applied).  As a simplified example of E(B$-$V) independent of  
  S$_{Pa\alpha, corr}$,   Figure~\ref{fig14}  (right) 
  shows the case of constant value E(B$-$V)=1 for all Pa$\alpha$ LSDs. As long 
  as the color excess is sufficiently large that most of the UV stellar light is attenuated by dust, 
  the model's trend for high luminosity regions does not change. However, when extended to 
  low LSDs, the model with constant E(B$-V$) tends to exceed the low metallicity datapoints, and 
  to  show a flatter slope than the best fit line through the high--metallicity data. The latter is due to 
  the models having reached the single photon heating for the 24~$\mu$m emission, where there 
  is a linear  correlation between S$_{24~\mu m}$ and S$_{Pa\alpha, corr}$. 

\begin{figure}
\figurenum{9}
\plotone{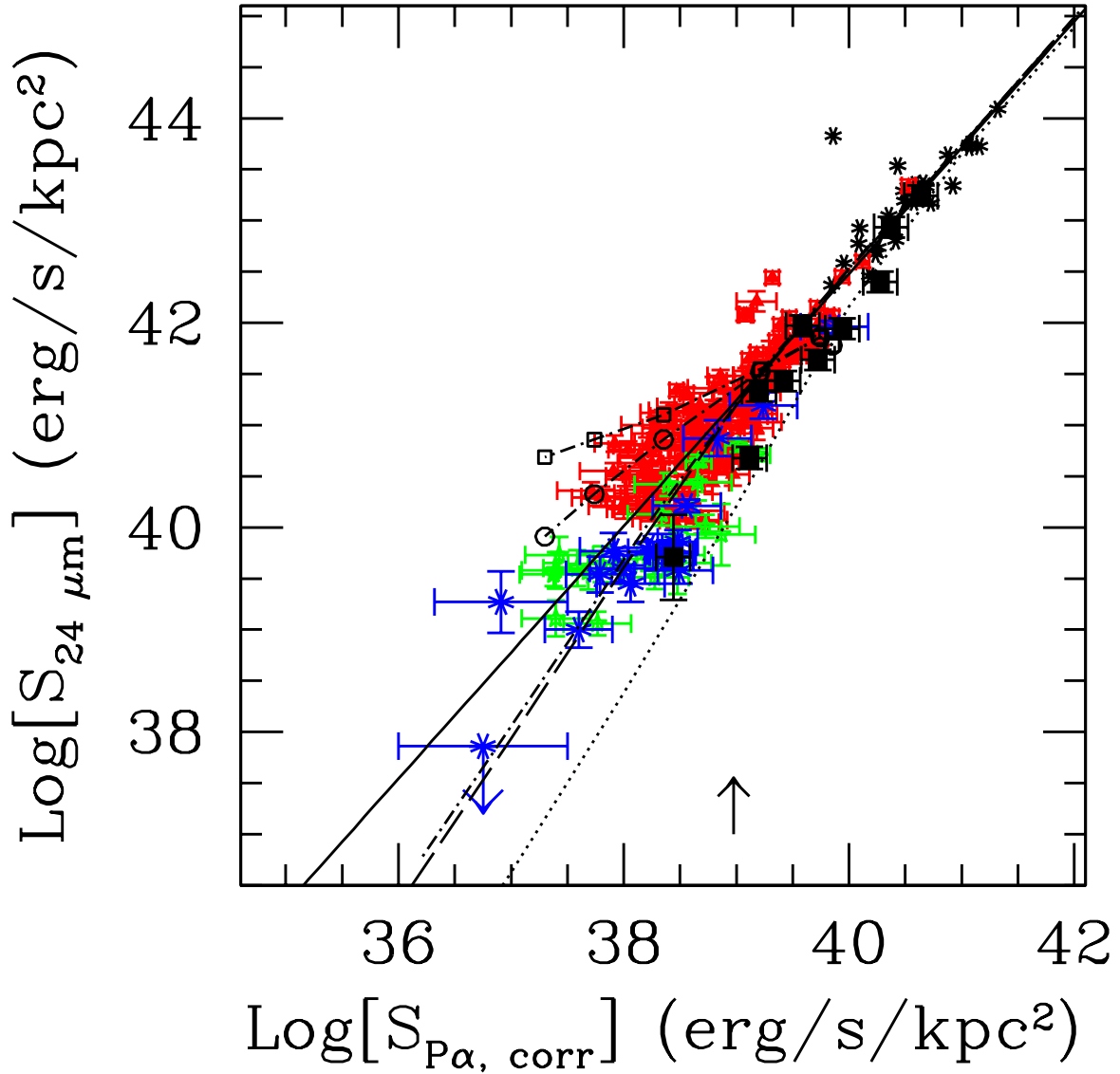}
\caption{Luminosity surface density at 24~$\mu$m as a function of the extinction--corrected  
Pa$\alpha$ LSD for the same datapoints as Figure~\ref{fig4} (after removal of 
the Sy~2 nuclei, the foreground and background sources, and the NGC5033 HII~knots data, leaving 
164 independent datapoints in the high metallicity HII~knot subsample). 
The continuous line shows the best linear fit through the high metallicity HII~knots, from 
Figure~\ref{fig4}. Models 
of infrared and ionized gas emission are superimposed on the data, for a variety of star formation 
histories, stellar population ages, and metallicity (see Appendix). Models with solar metallicity 
(Z=Z$_{\odot}$) ISM and stellar populations include: 
100~Myr--old constant star formation (SFR/area=4$\times$10$^{-5}$--4~M$_{\odot}$~yr$^{-1}$~kpc$^{-2}$, long--dash line);  instantaneous burst  with variable mass 
(10$^3$--10$^8$~M$_{\odot}$~kpc$^{-2}$) and color--excess, and 
constant age of 4~Myr (dot--dashed line); instantaneous bursts  with constant mass
 (10$^6$~M$_{\odot}$~kpc$^{-2}$) and variable age, and both variable color excess (dot--dashed line with  empty circles)  and constant color excess (E(B$-$V)=2~mag, dot--dashed line with empty 
 squares). The circle and square symbols mark the population ages, right--to--left: 0.01, 2, 4, 6, 8, 10~Myr.  
 The dotted line marks a 1/10~Z$_{\odot}$ model of constant star formation over the past 100~Myr.  
The upward--pointing arrow marks the approximate luminosity where the transition between single--photon 
heating and thermal equilibrium heating  for the dust begins to occur. 
\label{fig9}}
\end{figure}

\begin{figure}
\figurenum{10}
\plotone{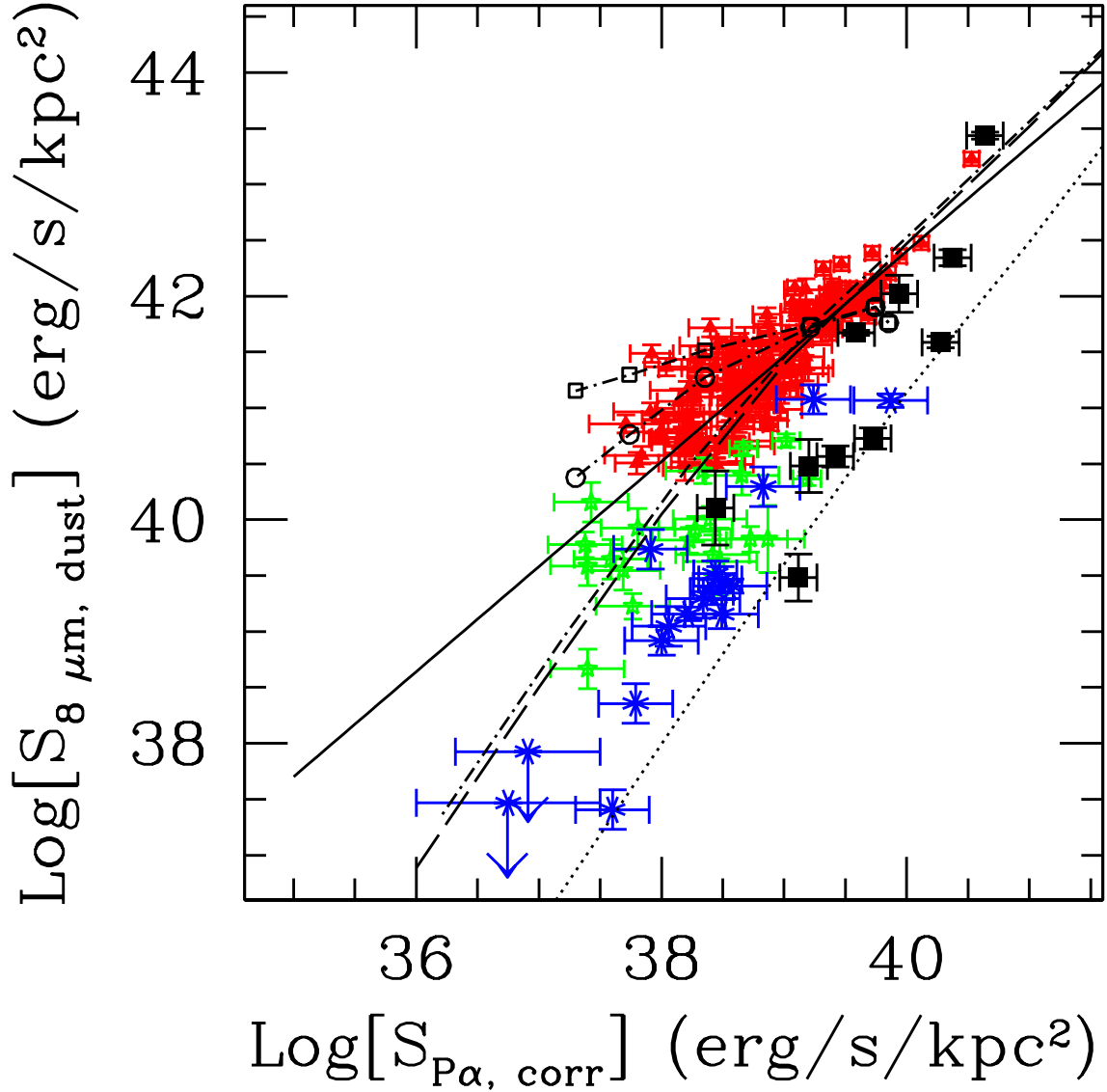}
\caption{Luminosity surface density at 8~$\mu$m  as a
function of the extinction--corrected Pa$\alpha$ LSD for the same regions/galaxies as Figure~\ref{fig9} 
(minus the LIRGs).  The continuous line is the best linear fit through the high metallicity HII~knots, as in Figure~\ref{fig3}. Models of infrared and ionized gas emission are the same as Figure~\ref{fig9}.
\label{fig10}}
\end{figure}

\begin{figure}
\figurenum{11}
\plotone{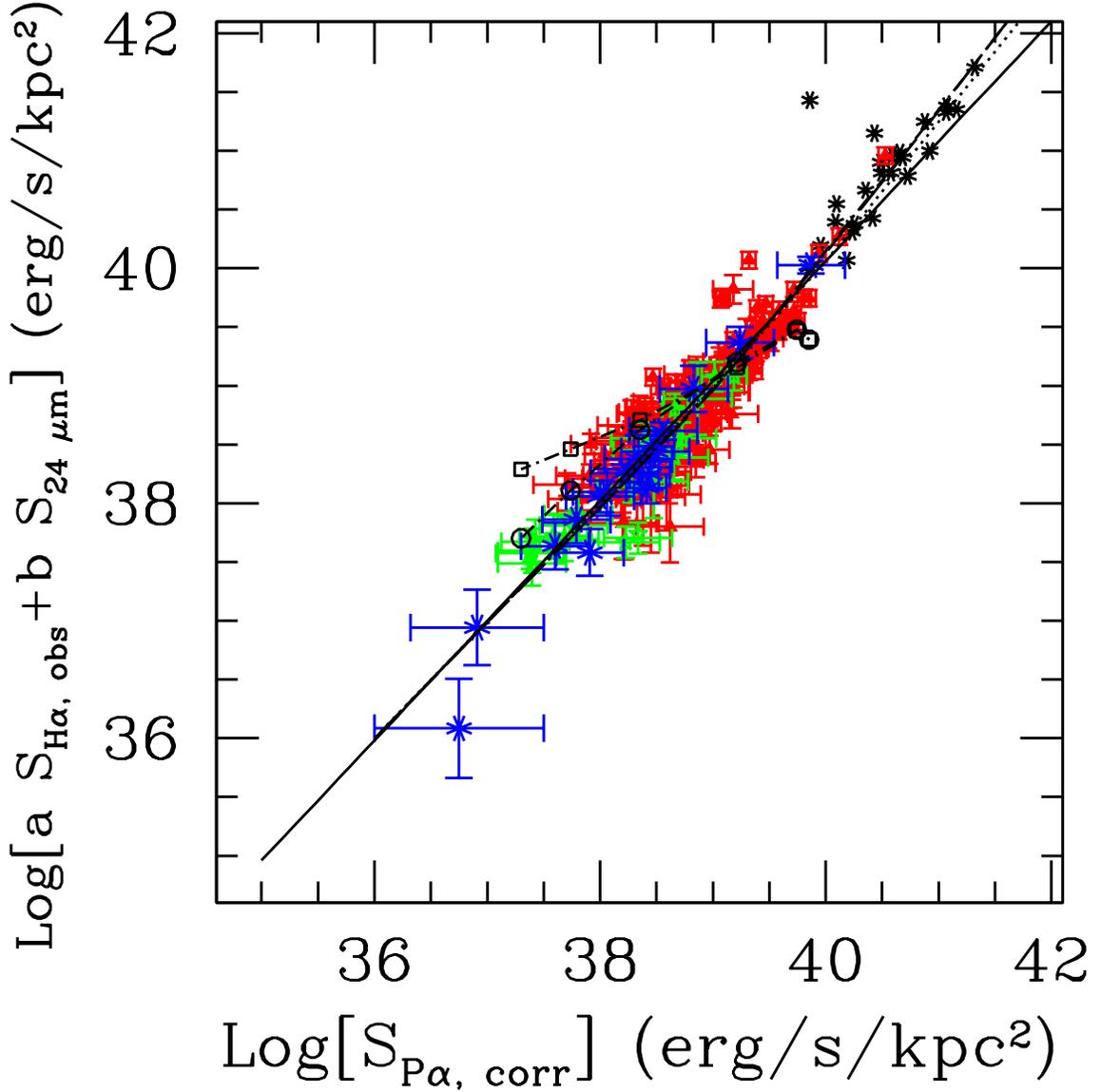}
\caption{Combined--optical/mid--IR LSD  as a
function of the extinction--corrected 
Pa$\alpha$ LSD for the same data as Figure~\ref{fig9}. The continuous line is the best linear fit through the high metallicity HII~knots, as in Figure~\ref{fig7}. Models of infrared and ionized gas emission are the same as Figure~\ref{fig9}.
\label{fig11}}
\end{figure}

\begin{figure}
\figurenum{12}
\plotone{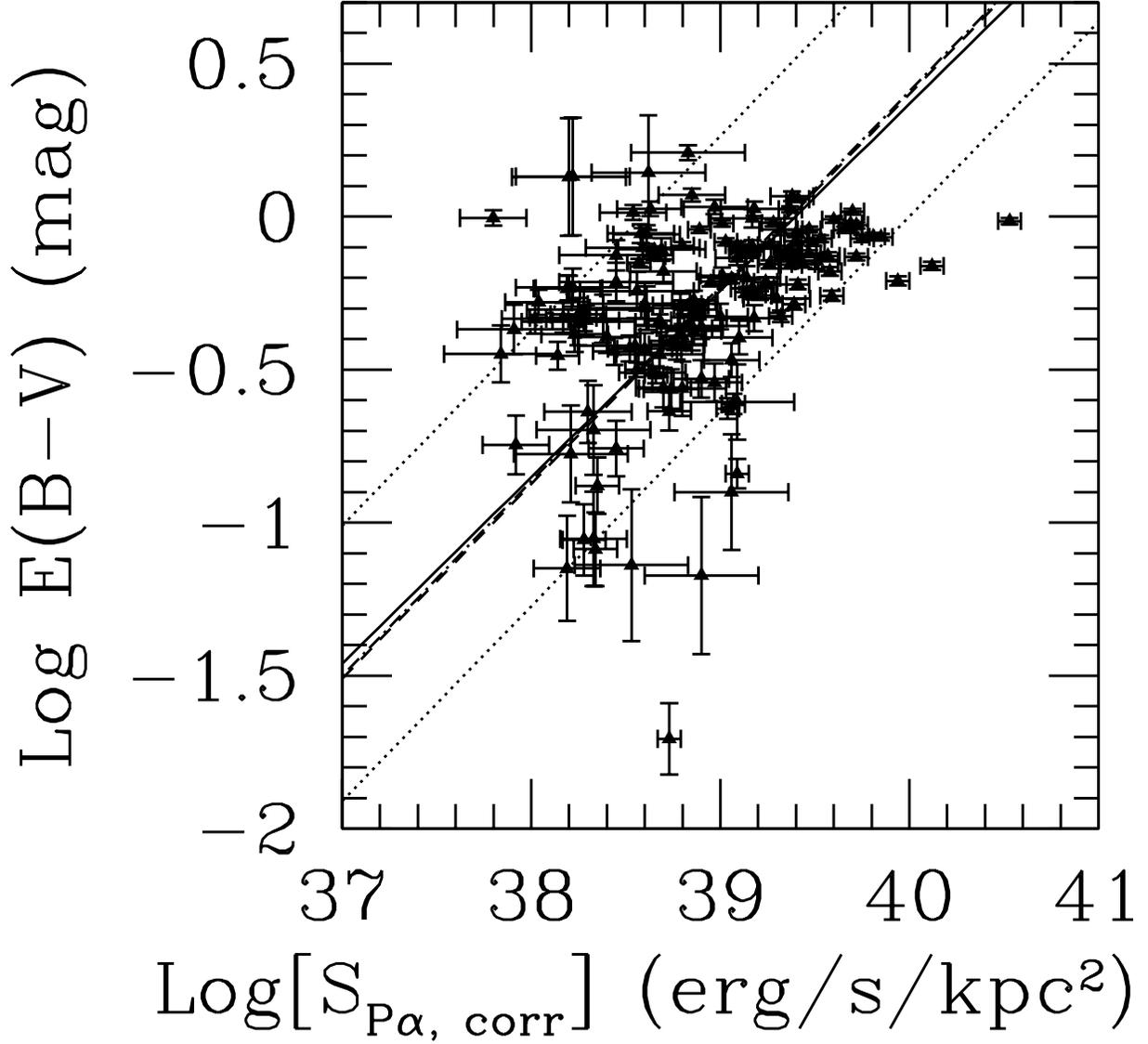}
\caption{The color excess E(B$-$V), measured from the H$\alpha$/Pa$\alpha$ ratio (section~3.2), 
as a function of  the Pa$\alpha$ LSD, for the 164 HII knots in the high metallicity subsample.
The best bi-linear fit through the data (continuous line) has slope $0.61$ (equation~A2). The dotted 
lines mark the region containing 90\% of the datapoints around the best fit line. 
The dash line is from equation~A3, where the slope is kept at the fixed 
value 0.64 \citep[from the Schmidt Law in NGC5194,][]{kenn06}.
\label{fig12}}
\end{figure}

\begin{figure}
\figurenum{13}
\plottwo{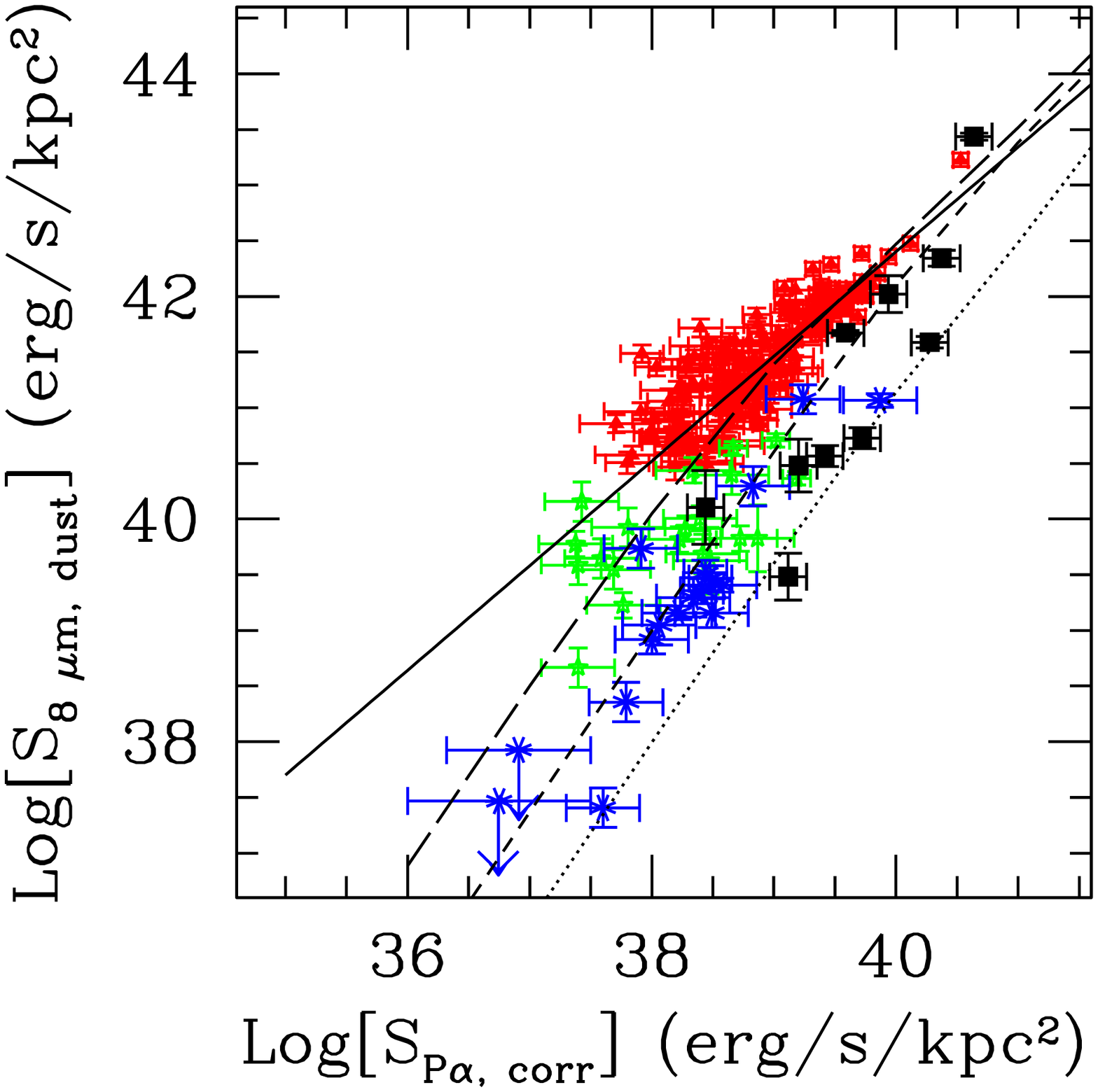}{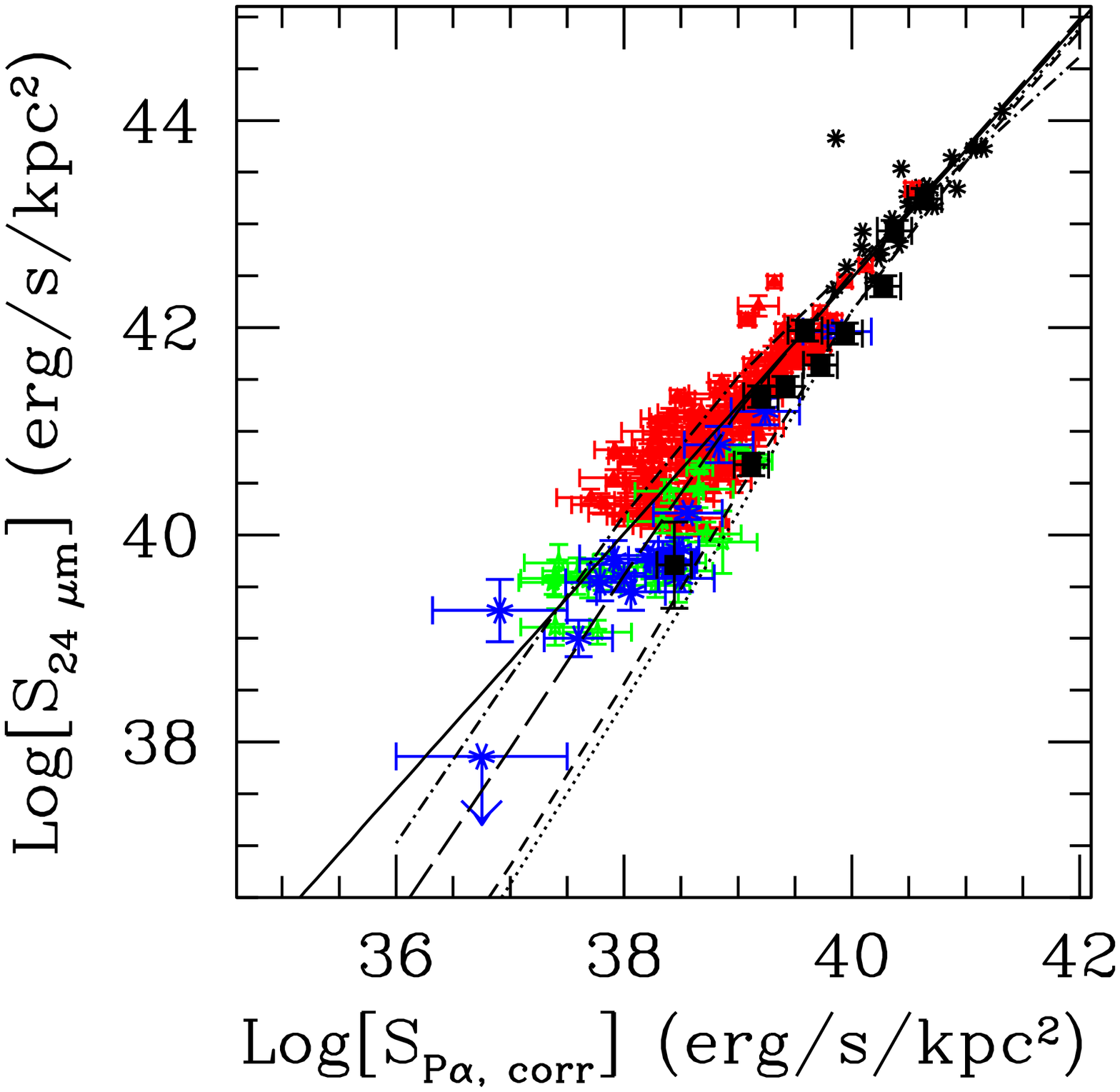}
\caption{Models of dust absorption and emission for constant star formation  
populations and variable metallicity are compared with data, for the 8~$\mu$m (left) and the 24~$\mu$m 
(right) LSD as a function 
of the Pa$\alpha$ LSD. Symbols for the HII~knots, the local  
starbursts, and the LIRGs are  as in Figures~\ref{fig9}--\ref{fig10}. Straight continuous lines 
are the best fits through the high--metallicity HII~knots, similar to the lines in Figures~\ref{fig3} and 
\ref{fig4}.  All models are for a 100~Myr~old  constant star formation population, and 
include: solar metallicity (Z=Z$_{\odot}$, long--dash line); 1/10~Z$_{\odot}$ and standard low--mass PAH molecules fraction (short--dash line); 1/10~Z$_{\odot}$ and depleted low--mass PAH molecules fraction \citep[dotted line][]{drai06}.  For the 24~$\mu$m--versus--Pa$\alpha$ plot, the effect of variations 
in the IR SED are also explored; in particular, our default assumption for the shape of the IR SED 
as a function of the starlight intensity \citep{drai06} is compared with 
the extreme assumption that the IR SED is constant, i.e., L(24)/L(IR)=const=0.3 (dot-dash line). 
\label{fig13}}
\end{figure}

\begin{figure}
\figurenum{14}
\plottwo{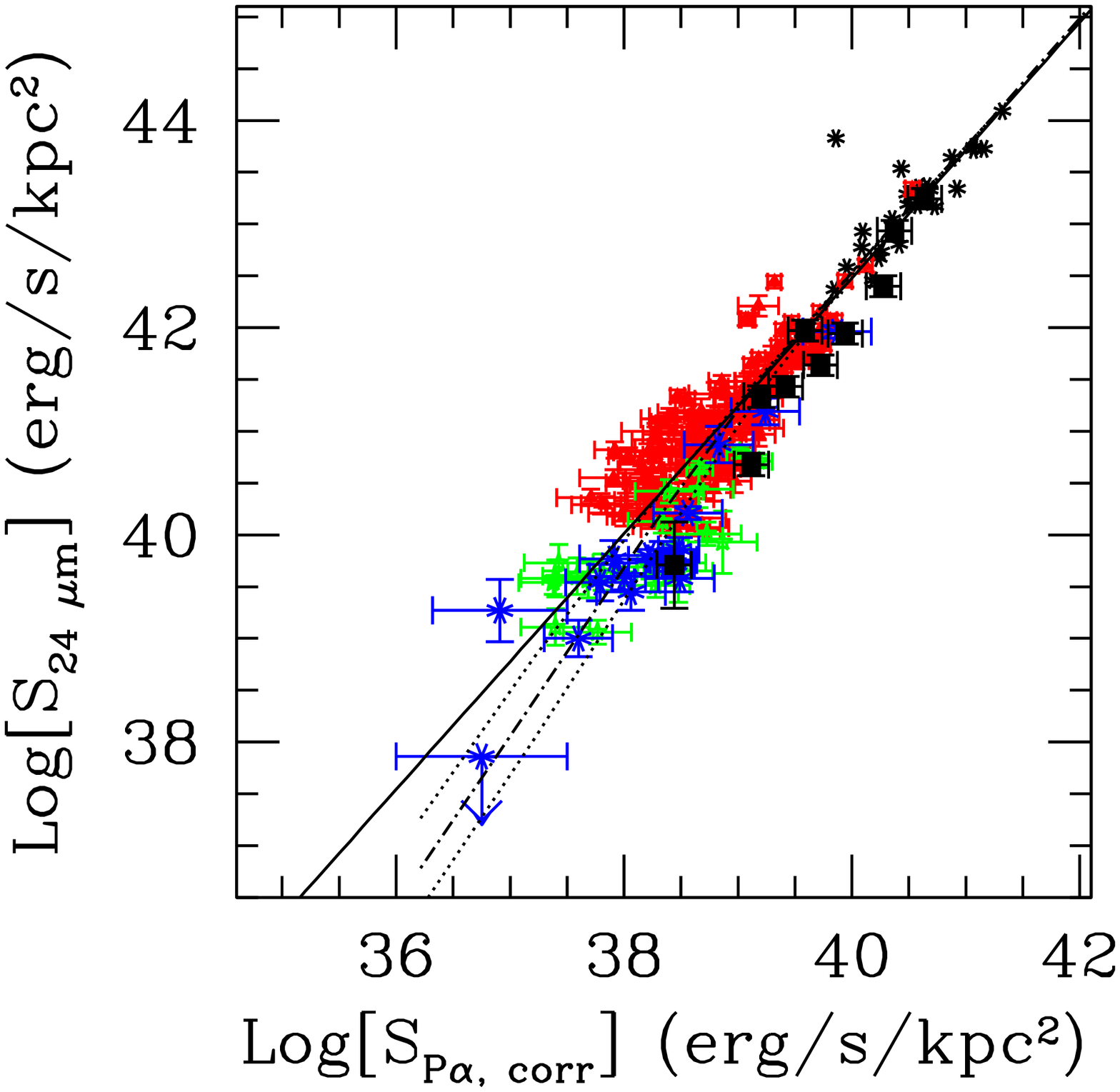}{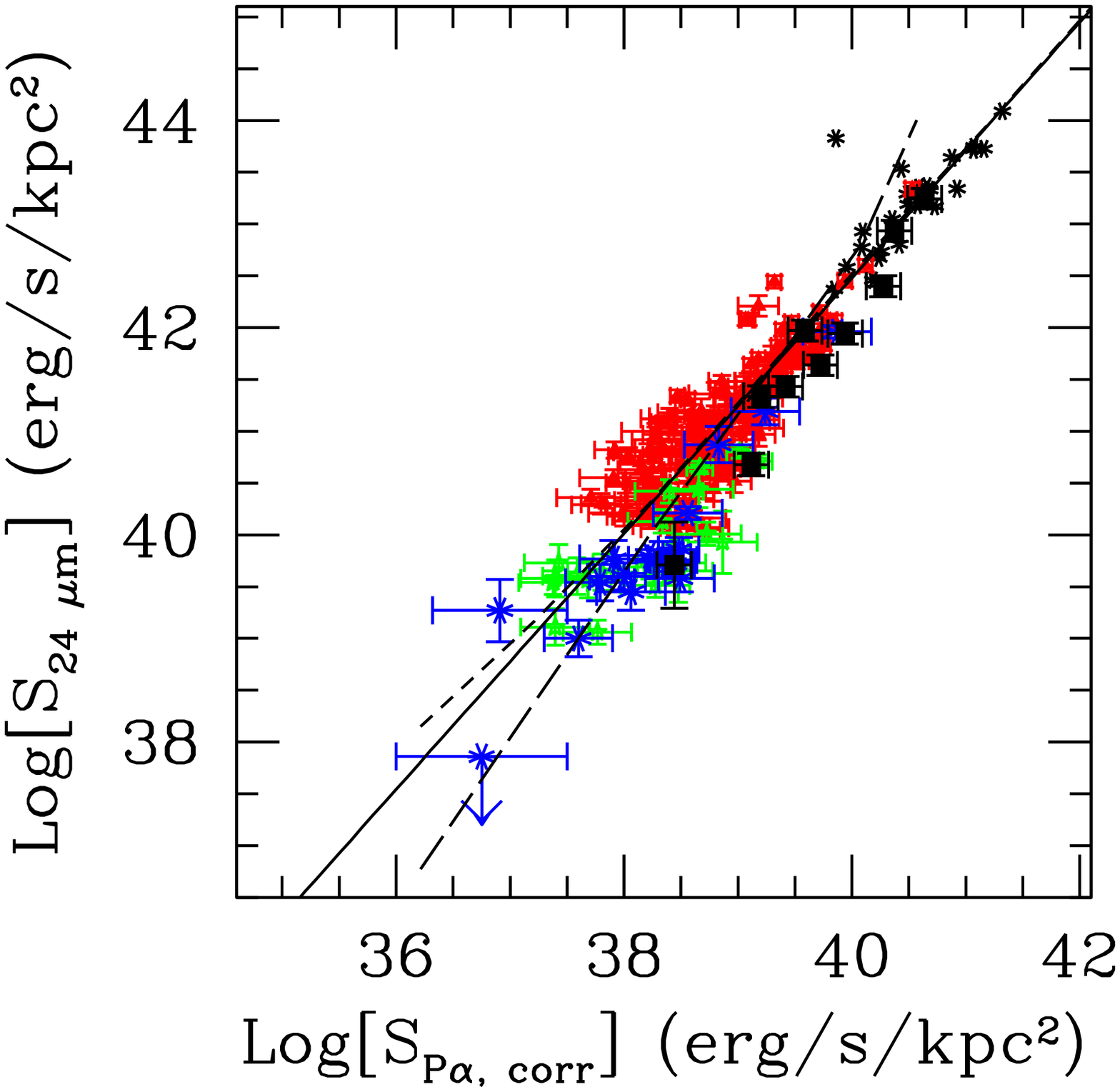}
\caption{The 24~$\mu$m LSD as a function 
of Pa$\alpha$  together with the best fit line (continuous 
straight line, Figure~\ref{fig4}). {\bf Left.}  
The fiducial model of a dusty 4~Myr~old  stellar cluster with increasing mass 
(dot-dashed line) is bracketed by the dispersion curves derived from including 
in equation~A1 the 90 percentile region of the E(B$-$V)--versus--S$_{Pa\alpha, corr}$ 
correlation (dotted lines and Figure~\ref{fig12}). {\bf Right}  Model lines for a 4~Myr~old 
stellar cluster of increasing mass, and two assumptions for the dust distribution that are 
different from our baseline model: (1) foreground dust geometry and constant E(B$-$V)=1 as a 
function of S$_{Pa\alpha, corr}$  (short--dash line), and (2) a homogeneously mixed dust--star geometry with no differential extinction between gas and 
stars, and variable E(B$-$V) according equation~A2 (long--dash line).   
\label{fig14}}
\end{figure}

Age variations among different regions have been discussed in section~7, and they effectively  
account for the spread around the mean trend for the infrared--versus--Pa$\alpha$ LSDs 
(Figures~\ref{fig9}--\ref{fig11}). Here we discuss the effect of adopting  different dust models than 
our default one. 

Most non--foreground dust geometries, e.g., mixtures of dust and 
gas/stars, will have the effect of `hiding' from direct detection a proportionally  larger fraction of the stellar 
emission than foreground geometries, 
thus increasing the observed infrared emission, and  S$_{24 \mu m}$ and S$_{8 \mu m,\ dust}$, relative to the ionizing photon rate (S$_{Pa\alpha, corr}$) that can be recovered with simple extinction 
correction methods (e.g., from H$\alpha$/Pa$\alpha$). Figure~\ref{fig14} (right)  shows the 
effect of replacing our foreground dust assumption with an homogeneously mixed dust/stars 
geometry. The new model follows closely our baseline model up to LSDs 
S$_{Pa\alpha, corr}\sim$3$\times$10$^{39}$~erg~s$^{-1}$~kpc$^{-2}$, and deviates upward of the 
baseline model for higher LSDs. This is the regime where the dust opacity is such that our standard 
procedure does not recover completely the number of ionizing photons and S$_{Pa\alpha, corr}$ 
is underestimated. If we were to allow for dusty 
cores with arbitrarily high attenuation values in our HII~knots, the mixed model would deviate from the baseline model at arbitrarily low Pa$\alpha$ LSDs, thus at least partially accounting  for the dispersion 
of the datapoints around the mean trend. 

The competition of the dust with the gas for absorption of the ionizing photons will become 
increasingly more effective as the density of the star forming regions increases, as in the case of 
LIRGs and brighter infrared galaxies. \citet{rigb04} proposes that the lack of high excitation 
infrared fine structure lines in infrared galaxies is due to much of the massive star formation 
occurring in ultracompact HII regions, where such competition would be significant. 
\citet{dopi06} have modeled the behavior of such regions, confirming that the infrared
continuum should get warmer, but also show the potential to suppress the emission
lines substantially.
  
\begin{figure}
\figurenum{15}
\plotone{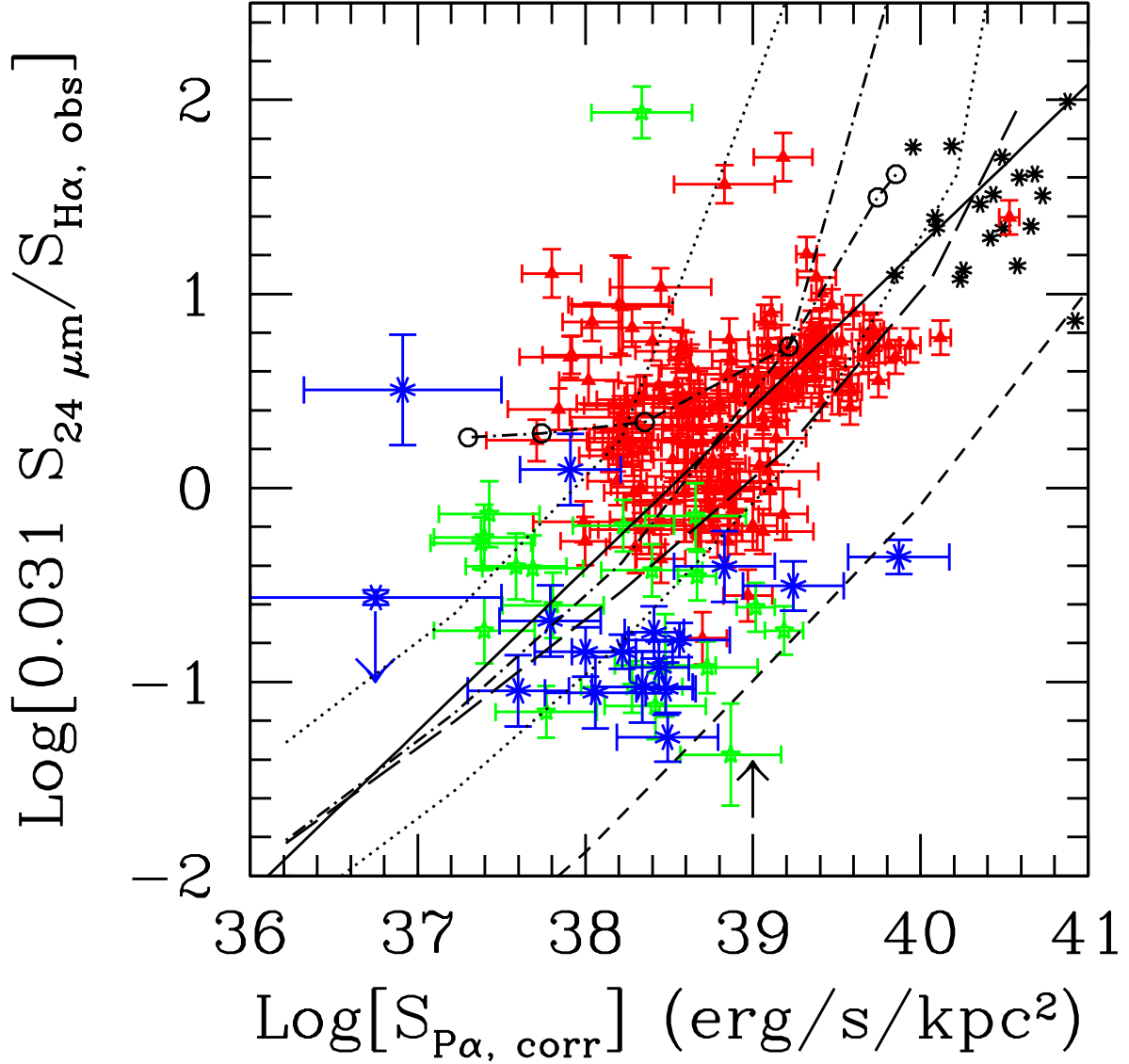}
\caption{The ratio of the 24~$\mu$m to the observed H$\alpha$ LSD as a function 
of Pa$\alpha$, for the HII~knots and LIRGs (Figure~\ref{fig9}), compared with  models.  
The datapoints are correlated with a $\sim$7~$\sigma$ significance; the continuous  
line is the best bi-linear  fit through the data, and has slope $0.83$. Solar metallicity 
models include (Figure~\ref{fig9}):  instantaneous burst  with variable mass 
(10$^3$--10$^8$~M$_{\odot}$) and color--excess, and constant age of 4~Myr (dot--dashed line); 
instantaneous burst  with constant mass (10$^6$~M$_{\odot}$), variable age and variable 
color excess (dot--dashed line with empty circles marking, right--to--left: 0.01, 2, 4, 6, 8, 10~Myr). 
The dotted lines mark the upper and lower boundaries to the 90-percentile region from Figure~\ref{fig12}. 
The long-dash line is the model of homogeneous dust--stars mixture, with no differential extinction 
between gas and stars, from Figure~\ref{fig14}.
The short-dash line is the 1/10th solar metallicity model, and marks to lower envelope to the datapoints. 
The upward--pointing arrow marks the approximate luminosity where the transition between single--photon 
heating and thermal equilibrium heating  for the dust begins to occur (Figure~\ref{fig9}). 
\label{fig15}}
\end{figure}

One of the main assumptions  in our analysis  is  that the `typical' HII knot in our 
 sample is described by a single mean stellar population. Realistically, variations are expected, not only 
 from galaxy to galaxy, but also from region to region. Under most circumstances, each of our 
 apertures will include multiple stellar populations covering a range of ages \citep{calz05}. 
 Furthermore, the dust covering factor may change from population to population within each 
 region, with older stellar populations often located in areas of lower extinction, or located further 
 away from the dust they heat.  The ratio of the 
24~$\mu$m LSD to the observed H$\alpha$ LSD, which by 
construction exacerbates any effect of dust extinction (Figure~\ref{fig15}) can provide insights 
into this effect. The ratio 24~$\mu$m/H$\alpha_{obs}$ spans more than two orders of
magnitude for our data, with values between $\sim$0.04 and $\sim$10.  The 
S$_{24 \mu m}$/S$_{H\alpha, obs}$ ratio is correlated (with a 7~$\sigma$ significance for a 
Spearman non--parametric test) with the Pa$\alpha$ LSD, with slope $0.83\pm0.03$ in a
log--log plane. This is a different way of expressing the correlation between 
star formation and dust extinction already shown in Figure~\ref{fig12}.  

The model lines for solar metallicity are within the general locus of the data--points at low values 
of the Pa$\alpha$ LSD, especially when the 90--percentile region from Figure~\ref{fig12} is included 
(Figure~\ref{fig15}).  The brightest regions in Pa$\alpha$ LSD, and most of the LIRGs, tend to be more consistent with a 
model where stars and dust are homogeneously mixed and there is no differential extinction between 
gas and stars \citep[as assumed in all our baseline models, from][]{calz94}. 
This is not unexpected, as strongly star forming regions are in general more heavily enshrouded 
in dust  than less active regions \citep{gold02}.  In addition, in this regime, small errors in the 
measurement of the (faint) H$\alpha$ line can produce large deviations of the data. 
The moderate discrepancy between the baseline models and data at the high luminosity 
end seen in Figure~\ref{fig15}  has negligible  impact on our results for the mid--IR LSDs as a 
function of the Pa$\alpha$ LSD (section~6): in this regime, over 80\% of the 
UV light is re-processed by dust into the infrared, and large fluctuations on the stellar light observed 
directly (including the nebular hydrogen lines) produce only small changes of L(IR).

Despite the potential shortcomings of our assumptions, the ability of the models to describe 
the overall trends of the data enables us to dissect the individual `ingredients' that produce the 
observed correlations to better understand their origin.

\clearpage

\begin{table}
\caption{Characteristics of the Sample Galaxies. \label{tab1}}
\scriptsize
\begin{tabular}{lllrrrrrrr}
\tableline\tableline
Name & Morph.\tablenotemark{a} & Nucleus\tablenotemark{a} & v$_H$\tablenotemark{a} & E(B$-$V)$_G$\tablenotemark{a} & Dist.\tablenotemark{b} & M$_V$\tablenotemark{c} & 12$+$log(O/H)\tablenotemark{d} & \# Regions\tablenotemark{e} & HST ID\tablenotemark{f} \\
       &     &  & (km s$^{-1}$)  & & (Mpc)  & & & & \\
\tableline
\multicolumn{9}{c}{High Metallicity Galaxies}\\
\tableline
NGC0925 & SAB(s)d   &        & 553 & 0.081 & 9.12 &$-$20.33  & 8.24--8.78 & 10 & 7919, 9360\\  
NGC1512 & SB(r)ab   &        & 898 & 0.011 & 10.5  &$-$19.90 & 8.37--9.05 & 3 & 9360\\  
NGC2403 & SAB(s)cd  &        & 131 & 0.043 & 3.5   &$-$19.68 & 8.31--8.81  & 8 & 7919\\  
NGC2841 & SA(r)b    & Sy1/LIN& 638 & 0.017 & 9.8   &$-$21.12 & 8.52--9.19  & 5 & 7919\\ 
NGC2976 & SAc       &        &   3  & 0.074  & 3.5 &$-$17.97 & 8.30--8.98 & 8 & 7919\\   
NGC3184 & SAB(rs)cd &        & 592 & 0.018 & 11.10 &$-$20.46 & 8.48--9.14 & 6 & 9360\\  
NGC3198 & SB(rs)c   &        & 663 & 0.013 & 13.68 &$-$20.90 & 8.32--8.87 & 9 & 9360\\  
NGC3351 & SB(r)b    &        & 778 & 0.030 & 10.1  &$-$20.48 & 8.60--9.22 & 2 & 9360\\  
NGC3627 & SAB(s)b   & Sy2/LIN& 727 & 0.035 & 8.7   &$-$21.17 & 8.49--9.10  & 6 & 7919\\  
NGC3938 & SA(s)c    &        & 809 & 0.023 & 12.2  &$-$20.10 & 8.35--9.07 & 9 & 9360\\  
NGC4125 & E6        & LIN    &1356 & 0.020 & 21.4  &$-$21.89 & 8.58--9.21 & 1 & 9360\\  
NGC4559 & SAB(rs)cd &        & 816 & 0.019 & 11.1  &$-$20.84 & 8.25--8.79  & 6 & 7919\\  
NGC4569 & SAB(rs)ab & Sy/LIN &$-$235&0.049 & 16.58 &$-$21.90 & 8.56--9.19 & 5 & 9360\\  
NGC4625 & SAB(rs)m  &        & 609 & 0.019 & 9.17  &$-$17.53 & 8.27--9.04 & 6 & 9360\\  
NGC4736 & (R)SA(r)ab& Sy2    & 308 & 0.019 & 5.3   &$-$20.59 & 8.31--9.01 & 5 & 9360\\  
NGC4826 & (R)SA(rs)ab& Sy2   & 408 & 0.044 & 5.6   &$-$20.63 & 8.59--9.24 & 8 & 9360\\  
NGC5033 & SA(s)c    & Sy1.9  & 875 & 0.012 & 13.28 &$-$20.87 & 8.27--8.90 & 9 & 9360\\  
NGC5055 & SA(rs)bc  & LIN    & 504 & 0.019 & 7.82  &$-$21.08 & 8.42--9.13 & 9 & 9360\\  
NGC5194 & SA(s)bc   & Sy2.5  & 463 & 0.037 & 8.2   &$-$21.43 & 8.54--9.18 & 43 &7237\\  
NGC5195 & SB0\_1     & LIN    & 465 & 0.038 & 8.2   &$-$19.99 & 8.28--8.83 & 1 & 9360\\  
NGC5866 & S0\_3      & LIN    & 672 & 0.014 & 12.1  &$-$20.52 & 8.43--9.02 & 3 & 9360\\  
NGC6946 & SAB(rs)cd &        &  48 & 0.365 & 5.0   &$-$21.11 & 8.40--9.04 & 9 & 9360\\  
NGC7331 & SA(s)b    & LIN    & 816 & 0.097 & 15.1  &$-$22.14 & 8.40--9.05 & 8 & 9360\\  
\tableline
\multicolumn{9}{c}{Medium Metallicity Galaxies}\\
\tableline
NGC1705 & SA0-      &        & 632 & 0.009 & 5.1  &$-$16.29  & 8.20--8.43  & 3 & 7919\\  
IC2574  & SAB(s)m   &        &  57 & 0.039 & 2.8   &$-$17.33 & 7.94--8.26 & 6 & 9360\\   
NGC4236 & SB(s)dm   &        &   0 & 0.015 & 4.45 &$-$19.12  & 8.07--8.56 & 3 & 9360\\  
IC4710  & SB(s)m    &        & 739 & 0.095 & 7.8   &$-$17.96 & 8.11--8.62  & 7 & 7919\\  
NGC6822 & IB(s)m    &        &$-$57& 0.253 & 0.47 &$-$14.97  & 8.04--8.67 & 3 & 7919\\  
\tableline
\multicolumn{9}{c}{Low Metallicity Galaxies}\\
\tableline
Ho II   & Im        &        & 142 & 0.035 &  3.5 &$-$17.25  & 7.68--8.07  & 3 & 9360\\  
DDO053  & Im        &        &  20 & 0.040 & 3.56 &$-$13.68  & 7.77--8.13  & 6 & 9360\\  
Ho IX   & Im        &        &  46 & 0.085 & 3.3  &$-$13.67   & 7.61--7.98 & 3 & 9360\\  
M81DwB  & Im        &        & 347 & 0.085 &  6.5 &$-$14.20 & 7.85--8.20 & 4 & 9360\\  
NGC5408 & IB(s)m    &        & 506 & 0.074 & 4.8  &$-$17.22 & 7.81--8.23 & 3 & 9360\\  
\tableline
\multicolumn{9}{c}{Discarded Galaxies}\\
\tableline
NGC0024 & SA(s)c    &        & 554 & 0.021 & 8.8  &$-$19.01  & 8.32--8.92  &\nodata&9360\\  
NGC1291 &(R\_1)SB(l)0/a&  & 839 & 0.014 & 10.5 &$-$21.69  & 8.42--9.01 &\nodata&9360\\ 
M81DwA   & I ?       &          &  113  & 0.022 & 3.55  & \nodata  &  7.34--7.64 & \nodata&9360\\  
NGC3034 & I0        &        & 203 & 0.170 & 5.2  &$-$20.51  & 8.36--9.09   &\nodata&7919\\ 
NGC4631 & SB(s)d    &        & 606 & 0.018 & 8.5  &$-$21.59  & 8.13--8.76  &\nodata&9360\\  
DDO154    & IB(s)m     &       &  374 & 0.010 & 4.3   & $-$14.73   &  7.54--8.02 & \nodata&9360\\  
\tableline
\end{tabular}

\tablenotetext{a}{Galaxy morphologies, nuclear activity, heliocentric
velocity, and foreground Galactic color excess are from the NASA/IPAC
Extragalactic Database (NED). The nuclear activity is reported for those cases where  
non--thermal emission dominates the emission in the central region. 
The Galactic color excesses, E(B$-$V)$_G$, are from \citet{schl98}.}
\tablenotetext{b}{Adopted distances, in Mpc, as derived by \citet{mast05}. Note that 
changes in the galaxy distances do not affect the analysis, which is 
based upon luminosity surface densities (luminosity/area).}
\tablenotetext{c}{Galaxy's absolute magnitude,  
based on V$_T^0$ from the RC3 \citep{deva91}, as retrieved from NED. For NGC6822 
the B$_T^0$ and for M81DwB the m$_B^0$ values have been used.}
\tablenotetext{d}{Oxygen abundances; the two columns of values are from \citet{mous07}, 
see section~2 for a brief description of their derivation.}
\tablenotetext{e}{Number of independent regions measured in the galaxy.}
\tablenotetext{f}{HST Program ID for the NICMOS observations used in this 
paper.}

\end{table}

\clearpage
\thispagestyle{empty}
\setlength{\voffset}{-5mm}
\begin{table}
\caption{Galaxies Measurements. \label{tab2}}
\scriptsize
\begin{tabular}{lllllll}
\tableline\tableline
Name& log S$_{Pa\alpha, corr}$\tablenotemark{a} & log S$_{H\alpha, obs}$\tablenotemark{a} & [NII]/H$\alpha$\tablenotemark{b} & A$_V$\tablenotemark{a} & log S$_{8 \mu m,\ dust}$\tablenotemark{a} & log S$_{24 \mu m}$\tablenotemark{a} \\
           & (erg~s$^{-1}$~kpc$^{-2}$) &   (erg~s$^{-1}$~kpc$^{-2}$) & &  (mag) & (erg~s$^{-1}$~kpc$^{-2}$) 
 & (erg~s$^{-1}$~kpc$^{-2}$) \\
\tableline
\multicolumn{7}{c}{High Metallicity Galaxies}\\
\tableline
NGC0925&38.44$\pm$0.18&39.18$\pm$0.08&0.24&0.45$\pm$0.19&41.02$\pm$0.11&40.48$\pm$0.11\\
NGC1512&38.55$\pm$0.06&38.80$\pm$0.02&0.47&1.94$\pm$0.06&41.09$\pm$0.06&40.75$\pm$0.06\\
NGC2403&38.73$\pm$0.15&39.27$\pm$0.08&0.62&1.07$\pm$0.17&41.24$\pm$0.11&40.35$\pm$0.11\\
NGC2841&37.83$\pm$0.30&38.03$\pm$0.30&0.55A&2.12$\pm$0.43&40.43$\pm$0.08&40.08$\pm$0.08\\
NGC2976&38.50$\pm$0.15&39.12$\pm$0.08&0.30&0.83$\pm$0.17&41.12$\pm$0.10&40.59$\pm$0.10\\
NGC3184&38.59$\pm$0.18&38.72$\pm$0.11&0.44&2.31$\pm$0.21&40.97$\pm$0.11&40.69$\pm$0.11\\
NGC3198&38.56$\pm$0.18&38.42$\pm$0.08&0.36&3.16$\pm$0.19&41.20$\pm$0.10&41.09$\pm$0.10\\
NGC3351&39.24$\pm$0.06&39.48$\pm$0.04&0.40&1.97$\pm$0.07&41.72$\pm$0.06&41.68$\pm$0.06\\
NGC3627&38.92$\pm$0.23&39.18$\pm$0.04&0.55A&1.94$\pm$0.23&41.72$\pm$0.11&41.33$\pm$0.11\\
NGC3938&37.99$\pm$0.30&38.76$\pm$0.06&0.54&0.38$\pm$0.31&41.08$\pm$0.08&40.46$\pm$0.08\\
NGC4125\tablenotemark{c}&37.13U &36.00U &\nodata& \nodata &40.15$\pm$0.06&39.95$\pm$0.06\\
NGC4559&38.51$\pm$0.30&39.33$\pm$0.08&0.32&0.24$\pm$0.31&41.30$\pm$0.08&40.75$\pm$0.08\\
NGC4569&38.29$\pm$0.30&38.84$\pm$0.04&0.50A&1.05$\pm$0.30&41.56$\pm$0.08&41.35$\pm$0.08\\
NGC4625&38.06$\pm$0.11&38.95$\pm$0.08&0.46&0.03$\pm$0.14&40.79$\pm$0.08&40.39$\pm$0.07\\
NGC4736&37.96$\pm$0.18&38.88$\pm$0.04&0.50\tablenotemark{d}&0.01$\pm$0.18&41.96$\pm$0.08&41.54$\pm$0.08\\
NGC4826&39.28$\pm$0.06&39.53$\pm$0.02&0.55\tablenotemark{d}&1.97$\pm$0.06&42.12$\pm$0.06&41.69$\pm$0.06\\
NGC5033\tablenotemark{e}&38.57$\pm$0.08\nodata&\nodata&\nodata&41.88$\pm$0.06&41.35$\pm$0.06\\
NGC5055&38.61$\pm$0.11&39.28$\pm$0.08&0.50\tablenotemark{d}&0.68$\pm$0.14&41.74$\pm$0.06&41.17$\pm$0.06\\
NGC5194&38.95$\pm$0.06&39.26$\pm$0.04&0.50&1.78$\pm$0.07&41.54$\pm$0.06&41.19$\pm$0.06\\
NGC5195\tablenotemark{c}&36.89U &37.83U &\nodata& \nodata &42.39$\pm$0.06&41.35$\pm$0.06\\
NGC5866&37.64$\pm$0.30&38.19$\pm$0.11&0.55A&1.04$\pm$0.32&40.04$\pm$0.06&39.47$\pm$0.06\\
NGC6946&39.44$\pm$0.06&39.77$\pm$0.04&0.56&1.72$\pm$0.07&42.34$\pm$0.06&42.20$\pm$0.06\\
NGC7331&39.08$\pm$0.11&39.14$\pm$0.08&0.32&2.54$\pm$0.14&41.69$\pm$0.08&41.20$\pm$0.08\\
\tableline
\multicolumn{7}{c}{Medium Metallicity Galaxies}\\
\tableline
NGC1705&38.56$\pm$0.11&39.48$\pm$0.11&0.09&0.03$\pm$0.16&40.09$\pm$0.06&40.06$\pm$0.06\\
IC2574&37.55$\pm$0.30&38.44$\pm$0.08&0.12&0.14$\pm$0.31&39.17$\pm$0.18&39.18$\pm$0.18\\
NGC4236&37.57$\pm$0.30&38.15$\pm$0.08&0.17\tablenotemark{f}&1.09$\pm$0.31&39.76$\pm$0.18&39.42$\pm$0.18\\
IC4710&37.81$\pm$0.30&38.56$\pm$0.08&\nodata\tablenotemark{g}&0.57$\pm$0.31&39.64$\pm$0.11&39.10$\pm$0.11\\
NGC6822&37.85$\pm$0.30&38.74$\pm$0.11&\nodata\tablenotemark{g}&0.12$\pm$0.32&39.46$\pm$0.30&39.13$\pm$0.30\\
\tableline
\multicolumn{7}{c}{Low Metallicity Galaxies}\\
\tableline
Ho~II&38.37$\pm$0.30&39.31$\pm$0.04&0.10A&0.01$\pm$0.30&40.01$\pm$0.13&40.06$\pm$0.13\\
DDO053&37.97$\pm$0.30&38.91$\pm$0.11&0.04&0.01$\pm$0.32&39.33$\pm$0.18&39.82$\pm$0.18\\
Ho~IX\tablenotemark{h}&36.12$\pm$0.73&36.32$\pm$0.2&0.05A& \nodata  &36.92U&38.14$\pm$0.30\\ 
M81DwB&37.76$\pm$0.18&38.60$\pm$0.08&0.04&0.29$\pm$0.19&39.03$\pm$0.11&39.15$\pm$0.11\\
NGC5408&38.26$\pm$0.30&39.20$\pm$0.04&0.02\tablenotemark{f}&0.01$\pm$0.30&39.29$\pm$0.06&40.05$\pm$0.06\\
\end{tabular}

\tablenotetext{a}{Average luminosity surface density and extinction in the central 
$\sim$50$^{\prime\prime}\times$50$^{\prime\prime}$ of each galaxy 
($\sim$144$^{\prime\prime}\times$144$^{\prime\prime}$ for NGC 5194). The 
extinction--corrected and  the `observed' (not extinction corrected) 
values are listed for Pa$\alpha$ and H$\alpha$, respectively.}
\tablenotetext{b}{[NII]($\lambda$ 6584 \AA)/H$\alpha$, as obtained from spectroscopy \citep{mous07}, 
and before convolution with the narrowband filters' transmission curves. An `A' after a value 
indicates an adopted (non measured) value, for those cases where a spectrum is not available or  
the available spectrum is dominated by a central non--thermal source. Adopted values come from 
galaxies of comparable metallicity to the target ones.}
\tablenotetext{c}{For NGC4125 and NGC5195,  the only detected sources are the central Sy2 nuclei; 
for these, only 8~$\mu$m and 24~$\mu$m emission is detected, while H$\alpha$ and Pa$\alpha$ are upper limits.}
\tablenotetext{d}{For NGC4736, NGC4826, and NGC5055, the [NII]/H$\alpha$ values are 
derived from the comparison of the HST and ground--based narrowband images. The HST 
narrowband filters centered on H$\alpha$ reject almost completely [NII], thus providing a reference 
for the ground--based images.}
\tablenotetext{e}{For NGC5033, no H$\alpha$ data are available.}
\tablenotetext{f}{Values from the spectrum of \citet{ho97} (NGC4236) and \citet{mase94} (NGC5408).}
\tablenotetext{g}{For IC4710 and NGC6822, no ratios  are available. However the [NII] 
contribution to the narrowband H$\alpha$ filter is negligible for these two galaxies, since both galaxies have low metallicity and  the narrowband filters  transmit less than 3\% and 4\% of the light from 
the 6548 \AA\  and 6584 \AA\ [NII] lines, respectively.}
\tablenotetext{h}{The Pa$\alpha$ emission for HoIX is the middle value between the upper 
limit measured from the HST/NICMOS image and the lower limit represented by the detected  
H$\alpha$ emission at zero extinction (section~4.2);  the emission at 8~$\mu$m for the galaxy is an upper limit.}
\end{table}

\clearpage
\setlength{\voffset}{0mm}

\begin{table}
\caption{Characteristics of the Starburst Galaxies. \label{tab3}}
\scriptsize
\begin{tabular}{lllrrr}
\tableline\tableline
Name & Morph.\tablenotemark{a} & v$_H$\tablenotemark{a} & E(B$-$V)$_G$\tablenotemark{a}  & R\tablenotemark{b} & 12$+$log(O/H)\tablenotemark{c} \\
       &     & (km s$^{-1}$)  & & (arcsec) &  \\
\tableline
UGCA292& ImIV-V & 308 & 0.016 & 5.1 & 7.2\\
SBS0335$-$052& BCG & 4043 &0.047& 4.1 & 7.3\\
HS0822$+$3542& BCG & 732 & 0.047 & 4.1 & 7.4\\
VIIZw403 & Pec. & $-$103 &0.036 & 6.1 & 7.7\\
UM461& BCD/Irr & 1039 & 0.018 & 5.1 & 7.8\\
Mrk1450 & Comp. &946 &0.011 & 5.1 & 8.0\\
IIZW40 & BCD & 789 & 0.820 & 11.2 & 8.1\\
NGC5253 & Im pec & 407 & 0.056 & 14.2 & 8.2\\
NGC2537& SB(s)m & 431 & 0.054 & 8.1 & 8.7\\
NGC2146& SB(a)ab & 893 & 0.096 & 13.8 & 8.4--9.0\\
\tableline
\end{tabular}
\tablenotetext{a}{Galaxy morphology, heliocentric
velocity, and foreground Galactic color excess are from the NASA/IPAC
Extragalactic Database (NED). The Galactic color excess,
E(B$-$V)$_G$, is from \citet{schl98}.}
\tablenotetext{b}{Radius, in arcseconds, of the region of active star formation, as measured 
in the NICMOS images.}
\tablenotetext{c}{Oxygen abundances, reproduced from Table~1 of \citet{enge05}. For UGCA292, the oxygen abundance is from \citet{Pilyu01}, for NGC5253  from \citet{mart97}; for NGC2146 the line 
ratios of \citet{ho97} have 
been converted to a range of possible oxygen abundances using the strong lines method of \citet{kewl02}.}

\end{table}

\begin{table}
\caption{Measurements of the Starburst Galaxies. \label{tab4}}
\scriptsize
\begin{tabular}{lrrr}
\tableline\tableline
Name & log S$_{Pa\alpha}$\tablenotemark{a} & log S$_{8 \mu m,\ dust}$\tablenotemark{a} & log S$_{24 \mu m}$\tablenotemark{a}  \\
             & (erg~s$^{-1}$~kpc$^{-2}$)  & (erg~s$^{-1}$~kpc$^{-2}$)  & (erg~s$^{-1}$~kpc$^{-2}$) \\
\tableline
UGCA292      &     38.44      &     40.11$\pm$0.33\tablenotemark{b}     &    39.71$\pm$0.42\\
SBS0335$-$052   &         39.59      &     41.67$\pm$0.02    &     41.97$\pm$0.10\\
HS0822$+$3542      &    39.12       &    39.48 $\pm$0.21    &     40.68$\pm$0.11\\
VIIZw403    &        39.2      &     40.48$\pm$0.24    &    41.34$\pm$0.11\\
UM461     &     39.42      &     40.57$\pm$0.09    &     41.43 $\pm$0.10\\
Mrk1450   &   39.72      &     40.73$\pm$0.09     &    41.64$\pm$0.10\\
IIZw40      &     40.28      &     41.59$\pm$0.05      &    42.4$\pm$0.10\\
NGC5253     &      40.37      &     42.34$\pm$0.07     &    42.93$\pm$0.10\\
NGC2537     &     39.94     &      42.02$\pm$0.16     &    41.94$\pm$0.10\\
NGC2146      &     40.64       &    43.44$\pm$0.03     &    43.24$\pm$0.10\\
\tableline
\end{tabular}
\tablenotetext{a}{The luminosity surface density at Pa$\alpha$, 8~$\mu$m, and 24~$\mu$m in 
the starburst regions, with radius listed in Table~\ref{tab3}. The Pa$\alpha$ LSDs are from 
the HST SNAP program 9360, and are only corrected 
for foreground Galactic extinction (Table~\ref{tab3}). Measurement uncertainties for S$_{Pa\alpha}$ 
are around 15\%--20\%. The LSDs in the Spitzer bands 
are `whole galaxy' measurements, corrected to infinite aperture, although in all cases the 
central starburst (measured in the NICMOS images) is the dominant contributor to the flux.} 
\tablenotetext{b}{The 8~$\mu$m emission from this galaxy is affected by a latent image from a 
previous observation. Every effort has been made to remove the contaminating latent image from 
the measurement, but the presence of some small remnant contamination cannot be excluded.}
\end{table}

\clearpage


\clearpage 

\clearpage 

\clearpage 

\clearpage 

\clearpage 

\clearpage 

\clearpage 

\clearpage 
\thispagestyle{empty}
\setlength{\voffset}{-22mm}

\clearpage 
\setlength{\voffset}{0mm}

\clearpage 

\clearpage 

\clearpage 

\clearpage 

\clearpage 
\thispagestyle{empty}
\setlength{\voffset}{-12mm}
\clearpage
\setlength{\voffset}{0mm}


\end{document}